\documentclass[12pt,twoside]{article}

\usepackage[whole]{bxcjkjatype}
\usepackage[height=8.85in,width=6.45in]{geometry}
\usepackage{fancyhdr}

\usepackage[utf8]{inputenc}
\usepackage[T1]{fontenc}
\usepackage{amsmath}
\usepackage{amssymb}
\usepackage{mathtools}
\numberwithin{equation}{section}
\allowdisplaybreaks

\usepackage{braket}
\usepackage[usenames,dvipsnames,svgnames,table]{xcolor}
\usepackage[colorlinks,citecolor=DarkGreen,linkcolor=FireBrick,urlcolor=FireBrick,
linktocpage,bookmarks=true,pdftitle={An Automated Generation of Bootstrap Equations for Numerical Study of Critical Phenomena},
pdfauthor={Mocho Go}]{hyperref}
\usepackage{cite}
\usepackage{graphicx}
\usepackage{tikz}
\usepackage{tikz-cd}
\usepackage{times}
\usepackage[scaled]{couriers}

\usepackage{bm}
\usepackage{subfig}

\usepackage{mdframed}

\usepackage{colonequals}
\usepackage{cleveref}
\usepackage[version=3]{mhchem}

\renewenvironment{figure}[1][]{
\begin{originalfigure}[#1]
	\begin{mdframed}[linecolor=black!0,backgroundcolor=black!1]
}{
	\end{mdframed}
\end{originalfigure}
}

\definecolor{identifiercolor}{rgb}{.4,.6,.56}
\definecolor{stringcolor}{gray}{0.5}
\definecolor{inactivecolor}{rgb}{0.15,0.15,0.5}
\usepackage{listings}
\newcommand{\code}[1]{\lstinline{#1}}
\newcommand{\ptext}[1]{\left(\text{#1}\right)}
\newcommand{\op}[1]{\mathcal{#1}}
\newcommand{\swap}{\leftrightarrow}
\newcommand{\abs}[1]{\left\lvert#1\right\rvert}
\newcommand{\Abs}[1]{\left\lVert#1\right\rVert}
\newcommand{\id}[0]{\operatorname{id}}
\newcommand{\Irr}[1]{\operatorname{Irr}\left(#1\right)}
\newcommand{\inv}[1]{\operatorname{inv}\braket{#1}}

\newcommand{\sud}[2]{^{#1}\!{}_{#2}}

\begin{document}
\pagestyle{fancy}
\fancyfoot{}
\fancyfoot[LE,RO]{\thepage}

\begin{titlepage}
	\begin{center}
		{\Large Doctoral Dissertation}

		\vskip 3cm

		{\Huge \bfseries An Automated Generation of Bootstrap Equations for Numerical Study of Critical Phenomena}

		\bigskip
		\bigskip

		Mocho Go

		\bigskip

		\begin{tabular}{ll}
		& Kavli Institute for the Physics and Mathematics of the Universe (WPI), \\
		& University of Tokyo, Kashiwa, Chiba 277-8583, Japan
		\end{tabular}

		\vskip 2cm
	\end{center}

	\noindent In this thesis, we introduce new tools for the conformal bootstrap, \code{autoboot} and \code{qboot}.
	Each tool solves a different step in the whole computational stack, and combined with an existing efficient tool \code{SDPB}
	which solves semidefinite programming (SDP), our tools make it easier to study conformal field theories
	with more complicated global symmetries and with more general spectra.
	In the introduction, we review how the conformal bootstrap method gives rich information about the theory at
	the fixed point of renormalization group, or in other words, the critical phenomena such as the Ising model at criticality.
	The following three sections focus on the theories behind the implementation of \code{autoboot} and \code{qboot},
	and the explicit implementation, freely available at \url{https://github.com/selpoG/autoboot/} and \url{https://github.com/selpoG/qboot/},
	is discussed in \autoref{sec:implementation}.
	We also have two applications in the last section, the Ising model and the $O(2)$ vector model in three dimensions,
	each of them has close relationship with a physical system in the real world.
\end{titlepage}

\null\vspace{5cm}
\begin{abstract}
	\noindent In this thesis, we introduce new tools for the conformal bootstrap, \code{autoboot} and \code{qboot}.
	Each tool solves a different step in the whole computational stack, and combined with an existing efficient tool \code{SDPB}
	which solves semidefinite programming (SDP), our tools make it easier to study conformal field theories
	with more complicated global symmetries and with more general spectra.
	In the introduction, we review how the conformal bootstrap method gives rich information about the theory at
	the fixed point of renormalization group, or in other words, the critical phenomena such as the Ising model at criticality.
	The following three sections focus on the theories behind the implementation of \code{autoboot} and \code{qboot},
	and the explicit implementation, freely available at \url{https://github.com/selpoG/autoboot/} and \url{https://github.com/selpoG/qboot/},
	is discussed in \autoref{sec:implementation}.
	We also have two applications in the last section, the Ising model and the $O(2)$ vector model in three dimensions,
	each of them has close relationship with a physical system in the real world.
\end{abstract}
\newpage

\setcounter{tocdepth}{2}
\tableofcontents
\newpage

\null\vskip 5cm
This thesis is based on a published paper \cite{autoboot} collaborated with Yuji Tachikawa,
and a single-authored paper \cite{qboot} to appear.
The code is freely available at \url{https://github.com/selpoG/autoboot/} and \url{https://github.com/selpoG/qboot/}.

\newpage
\section{Introduction}
\label{sec:intro}

\subsection{General Introduction}
\label{sec:intro_gen}
A conformal field theory (CFT) is a quantum field theory (QFT) with conformal symmetry,
which has Euclidean or Poincar\'{e} group as a subgroup and is enhanced by scale invariance.
A conformal transformation $x\to x'$ is an $x$-dependent local rescaling $\omega(x)>0$ with a rotation $\Lambda\sud{\mu}{\nu}(x)\in O(d)$
\begin{equation}
	\frac{\partial x'^\mu}{\partial x^\nu} = \omega(x)\Lambda\sud{\mu}{\nu}(x),
\end{equation}
and the conformal symmetry is a natural generalization of the Euclidean symmetry, which is a fundamental spacetime symmetry of a usual QFT.
The scale invariance arises, e.g., in the critical phenomena of lattice theory like the $O(N)$ vector model or the Ising model
on the $d$-dimensional lattice.
Rescaling the momentum cutoff $\Lambda$ of a QFT with a normalized kinetic term
\begin{equation}
	\op{H}=\partial_{\mu}\phi\partial^{\mu}\phi+\sum_i g^i\op{O}^i,
\end{equation}
where $\op{O}^i$ is some interaction term and $g^i$ is the coupling constant,
generates a nonlinear flow of Wilson's renormalization group (RG) \cite{Wilson:1974,Wilson:1983}:
\begin{equation}
	\frac{\partial g^i}{\partial\log\Lambda} = \beta^i(g).
\end{equation}
Around a fixed point of the flow which is defined by $\beta^i(g_*)=0$, the flow can be linearized as
\begin{equation}
	\beta^i(g) = -\gamma^i_j(g-g_*)^j,
\end{equation}
and diagonalizing $\gamma$ matrix gives
\begin{equation}
	\beta^a(g) = -(d-\Delta_a)(g-g_*)^a, \label{eq:rg_eigen}
\end{equation}
where $d-\Delta_a$ is the eigenvalue of the eigen operator $\op{O}^a$.
The critical point is a fixed point of the flow, and thus scale invariant.
In general, a scale invariant theory is also conformal invariant
as conjectured in \cite{Polyakov:1970xd} and discussed in \cite{Polchinski:1987dy,Nakayama:2013is,Dymarsky:2013pqa}.

It is well-known (or widely believed) that the RG fixed point of $\phi^4$ theory in 3 dimensions
\begin{equation}
	\op{H}=\partial_\mu\phi\partial^\mu\phi+m^2\phi^2+\lambda\phi^4, \quad Z=\int\op{D}\phi\ e^{-\int\! d^3x\ \mathcal{H}},
	\label{eq:ising_hamiltonian}
\end{equation}
the Ising model in 3 dimensions at the critical temperature
\begin{equation}
	H=-J\sum_{(ij)}\sigma_i\sigma_j-h\sum_i \sigma_i, \quad Z=\sum_{\sigma\colon \mathbb{Z}^3\to\set{1,-1}} e^{-\beta H},
\end{equation}
and the behavior of water at the critical point (more generally, the critical point of the liquid-vapor transition)
are described by the same CFT \cite{Pelissetto:2000ek}.
This property at the IR limit is called the critical universality, and it is conjectured that
any theory with the same symmetry and the same number of relevant operators
reaches the same CFT at the continuum limit regardless of its microscopic details.
The Ising model has $\mathbb{Z}_2$ symmetry which flips all spins and two relevant parameters $\beta,h$.
Other important examples are the $O(N)$ vector models, in the control of $O(N)$ symmetry which rotates all spins.
The $O(2)$ model defined by two component real scalar fields $(\phi_1,\phi_2)$ and the Hamiltonian
\begin{equation}
	\op{H}=\partial_\mu\phi_i\partial^\mu\phi^i+m^2\phi_i\phi^i+\lambda(\phi_i\phi^i)^2,
\end{equation}
is known as the XY universality class, and describes ferromagnets with easy-plane anisotropy and the superfluid transition of \ce{^{4}He}.
One aspect of the universality can be shown from the RG as follows.
One start from some UV theory
\begin{align}
	\op{H} &= \partial_{\mu}\phi\partial^{\mu}\phi+\sum_i g^a\op{O}^a \\
	&= \op{H}_*+\sum_i (g-g_*)^a\op{O}^a,
\end{align}
written in the diagonalized basis around a fixed point $g_*$, and chase the RG flow to get an IR theory.
Operators are classified by the sign of $d-\Delta_a$;
if $d-\Delta_a<0$, $\op{O}^a$ is called an irrelevant operator and $(g-g_*)^a$ goes to zero in the IR theory, and
if $d-\Delta_a>0$, $\op{O}^a$ is called a relevant operator and $(g-g_*)^a$ increases in the IR theory.
The critical phenomena occurs when the resulting IR theory $\op{H}_*$ is interacting,
and one needs to tune all (in general, finite number of) relevant coupling constants to reach $g_*$
(the irrelevant coupling vanishes in IR automatically).
This argument proves one aspect of the universality
that we can drop all irrelevant interactions and start from a theory only with relevant operators to study the criticality
(just what we did to write the simple Hamiltonian \eqref{eq:ising_hamiltonian} of the Ising model without
higher order couplings $\phi^6,\phi^8,\ldots$),
but does not show the reason that the fixed point $g_*$ or the spectrum $\set{d-\Delta_a}$ is uniquely determined
among field theories with different set of fundamental fields only by the symmetry.

The scaling dimensions $\Delta$ of operators in a CFT are related with the critical exponents\cite{Pelissetto:2000ek}.
The scaling dimensions are defined so that $\op{O}$ with scaling dimension $\Delta$ behaves as $\op{O}\to a^{-\Delta}\op{O}$ under rescaling $x\to ax$,
and consistent with RG \eqref{eq:rg_eigen};
if $\op{O}^a$ has scaling dimension $\Delta_a$, the coupling constant $g^a$ has scaling dimension $d-\Delta_a$.
Take a lattice theory with spin field $\phi$ around the critical temperature $T\approx T_c$.
The critical exponents $\alpha,\beta,\gamma,\delta,\eta,\nu$ are defined by the power law of the following quantities:
\begin{itemize}
	\item the specific heat: $C\sim\abs{t}^{-\alpha}$,
	\item the magnetization: $m\sim(-t)^{\beta}$, $m\sim\abs{h}^{1/\delta}$,
	\item the magnetic susceptibility: $\chi\sim t^{-\gamma}$,
	\item the correlation function: $\Braket{\phi(x)\phi(0)}-\Braket{\phi(0)}^2\sim\abs{x}^{2-d-\eta}$,
	\item the correlation length: $\xi\sim\abs{t}^{-\nu}$,
\end{itemize}
where $t=(T-T_c)/T_c$ is the reduced temperature, $h$ is the magnetic field coupling with $\phi$ \cite{Pelissetto:2000ek}.
Let the deformation of the Hamiltonian from the critical point be described by primary operators $\sigma$, $\epsilon$
as $\delta S=\int d^d x (h\sigma+t\epsilon)$.
From the dimensional analysis, the scalings for $t,h$ are $\Delta_t=d-\Delta_\epsilon$, $\Delta_h=d-\Delta_\sigma$
and the critical exponents are related with the scaling dimensions of $\sigma,\epsilon$ by \cite{Kos:2015mba}
\begin{align}
	\alpha &= \frac{d-2\Delta_\epsilon}{d-\Delta_\epsilon}, & \beta &= \frac{\Delta_\sigma}{d-\Delta_\epsilon}, \\
	\gamma &= \frac{d-2\Delta_\sigma}{d-\Delta_\epsilon}, & \delta &= \frac{d-\Delta_\sigma}{\Delta_\sigma}, \\
	\eta &= 2\Delta_\sigma-d+2, & \nu &= \frac{1}{d-\Delta_\epsilon}.
\end{align}

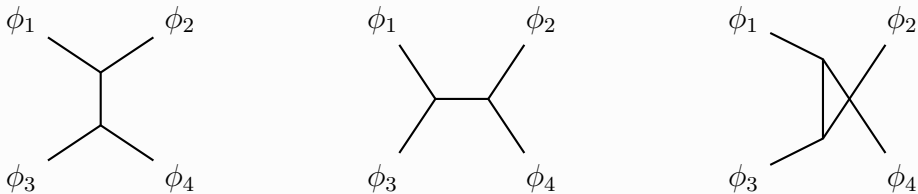
\begin{figure}[htpb]
	\centering
	\begin{tabular}{c}
		\begin{minipage}{0.3\hsize}
			\centering
			\begin{tikzpicture}
				\node (1) at (-30pt,+30pt) {$\phi_1$};
				\node (2) at (+30pt,+30pt) {$\phi_2$};
				\coordinate (a) at (0,+10pt);
				\coordinate (b) at (0,-10pt);
				\node (3) at (-30pt,-30pt) {$\phi_3$};
				\node (4) at (+30pt,-30pt) {$\phi_4$};
				\draw[thick] (1) -- (a) -- (b) -- (3);
				\draw[thick] (2) -- (a);
				\draw[thick] (b) -- (4);
			\end{tikzpicture}
		\end{minipage}
		\begin{minipage}{0.3\hsize}
			\centering
			\begin{tikzpicture}
				\node (1) at (-30pt,+30pt) {$\phi_1$};
				\node (2) at (+30pt,+30pt) {$\phi_2$};
				\coordinate (a) at (-10pt,0);
				\coordinate (b) at (+10pt,0);
				\node (3) at (-30pt,-30pt) {$\phi_3$};
				\node (4) at (+30pt,-30pt) {$\phi_4$};
				\draw[thick] (1) -- (a) -- (b) -- (2);
				\draw[thick] (3) -- (a);
				\draw[thick] (b) -- (4);
			\end{tikzpicture}
		\end{minipage}
		\begin{minipage}{0.3\hsize}
			\centering
			\begin{tikzpicture}
				\node (1) at (-30pt,+30pt) {$\phi_1$};
				\node (2) at (+30pt,+30pt) {$\phi_2$};
				\coordinate (a) at (0,+15pt);
				\coordinate (b) at (0,-15pt);
				\node (3) at (-30pt,-30pt) {$\phi_3$};
				\node (4) at (+30pt,-30pt) {$\phi_4$};
				\draw[thick] (1) -- (a) -- (b) -- (3);
				\draw[thick] (2) -- (b);
				\draw[thick] (a) -- (4);
			\end{tikzpicture}
		\end{minipage}
	\end{tabular}
	\caption{OPE in a four-point function}
	\label{fig:crossing}
\end{figure}

The observable quantities in a CFT are correlation functions $\Braket{\op{O}_1(x_1)\cdots\op{O}_n(x_n)}$;
the $S$ matrix is not well-defined in a CFT due to the Coleman-Mandula's theorem \cite{Coleman:1967ad, Haag:1974qh}.
We later show that all correlation functions are determined by the CFT data $\Set{\Delta_i,\lambda_{ijk}}$ in a unitary CFT,
and a four-point function $\Braket{\phi_1\phi_2\phi_3\phi_4}$ in a CFT
can be constructed from three-point functions, but in more than one way,
depending on how to group the four operators for the operator product expansions (OPEs):
$(\phi_1\phi_2)(\phi_3\phi_4)$, $(\phi_1\phi_3)(\phi_2\phi_4)$ or $(\phi_1\phi_4)(\phi_2\phi_3)$ shown in \autoref{fig:crossing}.
The bootstrap equation expresses the equality of the four-point function computed in these different decompositions,
and is one of the fundamental consistency conditions of a CFT.
The conformal bootstrap program aims to solve the CFT data non-perturbatively from the bootstrap equation
with the knowledge about the global symmetry of the theory and without a microscopic description such as a Lagrangian.

The bootstrap equation has been known for almost a third of a century, see e.g.~\cite{Dobrev:1975ru,Dobrev:1977qv};
for other early papers, we refer the reader to the footnote 5 of \cite{Poland:2018epd}.
It is particularly powerful in $d=2$ where the conformal group is infinite dimensional and generated by the Virasoro algebra,
and was already successfully applied for the study of 2d CFT in 1984 in the paper \cite{Belavin:1984vu}.
Its application to CFTs in dimension higher than two had to wait until 2008,
where the seminal paper \cite{Rattazzi:2008pe} showed that a clever rewriting into a form
where linear programming was applicable allowed us to extract detailed numerical information from the bootstrap equation.
The technique was rapidly developed by many groups and has been applied to many systems
(\cite{Rychkov:2009ij,Caracciolo:2009bx,Poland:2010wg,Rattazzi:2010gj,Rattazzi:2010yc,Vichi:2011ux,Poland:2011ey,Rychkov:2011et,
ElShowk:2012ht,Liendo:2012hy,ElShowk:2012hu,Beem:2013qxa,Kos:2013tga,Alday:2013opa,Gaiotto:2013nva,Berkooz:2014yda,El-Showk:2014dwa,
Nakayama:2014lva,Nakayama:2014yia,Alday:2014qfa,Chester:2014fya,Kos:2014bka,Caracciolo:2014cxa,Nakayama:2014sba,Bae:2014hia,
Beem:2014zpa,Chester:2014gqa,sdpb,Bobev:2015jxa,Kos:2015mba,Chester:2015qca,Beem:2015aoa,Iliesiu:2015qra,Poland:2015mta,
Lemos:2015awa,Lin:2015wcg,Chester:2015lej,Chester:2016wrc,Nakayama:2016cim,pycftboot,Nakayama:2016jhq,Iha:2016ppj,Kos:2016ysd,
Nakayama:2016knq,Echeverri:2016ztu,Li:2016wdp,Lin:2016gcl,Bae:2016yna,Bae:2016jpi,Lemos:2016xke,Beem:2016wfs,Li:2017ddj,
Cornagliotto:2017dup,Nakayama:2017vdd,Dymarsky:2017xzb,Chang:2017xmr,Cuomo:2017wme,Keller:2017iql,Cho:2017fzo,Li:2017kck,
Dymarsky:2017yzx,Bae:2017kcl,Dyer:2017rul,Chang:2017cdx,Cornagliotto:2017snu,Agmon:2017xes,Rong:2017cow,Baggio:2017mas,
Stergiou:2018gjj,Hasegawa:2018yqg,Liendo:2018ukf,Rong:2018okz,Atanasov:2018kqw,Behan:2018hfx,Kousvos:2018rhl,Cappelli:2018vir,
Gowdigere:2018lxz,Li:2018lyb,Karateev:2019pvw}
and more).
By now, we have many good introductory review articles on this approach,
see e.g.~\cite{El,Qualls:2015qjb,Rychkov:2016iqz,Simmons-Duffin:2016gjk,Antunes:2017vbx,Poland:2018epd,NakayamaSGC},
which make the entry to this fascinating and rapidly growing subject easier.

Some of the highlights in these developments for the purposes of this thesis are:
the consideration of constraints from the global symmetry in \cite{Rattazzi:2010yc},
the introduction of the semidefinite programming in \cite{Poland:2011ey},
and the extension of the analysis to the mixed correlators in \cite{Kos:2014bka}.
These techniques can now constrain the scaling dimensions of operators of 3d Ising and $\mathrm{O}(N)$ models within precise islands.

Previously, the study of a CFT has been done by the Monte Carlo (MC) simulation on lattice theory,
the direct measurements of physical system at critical point and perturbative loop diagram computations.
While the MC simulations cannot handle systems with infinite size directly,
at the critical point, the correlation length diverges and it is impossible to keep the system size sufficiently larger than the correlation length.
To obtain reliable results at infinite size limit from results in finite systems,
we need to apply the finite size scaling, which is a more systematic method than a naive extrapolation \cite{PhysRevB.82.174433,Hasenbusch:2011yya}.
Physical experiments can measure a large system with $O(10^{23})$ d.o.f.,
but generally do not give relatively accurate critical exponents because
reduced temperature can only reach $\abs{t}\sim 10^{-4}$ due to gravitational effects,
which cause a density gradient or make the effective correlation length finite (called gravitational rounding).
One of the most accurate measurements was done in the space shuttle STS-52 \cite{Lipa:2003},
and the results about the specific heat of \ce{^{4}He} near the lambda point
tell us the scaling dimension $\Delta_\epsilon$ of the first $O(2)$ invariant primary scalar $\epsilon$
in the three dimensional $O(2)$ vector model\cite{Kos:2015mba,Kos:2016ysd}:
\begin{equation}
	\Delta_\epsilon = 1.50946(22).
\end{equation}
This accuracy owes to the superfluidity to reach $\abs{t}\sim 10^{-7}$,
and less gravitation in the space shutle allows us to have $\abs{t}\approx 5\times 10^{-9}$ \cite{Chester:2019ifh}.
Two primary examples of perturbation methods are the large-$N$ expansion \cite{Lang:1991kp,Petkou:1994ad,Petkou:1995vu,Gracey:2002qa,Moshe:2003xn}
and the $\epsilon$-expansion \cite{Wilson:1971dc,ZinnJustin,Dey:2017oim}.
The $\epsilon$-expansion is motivated by the fact that in $d=4-\epsilon$ $\phi^4$ theory,
the nontrivial fixed point $(m^2_*,\lambda_*)$, called the Wilson-Fisher fixed point,
is $O(\epsilon)$ from the trivial fixed point $(m^2,\lambda)=(0,0)$.
Then the properties of the Wilson-Fisher fixed point can be perturbed from the trivial fixed point as a series of $\epsilon$,
and the physical limit $\epsilon\to 1$ would give the critical exponents of the three dimensional Ising model.
This argument has a subtlety that the series is asymptotic and its convergence radisu is zero,
and we need an appropriate resummation method to take $\epsilon\to 1$.

Summarizing previous results on the three dimensional Ising universality class, we have \cite{Pelissetto:2000ek,ElShowk:2012ht}
\begin{equation}
	\Delta_{\sigma} = 0.5182(3),\quad \Delta_{\epsilon} = 1.413(1). \label{eq:dim_ising_mc}
\end{equation}
Note that each of the MC simulation or the $\epsilon$-expansion cannot assign rigorous error bounds on exponents.
The conformal bootstrap applied for the same CFT shows that the relevant scaling dimensions
$(\Delta_\sigma,\Delta_\epsilon)$ must reside in a small island
which is calculated the most precisely in \cite{Kos:2016ysd} and whose bounding box is,
\begin{equation}
	\Delta_{\sigma} = 0.5181489(10),\quad \Delta_{\epsilon} = 1.412625(10). \label{eq:dim_ising_boot}
\end{equation}
The assumption used is the unitarity of the theory, $\mathbb{Z}_2$ invariance and
that $\sigma,\epsilon$ are the only relevant primary scalars, i.e., other primary scalars have $\Delta\geq d=3$.
The tininess of the island \eqref{eq:dim_ising_boot} explains another aspect of the critical universality;
any CFT with $\mathbb{Z}_2$ symmetry and unique $\mathbb{Z}_2$-even and odd relevant operators
must have quite close scaling dimensions to those of the Ising model.
The universality could be proven by showing that the island shrinks to the Ising point.
This is one of the advantages of the conformal bootstrap;
we have to assume some Hamiltonian in the MC simulation or $\epsilon$-expansion
and without the universality, cannot get results universal among all Hamiltonians by just one Hamiltonian.
The conformal bootstrap starts from an exact fixed point and only assumes the consistency of the CFT itself.

The developments in the conformal bootstrap have been helped by various computing tools.
For example, we have \code{SDPB}, an efficient solver of semidefinite programming (SDP) designed for the conformal bootstrap \cite{sdpb,sdpb2},
and SDP generators written in \code{Python}, namely
\code{PyCFTBoot} \cite{pycftboot}
and \code{cboot} \cite{Nakayama:2016jhq},
and in \code{Julia}, \code{JuliBootS} \cite{JuliBootS}.
There is also a \code{Mathematica} package to generate 4d bootstrap equations of arbitrary spin \cite{Cuomo:2017wme}.

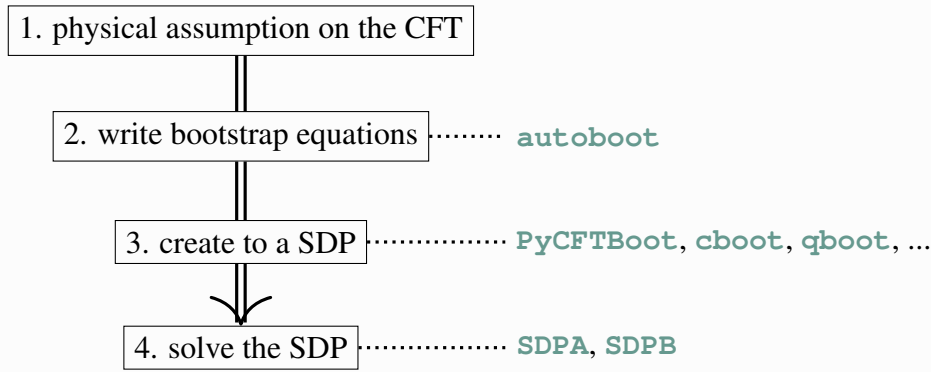
\begin{figure}[htpb]
	\centering
	\begin{tikzpicture}
		\node[draw] (a) at (0,0) {1. physical assumption on the CFT};
		\node[draw] (b) at (0,-40pt) {2. write bootstrap equations};
		\node[draw] (c) at (0,-80pt) {3. create to a SDP};
		\node[draw] (d) at (0,-120pt) {4. solve the SDP};
		\node[right] (x) at (+100pt,-40pt) {\code{autoboot}};
		\node[right] (y) at (+100pt,-80pt) {\code{PyCFTBoot}, \code{cboot}, \code{qboot}, ...};
		\node[right] (z) at (+100pt,-120pt) {\code{SDPA}, \code{SDPB}};
		\draw[->,very thick,double distance=2pt] (a) -- (b) -- (c) -- (d);
		\draw[dotted,distance=4pt,very thick] (b) -- (x);
		\draw[dotted,very thick] (c) -- (y);
		\draw[dotted,very thick] (d) -- (z);
	\end{tikzpicture}
	\caption{Relations between libraries used in the conformal bootstrap}
	\label{fig:libraries}
\end{figure}

To study some CFT with bootstrap, the first step is to enumerate the physical properties of the system
such as the global symmetry or the number of relevant primaries and then proceed along the path in \autoref{fig:libraries}.
The second step to write bootstrap equations has been ad hoc and done only by hand or with some help of \code{Mathematica},
and is automated by our \code{autoboot}.
We have some libraries in some programming languages for the third step to convert to a computable optimization problem,
and our \code{qboot} aims to handle more generic assumption on the spectrum.
SDP solvers such as \code{SDPA} \cite{SDPA6,SDPA7,SDPA-GMP} or \code{SDPB} \cite{sdpb} complete the last step
and from their results, we can get information on the CFT.
Now we can use free software in step 2, 3 and 4, and since our \code{autoboot}, \code{qboot} are designed to work smoothly with \code{SDPB},
a researcher can concentrate on the first step.

\subsection{autoboot}
\label{sec:intro_autoboot}
Given that the numerical bootstrap of precision islands of the 3d Ising and $O(N)$ models \cite{Kos:2014bka} was done in 2014,
it would not have been strange if there had been many papers studying CFTs with other global symmetries, especially with finite groups.
But this has not been the case, with a couple of exceptions e.g.~\cite{Rong:2017cow,Baggio:2017mas,Stergiou:2018gjj,Kousvos:2018rhl}.
We believe that this shortage of studies of CFTs with global symmetry is due to
the complexity in writing down the mixed-correlator bootstrap equations, with the constraints coming from the symmetry.
To solve this problem we developed a software \code{autoboot} in \code{Mathematica}
to automate the whole tedious process with the global symmetry \cite{autoboot}.
\code{autoboot} is an automatic generator of mixed-correlator bootstrap equations of scalar operators with global symmetry.

Let us illustrate the use with an example.
Suppose we would like to perform the numerical bootstrap of a CFT invariant under $D_8$, the dihedral group with eight elements.
Let us assume the existence of two scalar operators, \code{e} in the singlet and \code{v} in the doublet of $D_8$.
The mixed-correlator bootstrap equations can be generated by the following \code{Mathematica} code, after loading our package:
\begin{lstlisting}
	g=getGroup[8,3];
	setGroup[g];
	setOps[{op[e,rep[1]], op[v,rep[5]]}];
	eq=bootAll[];
	sdp=makeSDP[eq];
	py=toCboot[sdp];
	WriteString["D8.py",py];
\end{lstlisting}
Let us go through the example line by line:
\begin{enumerate}
	\item \lstinline{g=getGroup[8,3]} sets the group $D_8$ to $g$.
	This line illustrates the ability of \code{autoboot} to obtain the group theory data from the \code{SmallGrp} library \cite{SmallGrp}
	of the computer algebra system \code{GAP} \cite{GAP},
	which contains the necessary data of finite groups of order less than $2000$ and many others.
	The pair $(8,3)$ is a way to specify a finite group in \code{SmallGrp}.
	It simply says that $D_8$ is the third group in their list of groups of order 8.
	\item \lstinline{setGroup[g]} tells \code{autoboot} that the CFT is symmetric under $D_8$.
	\item \lstinline{setOps[...]} adds operators to \code{autoboot}.
	\lstinline{rep[n]} means the $n$-th representation of the group in the \code{SmallGrp} library;
	we set the operator \lstinline{e} to be a singlet and the operator \lstinline{v} to be a doublet.
	\item \lstinline{eq=bootAll[]} creates the bootstrap equations for all four-point functions with added operators
	in a symbolic form and sets them into \lstinline{eq}.
	\item \lstinline{sdp=makeSDP[eq]} converts the bootstrap equations into the form of a SDP.
	\item \lstinline{py=toCboot[sdp]} further rewrites it into an actual \code{Python} program,
	which uses \code{cboot} \cite{Nakayama:2016jhq} which runs in a \code{Python}-based system \code{Sage} \cite{Sage}.
	\item The last line simply writes the \code{Python} code into an external file.
\end{enumerate}

All what remains is to make a small edit of the resulting file, to set up the dimensions and gaps of the operators.
The \code{Python} code then generates the XML input file for \code{SDPB}.

\subsection{QBoot}
\label{sec:intro_qboot}
The SDP related with the conformal bootstrap has real variables $y_n$, linear objectives $\sum_n b_ny_n$
and positive semidefinite constraints on $y$: for all $\Delta\geq\Delta_0$,
\begin{equation}
	M^0(\Delta)+\sum_n y_n M^n(\Delta)\succeq 0, \label{eq:semidef}
\end{equation}
in which $M^n(\Delta)$ is a symmetric matrix whose elements are polynomials of $\Delta$ and $M\succeq 0$
means that $M$ is positive semidefinite.
$\Delta\geq\Delta_0$ means that all primary operators in the sector must have scaling dimension $\Delta\in[\Delta_0,\infty)$.
To solve the SDP by \code{SDPB}, we need to take a variable $x\in[0,\infty)$ in \eqref{eq:semidef},
which is easily achieved by $x=\Delta-\Delta_0\geq 0$.
Let us think about a constraint that the positive-semidefiniteness holds for all $\Delta_0\leq \Delta<\Delta_1$.
A change of variables $x=\frac{\Delta-\Delta_0}{\Delta_1-\Delta}$, $\Delta=\frac{\Delta_0+\Delta_1 x}{1+x}$ transforms
the inequality $\Delta_0\leq \Delta<\Delta_1$ into $x\geq 0$, and
\eqref{eq:semidef} multiplied by $(1+x)^d>0$ where $d$ is the maximum order of polynomials of $M^n(\Delta)$
gives a semidefinite constraint of a polynomial matrix in $[0,\infty)$.
This simple discussion allows a generalization of the assumption on the spectrum of the CFT
from $[\Delta_*,\infty]$ to $[\Delta_0,\Delta_1)\cup[\Delta_2,\Delta_3)\cup\cdots\cup[\Delta_*,\infty]$.
This idea led us to write a new SDP generator \code{qboot} written in modern \code{C++}, which we introduce in this thesis.
\code{qboot} is based on \code{cboot} and has better performance with multiple CPU cores.
\code{autoboot} also generates a program for \code{qboot} just by changing the last two lines in the example above as
\begin{lstlisting}
	cpp=toQboot[sdp];
	WriteString["D8.cpp",cpp];
\end{lstlisting}
The \code{C++} code then generates the input directory for \code{SDPB}.

\subsection{Organization}
\label{sec:intro_org}
The rest of the thesis is organized as follows.

In \autoref{sec:basics}, we summarize common basics and notions of CFT sufficient to study the conformal bootstrap.

\textcolor{FireBrick}{Section} \ref{sec:sym} is the theoretical part of our \code{autoboot}.
We first explain our notations for the group theory constants in \autoref{sec:group},
and then in \autoref{sec:symcft} and \autoref{sec:boot} describe how the bootstrap equations can be obtained,
given the set of external scalar primary operators $\phi_i$ in the representation $r_i$ of the symmetry group $G$.

In \autoref{sec:numerical}, we discuss numerical methods used in \code{cboot} and our \code{qboot}.
Most of these methods has been known, except for a simple technique which we call \emph{hot-starting} and introduce in \autoref{sec:hot}.
This technique reduces the running time of the SDP solver significantly
by reusing parts of the computation for a given set of scaling dimensions of external operators to the computation of
another nearby set of scaling dimensions.
Our experience shows that it often gives an increase in the speed by about a factor of 10 to 20.

In \autoref{sec:implementation}, we discuss the implementation of our \code{autoboot} and \code{qboot},
and in \autoref{sec:group_impl}, \autoref{sec:cft_impl} and \autoref{sec:boot},
we discuss how our \code{autoboot} implements the procedure given in \autoref{sec:sym}.
\autoref{sec:sdp} and \autoref{sec:qboot_impl} describe the methods used to implement our \code{qboot}.

In \autoref{sec:examples}, we describe two examples using \code{autoboot} and \code{qboot}.
The first is to perform the mixed-correlator bootstrap of the 3d Ising model, with new rigorous results about irrelevant operators using \code{qboot}.
The second is to study the $O(2)$ model with three types of external scalar operators.
Without \code{autoboot}, it is a formidable task to write down by hand the set of bootstrap equations,
but with \code{autoboot}, it is immediate.

We conclude the thesis in \autoref{sec:conclusion} by discussing future directions.

\newpage
\section{Basics of CFT}
\label{sec:basics}

\subsection{Conformal symmetry}
\label{sec:cft}
In $d$ dimensional (Euclidean) field theory, the conformal transformation consists of
$x^\mu\mapsto x'^\mu$ which preserve the metric up to an overall factor $\Omega(x)$:
\begin{equation}
	g'_{\mu\nu}
	= g_{\rho\sigma}\frac{\partial x'^\rho}{\partial x^\mu}\frac{\partial x'^\sigma}{\partial x^\nu}
	= \Omega(x)g_{\mu\nu}, \label{eq:preserve_metric}
\end{equation}
and generated by
\begin{align}
	\op{P}_\mu &= \partial_\mu, \\
	\op{M}_{\mu\nu} &= -x_\mu\partial_\nu+x_\nu\partial_\mu, \\
	\op{D} &= x^\mu\partial_\mu, \\
	\op{K}_\mu &= \abs{x}^2\partial_\mu-2x_\mu x^\nu\partial_\nu.
\end{align}
From \eqref{eq:preserve_metric}, the Jacobian is decomposed as
\begin{equation}
	\frac{\partial x'^\mu}{\partial x^\nu} = \omega(x)\Lambda\sud{\mu}{\nu}(x), \label{eq:jacob}
\end{equation}
where $\omega(x)^2=\Omega(x)$ and $\Lambda\sud{\mu}{\nu}(x)\in O(d)$.
Translations $\op{P}_\mu$ and rotations $\op{M}_{\mu\nu}$ generate the Euclidean transformations
, $\op{D}$ is called the dilatation and $\op{K}_\mu$ is the special conformal transformations.
The inversion $I\colon x^\mu\mapsto x^\mu/\abs{x}^2$ is an involutive conformal transformation
disconnected from the identity, and relates $\op{K}_\mu$ with $\op{P}_\mu$ by $\op{K}_\mu=I\op{P}_\mu I$.
The conformal group is a Lie group which is isomorphic to the group of conformal transformations
and we take its generators $P_\mu, M_{\mu\nu}, D, K_\mu$ so that
$\op{P}_\mu, \op{M}_{\mu\nu}, \op{D}, \op{K}_\mu$ are representations in $d$ dimensional space.
It is easy to check that the commutation relations are the following:
\begin{align}
	\left[M_{\mu\nu},P_\rho\right] & = g_{\nu\rho}P_\mu-g_{\mu\rho}P_\nu, \\
	\left[M_{\mu\nu},K_\rho\right] & = g_{\nu\rho}K_\mu-g_{\mu\rho}K_\nu, \\
	\left[M_{\mu\nu},M_{\rho\sigma}\right]
	& = \left(g_{\nu\rho}M_{\mu\sigma}-g_{\mu\rho}M_{\nu\sigma}\right)-\left(\rho\swap\sigma\right), \\
	\left[D,P_\mu\right] & = P_\mu, \\
	\left[D,K_\mu\right] & = -K_\mu, \\
	\left[P_\mu,K_\nu\right] & = 2g_{\mu\nu}D+2M_{\mu\nu}, \\
	\left[P_\mu,P_\nu\right] & = \left[K_\mu,K_\nu\right] = \left[D,M_{\mu\nu}\right] = 0.
\end{align}
The conformal algebra is isomorphic to that of $d+2$ dimensional rotation group with a metric $G=g\oplus\operatorname{diag}(1,-1)$
by the following identification:
\begin{align}
	J_{\mu\nu} &= M_{\mu\nu}, \\
	J_{d+1,d+2} &= D, \\
	J_{\mu,d+1} &= (P_\mu-K_\mu)/2, \\
	J_{\mu,d+2} &= -(P_\mu+K_\mu)/2.
\end{align}

The actions of the conformal group to a field operator $\op{O}(x)$ can be determined by those at the origin $x=0$, since
\begin{equation}
	\op{O}(x) = e^{x^\mu P_\mu}\op{O}(0)e^{-x^\mu P_\mu},
\end{equation}
and actions at $x\neq 0$ can be recovered from conjugation by $e^{x^\mu P_\mu}$ using the conformal algebra.
The subgroup which stabilizes the origin is generated by $M_{\mu\nu},D,K_{\mu}$
and since we can take a maximal torus generated by that of $SO(d)$ and $D$,
$\op{O}(0)$ is classified by an irreducible representation (irrep) $r$ of $SO(d)$ and an eigenvalue $\Delta$ of $D$ (which called the scaling dimension):
\begin{align}
	\left[D,\op{O}^i(0)\right] &= \Delta\op{O}^i(0), \\
	\left[M_{\mu\nu},\op{O}^i(0)\right] &= -\left(S^r_{\mu\nu}\right)\sud{i}{j}\op{O}^j(0),
\end{align}
where $i$ is an index of $r$ and $S^r_{\mu\nu}$ is a representation matrix of $M_{\mu\nu}$.
From the commutation relation $[D,K_\mu]=-K_\mu$, $K_\mu$ lowers the scaling dimension and, in unitary CFT, scaling dimensions is bounded from below.
Then we must have the lowest eigenstate which is annihilated by $K_\mu$:
\begin{equation}
	\left[K_\mu,\op{O}^i(0)\right] = 0.
\end{equation}
Such an operator is called primary and a primary operator is characterized by two labels: an irreducible representation $r$ of $SO(d)$
and the scaling dimension $\Delta$.
$P_\mu$ acts as a differential operator $[P_\mu,\op{O}^i(0)]=\partial_\mu\op{O}^i(0)$
and if $\op{O}$ is a primary operator, $\op{O}$ and its derivatives $\partial_\mu\op{O},\partial_\mu\partial_\nu\op{O},\ldots$
form a conformal multiplet, called the conformal tower of $\op{O}$. Operators in the multiplet but $\op{O}$ are called descendants of $\op{O}$.

The rank $l$ traceless symmetric representation of $SO(d)$ is called the spin-$l$ representation.
In most of this thesis, a primary operator is assumed to belong to some spin-$l$ representation (the spin of a scalar is 0).
The action of a finite conformal transformation $S\colon x\mapsto S(x)$ on a spin-$l$ primary operator $\op{O}$ is\cite{Rychkov:2016iqz}
\begin{equation}
	S^{-1}\op{O}^{\mu_1\cdots\mu_l}(S(x))S
	= \omega(x)^{-\Delta}\Lambda(x)\sud{\mu_1}{\nu_1}\cdots\Lambda(x)\sud{\mu_l}{\nu_l}\op{O}^{\nu_1\cdots\nu_l}(x),
\end{equation}
where the Jacobian is decomposed as $\partial S(x)^\mu/\partial x^\nu=\omega(x)\Lambda\sud{\mu}{\nu}(x)$ in \eqref{eq:jacob}.
In case $S$ is the inversion $I$, the action of $I$ together with the complex conjugate defines the Hermitian conjugate:
\begin{equation}
	\op{O}^{\mu_1\cdots\mu_l}(x)^\dagger
	= I^{-1}\overline{\op{O}^{\mu_1\cdots\mu_l}(x)}I
	= \abs{x}^{-2\Delta}I(x)\sud{\mu_1}{\nu_1}\cdots I(x)\sud{\mu_l}{\nu_l}\overline{\op{O}^{\nu_1\cdots\nu_l}(x/\abs{x}^2)},
	\label{eq:dual_op}
\end{equation}
where $I_{\mu\nu}(x)=g_{\mu\nu}-2x_{\mu}x_{\nu}/\abs{x}^2$.

\subsection{State-operator correspondence}
\label{sec:st_op}
CFT has a crucial property that there is a correspondence between states and operators\cite{Polchinski:1998rq,Simmons-Duffin:2016gjk}.
We describe the correspondence in the picture of path integral.

First, with the scale invariance it is natural to take a $d-1$ dimensional sphere around the origin
and identify the radius of the sphere with time.
The state at time $t=\log r$ is specified by the configuration on the sphere $\abs{x}=r$.
Let the theory be described by path integral of action $S[\phi]$ and $\Ket{\phi_b}$ be an eigenstate
defined on the sphere $\abs{x}=1$.
A state at time $t=0$ is decomposed as a linear combination of eigenstates:
\begin{equation}
	\Ket{\psi} = \int\op{D}\phi_b\Ket{\phi_b}\Braket{\phi_b|\psi}.
\end{equation}
The vacuum is defined by
\begin{align}
	\Ket{0} &= \int_{\abs{x}\leq 1}\op{D}\phi e^{-S[\phi]}, \\
	\Braket{\phi_b|0} &= \int_{\substack{\abs{x}\leq 1\\\left.\phi\right|_{\abs{x}=1}=\phi_b}}\op{D}\phi e^{-S[\phi]}.
\end{align}
Then we can construct a state corresponding to a local operator $\op{O}(x)$ in the ball $\abs{x}<1$ as
\begin{equation}
	\Braket{\phi_b|\op{O}(x)|0} = \int_{\substack{\abs{x}\leq 1\\\left.\phi\right|_{\abs{x}=1}=\phi_b}}\op{D}\phi \op{O}(x) e^{-S[\phi]}.
	\label{eq:op_to_st}
\end{equation}

Next, we establish the opposite correspondence.
For a given state $\Ket{\op{O}}$, we define correlation functions with $\op{O}(x_0)$: $\Braket{\op{O}(x_0)\op{O}_1(x_1)\cdots\op{O}_n(x_n)}$.
Using the conformal invariance, we can assume that $x_0=0$ and all other $x_i$ satisfy $1<\abs{x_i}$,
and we have
\begin{equation}
	\Braket{\op{O}(0)\op{O}_1(x_1)\cdots\op{O}_n(x_n)}
	\colonequals \int\op{D}\phi_b\Braket{\phi_b|\op{O}} \int_{\substack{\abs{x}\geq 1\\\left.\phi\right|_{\abs{x}=1}=\phi_b}}
	\op{D}\phi \op{O}_1(x_1)\cdots\op{O}_n(x_n) e^{-S[\phi]}.
	\label{eq:st_to_op}
\end{equation}
The correspondence from operator to state \eqref{eq:op_to_st} and from state to operator \eqref{eq:st_to_op} just implies
that a correlation function is a path integral over all space:
\begin{equation}
	\Braket{\op{O}(x_0)\op{O}_1(x_1)\cdots\op{O}_n(x_n)}
	= \int_{x\in\mathbb{R}^d} \op{D}\phi \op{O}(x_0)\op{O}_1(x_1)\cdots\op{O}_n(x_n) e^{-S[\phi]},
\end{equation}
and we can identify states with operators:
\begin{equation}
	\Ket{\op{O}} \colonequals \op{O}(0)\Ket{0}.
\end{equation}
The vacuum is conformal invariant: $Q\Ket{0}=0$ for any conformal generator $Q$,
and we can see that a primary state which corresponds to a primary operator $\op{O}$ is an eigenstate of $D$ and annihilated by $K_\mu$:
\begin{align}
	D\Ket{\op{O}^i} &= \Delta\Ket{\op{O}^i}, \\
	M_{\mu\nu}\Ket{\op{O}^i} &= -\left(S^r_{\mu\nu}\right)\sud{i}{j}\Ket{\op{O}^j}, \\
	K_\mu\Ket{\op{O}^i} &= 0,
\end{align}
and $P_\mu$ generates descendant states of $\Ket{\op{O}^i}$.

We can take orthonormal basis of primary fields if the theory is unitary:
\begin{equation}
	\Braket{\op{O}|\op{O}'}=\delta_{\op{O}\op{O}'}.
\end{equation}
In case that $\op{O}$ is complex, we have $\delta_{\op{O}\overline{\op{O}}}$; we will discuss about this case with global symmetry later.
From the unitarity, the Gram matrices of descendants must be positive semidefinite.
This condition gives strong constraints on $\Delta$, which is called the unitarity bound.
For example, $\Abs{P_{\mu}\Ket{\op{O}^i}}\geq 0$ implies
\begin{equation}
	\Abs{P_{\mu}\Ket{\op{O}^i}} = \Braket{\op{O}^i|K_{\mu}P_{\mu}|\op{O}^i}
	= 2\Delta\Abs{\op{O}^i}\geq 0,
\end{equation}
and $\Delta\geq 0$.
The space of descendants is filtered by eigenvalue of $D$ and for each integer $N\geq 0$, the subspace with eigenvalue $\Delta+N$
is decomposed by $SO(d)$.
Thus the unitarity bounds are calculated by representation theory of $SO(d)$.
The unitarity bounds for the spin-$l$ representations are \cite{Minwalla:1997ka,Bourget:2017kik}
\begin{equation}
	\Delta\geq\begin{cases}
		\frac{d-2}{2} & \ptext{$l=0$}, \\
		l+d-2 & \ptext{$l>0$}. \\
	\end{cases}
	\label{eq:unit_bound}
\end{equation}
The bounds are saturated by a free scalar field $\partial^2\phi(x)=0$ with $\Delta=(d-2)/2$,
a spin-$1$ conserved current $\partial_\mu j^\mu(x)=0$ with $\Delta=d-1$
and the stress tensor $\partial_{\mu}T^{\mu\nu}(x)=0$ with $\Delta=d$.
If $d\geq 3$, conserved currents with higher spin are allowed only when the theory is free\cite{Maldacena:2011jn,Alba:2013yda,Alba:2015upa}.

Any conformal transformation $S\colon x^\mu\to x'^\mu$ gives another quantization.
One of the important examples is
\begin{equation}
	S\colon x^\mu\mapsto x'^\mu = \frac{x^\mu+c^\mu \abs{x}^2}{1+2c\cdot x+\abs{x}^2}-\frac{c^\mu}{2},
\end{equation}
where $c^\mu$ is some fixed vector with norm $\abs{c}^2=1$. This transformation maps $0$ to $-c^\mu/2$ and $\infty$ to $c^\mu/2$
and these two points $\pm c^\mu/2$ are called the north and south poles.
$S$ generates another quantization which is called N-S quantization \cite{Rychkov:2016iqz}.
The sphere of time $t=\log r$ in the radial quantization we discussed above
is mapped to a sphere which separates two poles.
The inversion $I$ is interpreted as a reflection on the plane $x\cdot c=0$:
$S(I(x))=\Theta_c(S(x))$ where $\Theta_c(x)=x^\mu-2(c\cdot x)c^\mu$.
Then the unitarity $\Braket{\psi|\psi}\geq 0$ for $\Ket{\psi}=\op{O}_1(x_1)\cdots\op{O}_n(x_n)\Ket{0}$ and $c\cdots x_j<0$ implies
\begin{equation}
	\Braket{0|\Theta\left[\op{O}_1(x_1)\cdots\op{O}_n(x_n)\right]\op{O}_1(x_1)\cdots\op{O}_n(x_n)|0}\geq 0,
\end{equation}
which is called the reflection positivity.

The correspondence gives the operator product expansion.
A product of two primary operators $\op{O}^i_1(x_1)\op{O}^j_2(x_2)$ defines some state $\Ket{\Psi}$ by path integral in a ball around $x_2$,
and the operator which corresponds to $\Ket{\Psi}$ can be expanded as a linear combination of primary operators at $x_2$ and their descendants:
\begin{equation}
	\op{O}^i_1(x_1)\op{O}^j_2(x_2) = \sum_{\op{O}}C_{12\op{O},k}^{ij}(x_{12},\partial_2)\op{O}^k(x_2),
	\label{eq:ope}
\end{equation}
or, in the words of states,
\begin{equation}
	\op{O}^i_1(x)\Ket{\op{O}^j_2} = \sum_{\op{O}}C_{12\op{O},k}^{ij}(x,P)\Ket{\op{O}^k},
\end{equation}
where $C_{12\op{O},k}^{ij}(x,P)$ describes the contribution of descendants of $\op{O}$.
We can take OPE in correlation functions if the ball we integrated out does not contain any other operator insertions $\op{O}_n(x_n)$,
and \eqref{eq:ope} converges if $\abs{x_{1}-x_2}<\abs{x_{n}-x_2}$ \cite{Pappadopulo:2012jk}.
In general, $C_{12\op{O},k}^{ij}$ is a linear combination of
finitely many (but possibly not the unique) functions which is completely fixed by conformal algebra:
\begin{equation}
	\op{O}^i_1(x_1)\op{O}^j_2(x_2) = \sum_{\op{O}}\sum_n \lambda_{12\op{O}}^nC_{12\op{O},n,k}^{ij}(x_{12},\partial_2)\op{O}^k(x_2),
	\label{eq:open}
\end{equation}
and $\lambda_{12\op{O}}^n$ is called OPE coefficients. This will be discussed below.

\subsection{Correlation functions}
\label{sec:cor}
The conformal symmetry is three-transitive: for any given three points $x_1,x_2,x_3$,
there exists a conformal transformation which sends them to $0,1,\infty$ ($1$ stands for some fixed unit vector; e.g., $(1,0,\ldots,0)$).
It is easy to check this property: send $x_1\to 0$ by $P_\mu$, $x_3\to\infty$ by $K_\mu=IP_\mu I$
and $x_2\to1$ by $D$ and $M_{\mu\nu}$.
As a consequence, a three-point function $\Braket{\op{O}^i_1(x_1)\op{O}^j_2(x_2)\op{O}^k_3(x_3)}$
among primary operators with irreps $r_1,r_2,r_3$ of $SO(d)$
is determined by $\Braket{\op{O}^i_1(0)\op{O}^j_2(1)\op{O}^k_3(\infty)}$, which does not have any continuous variables.
This tensor in $r_1\otimes r_2\otimes r_3$ must be invariant under the stabilizer $SO(d-1)$ of $(0,1,\infty)$,
so there is a finite-dimensional basis for the three-point function:
\begin{equation}
	\Braket{\op{O}^i_1(x_1)\op{O}^j_2(x_2)\op{O}^k_3(x_3)}
	= \frac{\sum_n \lambda^n_{\op{O}_1\op{O}_2\op{O}_3}Z_n^{ijk}(x_{13},x_{23})}
	{\abs{x_{12}}^{\Delta_1+\Delta_2-\Delta_3}\abs{x_{23}}^{-\Delta_1+\Delta_2+\Delta_3}\abs{x_{31}}^{\Delta_1-\Delta_2+\Delta_3}}.
	\label{eq:3pt}
\end{equation}
Here we factored out the obvious scaling factor. $\lambda^n_{\op{O}_1\op{O}_2\op{O}_3}$ are called OPE coefficients.

A similar argument shows that a two-point function survives only in case two operators have the same irrep $r$ of $SO(d)$ and scaling dimension:
\begin{equation}
	\Braket{\op{O}_1^i(x_1)\op{O}_2^j(x_2)}
	= \frac{\delta_{12}I^{ij}(x_{12})}{\abs{x_{12}}^{2\Delta}}, \label{eq:2pt}
\end{equation}
where $I^{ij}(x)$ is the unique invariant tensor. The uniqueness follows from that $(x_1,x_2)=(0,\infty)$ is stabilized by $SO(d)$ and dilatations.
In case $r$ is the spin-$l$ representation, we have\cite{Costa:2011mg}
\begin{align}
	I^{\mu\nu}(x) &= g^{\mu\nu}-\frac{2x^\mu x^\nu}{\abs{x}^2}, \\
	I^{\mu_1\cdots\mu_l;\nu_1\cdots\nu_l}
	&= \frac{1}{l!}\sum_{\tau\in\op{S}_l}I_{\mu_1\nu_{\tau(1)}}\cdots I_{\mu_l\nu_{\tau(l)}}-\ptext{traces}.
\end{align}
The state-operator correspondence shows that the overall factors in two-point functions can be diagonalized as above.
Let $\op{O}_1$ have spin $l$ and denote $\mu_1,\ldots,\mu_l$ by $i$ and $\nu_1,\ldots,\nu_l$ by $j$.
Taking conjugate of $\Ket{\op{O}^i}=\op{O}^i(0)\Ket{0}$ and \eqref{eq:dual_op} gives
\begin{align}
	\Bra{\op{O}^i}
	&= \lim_{x\to 0}\Bra{0}\op{O}^i(x)^\dagger \\
	&= \lim_{x\to \infty}\abs{x}^{2\Delta}I(x)\sud{\mu_1}{\nu_1}\cdots I(x)\sud{\mu_l}{\nu_l}\Bra{0}\op{O}^{\nu_1\cdots\nu_l}(x)
\end{align}
and if we have an overall factors $M_{12}$ instead of $\delta_{12}$ in the right hand of \eqref{eq:2pt}, the inner product is
\begin{align}
	\Braket{\op{O}_1^i|\op{O}_2^j}
	&= \lim_{x\to \infty}\abs{x}^{2\Delta}I(x)\sud{\mu_1}{\nu_1}\cdots I(x)\sud{\mu_l}{\nu_l}
	\Braket{0|\op{O}_1^{\nu_1\cdots\nu_l}(x)\op{O}_2^j(0)|0} \\
	&= \lim_{x\to \infty}M_{12}I(x)\sud{\mu_1}{\nu_1}\cdots I(x)\sud{\mu_l}{\nu_l} I^{ij}(x) \\
	&= M_{12}\delta^{ij},
\end{align}
and since we have taken an orthonormal basis, we have $M_{12}=\delta_{12}$.

Taking OPE of $\op{O}_1$ and $\op{O}_2$ in the three-point function \eqref{eq:3pt}, we obtain
\begin{align}
	\Braket{\op{O}^i_1(x_1)\op{O}^j_2(x_2)\op{O}^k_3(x_3)}
	&= C_{123,k'}^{ij}(x_{12},\partial_2)\Braket{\op{O}^{k'}_3(x_2)\op{O}^k_3(x_3)}, \\
	\frac{\sum_n \lambda^n_{123}Z_n^{ijk}(x_{13},x_{23})}
	{\abs{x_{12}}^{\Delta_1+\Delta_2-\Delta_3}\abs{x_{23}}^{-\Delta_1+\Delta_2+\Delta_3}\abs{x_{31}}^{\Delta_1-\Delta_2+\Delta_3}}
	&= C_{123,k'}^{ij}(x_{12},\partial_2)\frac{I^{k'k}(x_{23})}{\abs{x_{23}}^{2\Delta_3}}.
\end{align}
This equation determines $\lambda_{123}^n$ and tensor structures $Z_n^{ijk}$ from OPE structures $C_{123,k}^{ij}$.
We can construct $C_{123,k}^{ij}$ from a conformal invariant three-point function;
in
\begin{align}
	\Braket{\op{O}_3^k|e^{-y^\mu K_\mu}\op{O}_1^i(x)|\op{O}_2^j}
	&= \Braket{\op{O}_3^k|e^{-y^\mu K_\mu}C_{123,k'}^{ij}(x,P)|\op{O}_3^{k'}},
\end{align}
the left hand equals to $\Braket{\op{O}_1^i(x)\op{O}_2^j(0)\op{O}_3^k(y)^\dagger}$, which can be calculated by the three-point function,
and expanding this equation around $y=0$, the coefficients of order $N$ in the left hand
gives the coefficients of $C_{123,k'}^{ij}(x,P)$ with $P_{\mu_1}\cdots P_{\mu_N}$.
Thus there is a one-to-one correspondence between the OPE structure and the three-point function and the equation \eqref{eq:open} is established.

The three-point function with general irreps of $SO(d)$ is completely solved in case $d=3$ \cite{Costa:2011mg,Iliesiu:2015qra}
and $d=4$ \cite{Elkhidir:2014woa}.
We need only the three-point functions with two scalars later.
The OPE coefficients of two scalar $\lambda_{\phi_1\phi_2\op{O}}$ is nontrivial only if $\op{O}$ belongs to spin-$l$ representation,
because $\Braket{\phi_1(0)\phi_2(x)\op{O}^{\mu_1\cdots\mu_l}(\infty)}$ can be $SO(d-1)$ invariant
only when it is proportional to $x^\mu_1\cdots x^\mu_l$, which is a symmetric tensor (tracelessness comes from irreducibility).
There is only one scalar-scalar-spin $l$ OPE coefficient\cite{Costa:2011mg}:
\begin{equation}
	\Braket{\phi_1(x_1)\phi_2(x_2)\op{O}^{\mu_1\cdots\mu_l}(x_3)}
	= \frac{\lambda_{\phi_1\phi_2\op{O}}Z^{\mu_1\cdots\mu_l}(x_{13},x_{23})}
	{\abs{x_{12}}^{\Delta_1+\Delta_2-\Delta}\abs{x_{23}}^{-\Delta_1+\Delta_2+\Delta}\abs{x_{31}}^{\Delta_1-\Delta_2+\Delta}}
\end{equation}
where
\begin{align}
	Z^\mu(x,y) &= \frac{\abs{x}\abs{y}}{\abs{x-y}}\left[\frac{x_\mu}{\abs{x}^2}-\frac{y_\mu}{\abs{y}^2}\right], \\
	Z^{\mu_1\cdots\mu_l}(x,y) &= Z^{\mu_1}(x,y)\cdots Z^{\mu_l}(x,y)-\ptext{traces}.
\end{align}
OPE coefficients are real; we show this later in more general context. The leading term of the OPE structure of two scalars is
\begin{equation}
	C_{12\op{O},\mu_1\cdots\mu_l}(x,P) = \frac{\hat{x}_{\mu_1}\cdots \hat{x}_{\mu_l}+O(\abs{x})}{\abs{x}^{\Delta_1+\Delta_2-\Delta}}
	\label{eq:opelead}
\end{equation}
where $\hat{x}=x/\abs{x}$.

\subsection{Conformal blocks}
\label{sec:cfbl}
All $n$-point functions can be calculated by the OPE recursively.
Taking OPE of $\op{O}_1$ and $\op{O}_2$, we can reduce $n$-point functions to $(n-1)$-point functions,
\begin{align}
	& \Braket{\op{O}_1^{i_1}(x_1)\op{O}_2^{i_2}(x_2)\op{O}_3^{i_3}(x_3)\cdots\op{O}_n^{i_n}(x_n)} \nonumber\\
	&= \sum_{\op{O}}\sum_n \lambda_{12\op{O}}^n C_{12\op{O},n,j}^{i_1i_2}(x_{12},\partial_2)
	\Braket{\op{O}^{j}(x_2)\op{O}_2^{i_2}(x_2)\op{O}_3^{i_3}(x_3)\cdots\op{O}_n^{i_n}(x_n)}.
\end{align}
Then all correlation functions are reduced to one-point functions,
\begin{equation}
	\Braket{\op{O}^i(x)}=\begin{cases}
		1 & \ptext{$\op{O}=1$}, \\
		0 & \ptext{otherwise}.
	\end{cases}
\end{equation}
This procedure determines all correlation functions in the theory from the spectrum $(l_i,\Delta_i)$ of operators
(with spin $l_i$ and scaling dimension $\Delta_i$) and OPE coefficients $\lambda_{ijk}$ among them.
$\Set{\Delta_i, \lambda_{ijk}}$ is called the CFT data and defines the theory.
We must have the same result for a $n$-point function independent from the order to take OPE,
or in other words, OPE must be associative:
\begin{equation}
	\op{O}_1(\op{O}_2\op{O}_3)=(\op{O}_1\op{O}_2)\op{O}_3.
\end{equation}
Taking inner product with $\op{O}_4$, we get
\begin{equation}
	\Braket{(\op{O}_4\op{O}_1)(\op{O}_2\op{O}_3)}
	=\Braket{(\op{O}_4\op{O}_3)(\op{O}_1\op{O}_2)}
\end{equation}
(for simplicity, we neglected the negative sign which arises if $\op{O}_3$ and $\op{O}_1\op{O}_2$ are fermionic).
This is called the (conformal) bootstrap equation.
The bootstrap equations for all primaries mean the theory has consistent OPE, and put constraints on the CFT data.
For example, the spectrum is known to be discrete if $d\geq 3$ and a thermal partition function is finite \cite{Simmons-Duffin:2016gjk,Levy:2018bdc}.

Let us calculate a scalar four-point function
\begin{equation}
	\Braket{\phi_1(x_1)\phi_2(x_2)\phi_3(x_3)\phi_4(x_4)}. \label{eq:4ptsc}
\end{equation}
The conformal symmetry is not four-transitive, and
for any given four points $x_1,x_2,x_3,x_4$, there exists a conformal transformation which sends them to
$0,z,1,\infty$, where we selected a fixed plane as a complex plane.
We cannot reduce the number of continuous variables,
and the configuration of four points fully fixed by the conformal symmetry is called a conformal frame.
The four-point functions depends on two real variables $\Re z,\Im z$ and cannot be determined only by the conformal symmetry.
Taking OPE of $\phi_1\phi_2$ and $\phi_3\phi_4$ in \eqref{eq:4ptsc}, we obtain
\begin{equation}
	\Braket{\phi_1(x_1)\phi_2(x_2)\phi_3(x_3)\phi_4(x_4)}
	= \sum_{\op{O}} \lambda_{12\op{O}}\lambda_{34\op{O}}
	C_{12\op{O},i}(x_{12},\partial_2)C_{34\op{O},j}(x_{34},\partial_4)\Braket{\op{O}^i(x_2)\op{O}^j(x_4)},
\end{equation}
and the scalar four-point function is decomposed by conformal blocks $g_{\op{O}}^{\Delta_{12},\Delta_{34}}$, which is defined by
\begin{align}
	\Braket{\phi_1(x_1)\phi_2(x_2)\phi_3(x_3)\phi_4(x_4)}
	&= \frac{1}{\abs{x_{12}}^{\Delta_1+\Delta_2}\abs{x_{34}}^{\Delta_3+\Delta_4}}
	\left(\frac{\abs{x_{24}}}{\abs{x_{14}}}\right)^{\Delta_{12}}
	\left(\frac{\abs{x_{14}}}{\abs{x_{13}}}\right)^{\Delta_{34}}g(u,v), \\
	g(u,v) &= \sum_{\op{O}}\lambda_{\phi_1\phi_2\op{O}}\lambda_{\phi_3\phi_4\op{O}}g_{\op{O}}^{\Delta_{12},\Delta_{34}}(u,v), \\
	g_{\op{O}}^{\Delta_{12},\Delta_{34}}(u,v) &=\abs{x_{12}}^{\Delta_1+\Delta_2}\abs{x_{34}}^{\Delta_3+\Delta_4}
	\left(\frac{\abs{x_{24}}}{\abs{x_{14}}}\right)^{-\Delta_{12}}\left(\frac{\abs{x_{14}}}{\abs{x_{13}}}\right)^{-\Delta_{34}} \nonumber\\
	& \quad\times C_{12\op{O},i}(x_{12},\partial_2)C_{34\op{O},j}(x_{34},\partial_4)\frac{I^{ij}(x_{24})}{\abs{x_{24}}^{2\Delta}} \label{eq:cfbl}
\end{align}
where $u=\frac{\abs{x_{12}}^2\abs{x_{34}}^2}{\abs{x_{13}}^2\abs{x_{24}}^2}$ and $v=\frac{\abs{x_{14}}^2\abs{x_{23}}^2}{\abs{x_{13}}^2\abs{x_{24}}^2}$
are conformally invariant cross ratios.
In $z$ coordinate, the OPE converges if $\abs{z}<1$\cite{Pappadopulo:2012jk}.
The conformal block in \eqref{eq:cfbl} can be calculated by $C_{ijk,i}(x,P)$, which is fully determined by the conformal symmetry,
and thus does not depend on the details of the CFT we consider.
$(u,v)$ is related with $z$ as $u=\abs{z}^2$ and $v=\abs{1-z}^2$.
In this thesis, we use 4 coordinate systems listed in \autoref{tb:coord}.
\begin{table}[htbp]
	\centering
	\begin{tabular}{l|l|l}
		Coordinate system & Relations & $(x_1,x_2,x_3,x_4)$ \\ \hline
		$z$ & & $(0,z,1,\infty)$ \\ \hline
		$(x,y)$ & $z=x+\sqrt{y}$, $\bar{z}=x-\sqrt{y}$ & \\ \hline
		$(u,v)$ & $u=\abs{z}^2$, $v=\abs{1-z}^2$ & \\ \hline
		$\rho$ & $z=\frac{4\rho}{(1+\rho)^2}$, $\rho=\frac{z}{(1-\sqrt{1-z})^2}$ & $(-\rho,\rho,1,-1)$ \\ \hline
		$(r,\eta)$ & $\rho=re^{i\theta}$, $\eta=\cos\theta$ &
	\end{tabular}
	\caption{Coordinate systems which fix the configuration of four points upto conformal invariance.}
	\label{tb:coord}
\end{table}

The conformal block satisfies
\begin{equation}
	g_{\Delta,l}^{\Delta_{12},\Delta_{34}}(z,\bar{z})\approx\frac{l!}{(-2)^l(\epsilon)_l}
	\abs{z}^\Delta C_l^{(\epsilon)}\left(\frac{\Re z}{\abs{z}}\right)
	\quad\ptext{$z\to 0$}
	\label{eq:gnorm2}
\end{equation}
where $\epsilon=(d-2)/2$ $(x)_n=x(x+1)\cdots(x+n-1)$ is the Pochhammer symbol and $C_n^{(m)}(x)$ is the Gegenbauer polynomial\cite{Poland:2018epd}.
To prove this, we consider the limit $x_{12},x_{34}\to 0$ in \eqref{eq:cfbl}.
Let $(x_1,x_2,x_3,x_4)=(x+y,x,y',0)$ and $y,y'$ be small.
In this limit, we have
\begin{equation}
	u \approx \frac{\abs{y}^2\abs{y'}^2}{\abs{x}^4},\quad
	v \approx 1 - 2 u^{1/2} t
\end{equation}
where
\begin{equation}
	t = \hat{y}_{\mu}I^{\mu\nu}(x)\hat{y'}_\nu = \hat{y}\cdot \hat{y}'
	- \frac{2\left(x\cdot \hat{y}\right)\left(x\cdot \hat{y}'\right)}{\abs{x}^2},
\end{equation}
and $\Re z/\abs{z}\approx t$. Using the leading term of the OPE structure \eqref{eq:opelead} in \eqref{eq:cfbl}, we obtain
\begin{align}
	g_{\Delta,l}^{\Delta_{12},\Delta_{34}}(z,\bar{z})
	&\approx u^{\Delta/2}\hat{y}_{\mu_1}\cdots\hat{y}_{\mu_l}I^{\mu_1\cdots\mu_l;\nu_1\cdots\nu_l}(x)\hat{y}'_{\nu_1}\cdots \hat{y}'_{\nu_l} \\
	&= \abs{z}^{\Delta} \frac{l! C_l^{(\epsilon)}(t)}{(-2)^l(\epsilon)_l},
\end{align}
which shows \eqref{eq:gnorm2}. In the diagonal limit $\bar{z}\to z$, \eqref{eq:gnorm2} takes the form
\begin{equation}
	g_{\Delta,l}^{\Delta_{12},\Delta_{34}}(z,z)\approx\frac{(2\epsilon)_l}{(-2)^l(\epsilon)_l}\abs{z}^\Delta
	\quad\ptext{$z\to 0$}.
	\label{eq:gnorm}
\end{equation}

The conformal block has an algebraic expression
\begin{align}
	&\Braket{0|\phi_3(x_3)\phi_4(x_4)|\op{O}|\phi_1(x_1)\phi_2(x_2)|0}\nonumber\\
	&=\frac{\lambda_{12\op{O}}\lambda_{34\op{O}}}{\abs{x_{12}}^{\Delta_1+\Delta_2}\abs{x_{34}}^{\Delta_3+\Delta_4}}
	\left(\frac{\abs{x_{24}}}{\abs{x_{14}}}\right)^{\Delta_{12}}
	\left(\frac{\abs{x_{14}}}{\abs{x_{13}}}\right)^{\Delta_{34}}g_{\Delta,l}^{\Delta_{12},\Delta_{34}}(z,\bar{z}),
	\label{eq:proj}
\end{align}
where $\abs{\op{O}}$ is the projection operator to the conformal multiplet of $\op{O}$, which is defined by
\begin{equation}
	\abs{\op{O}} = \sum_{\Ket{\alpha},\Ket{\beta}=\Ket{\op{O}},P\Ket{\op{O}},PP\Ket{\op{O}},\ldots}
	\left(G^{-1}\right)^{\alpha\beta}\Ket{\alpha}\Bra{\beta},
\end{equation}
where $G^{\alpha\beta}=\Braket{\alpha|\beta}$ is the Gram matrix and $G^{-1}$ is its inverse.
$\abs{\op{O}}$ is conformal invariant; for any conformal generator $Q$, $\left[Q,\abs{\op{O}}\right]=0$.
From \eqref{eq:proj}, not only $g(u,v)$ but also the conformal block is conformal invariant\cite{Costa:2011dw}.
Acting the quadratic Casimir
\begin{equation}
	C_2 = -\frac{1}{2}J^{ab}J_{ab} = -\frac{1}{2}M^{\mu\nu}M_{\mu\nu}+D(D-d)+P^\mu K_\mu
\end{equation}
($J^{ab}$ is the generator of $SO(d+1,1)$, which is isomorphic to the conformal algebra), we have
\begin{equation}
	\Braket{0|\phi_3(x_3)\phi_4(x_4)|\op{O}|C_2\phi_1(x_1)\phi_2(x_2)|0}
	=\op{C}_2\Braket{0|\phi_3(x_3)\phi_4(x_4)|\op{O}|\phi_1(x_1)\phi_2(x_2)|0},
\end{equation}
where $\op{C}_2$ is the representation of $C_2$ as a differential operator acting on $x_1$ and $x_2$.
All descendants in $\op{O}$ has the same eigenvalue of $C_2$:
\begin{align}
	\abs{\op{O}}C_2 &= c_2\abs{\op{O}}, \\
	c_2 &= \frac{1}{2}\left[l(l+d-2)+\Delta(\Delta-d)\right],
\end{align}
so we obtain the quadratic Casimir equation
\begin{equation}
	\op{D}_2 g_{\Delta,l}^{\Delta_{12},\Delta_{34}} = c_2 g_{\Delta,l}^{\Delta_{12},\Delta_{34}},
	\label{eq:cas2}
\end{equation}
and the quartic Casimir $J_{ab}J^{bc}J_{cd}J^{da}$ gives
\begin{align}
	\op{D}_4 g_{\Delta,l}^{\Delta_{12},\Delta_{34}} &= c_4 g_{\Delta,l}^{\Delta_{12},\Delta_{34}}, \label{eq:cas4} \\
	c_4 &= l(l+d-2)(\Delta-1)(\Delta-d+1).
\end{align}
The explicit form of $\op{D}_2$, $\op{D}_4$ is\cite{Dolan:2003hv,Dolan:2011dv}
\begin{align}
	\op{D}_2
	&= D_z+D_{\bar{z}}+\frac{2\epsilon z\bar{z}}{z-\bar{z}}\left[(1-z)\frac{d}{dz}-(1-\bar{z})\frac{d}{d\bar{z}}\right], \\
	\op{D}_4
	&= \left(\frac{z\bar{z}}{z-\bar{z}}\right)^{2\epsilon}(D_z-D_{\bar{z}})\left(\frac{z\bar{z}}{z-\bar{z}}\right)^{-2\epsilon}(D_z-D_{\bar{z}}),
\end{align}
where $D_z$ is a differential operator
\begin{align}
	D_z &= (1-z)z^2\frac{d^2}{dz^2}-(S+1)z^2\frac{d}{dz}-\frac{P}{2}z, \\
	S &= \frac{-\Delta_{12}+\Delta_{34}}{2}, \\
	P &= -\frac{\Delta_{12}\Delta_{34}}{2}.
\end{align}
For even $d$, there is a closed form for conformal blocks \cite{Dolan:2003hv} in terms of
\begin{equation}
	k_\beta(z) = x^{\beta/2}{}_2 F_1\left(\frac{\beta-\Delta_{12}}{2},\frac{\beta+\Delta_{34}}{2},\beta;z\right),
\end{equation}
We have in $d=2$
\begin{equation}
	g_{\Delta,l}^{\Delta_{12},\Delta_{34}}
	= \frac{1}{(-2)^l\delta_{l0}}\left[k_{\Delta+l}(z)k_{\Delta-l}(\bar{z})+\ptext{$z\swap\bar{z}$}\right],
\end{equation}
and in $d=4$
\begin{equation}
	g_{\Delta,l}^{\Delta_{12},\Delta_{34}}
	= \frac{1}{(-2)^l}\frac{z\bar{z}}{z-\bar{z}}\left[k_{\Delta+l}(z)k_{\Delta-l-2}(\bar{z})-\ptext{$z\swap\bar{z}$}\right],
\end{equation}
and the recursion relation given in \cite{Dolan:2003hv} gives formula for general even $d$.
Odd $d$ is discussed in \autoref{sec:rational}.

\newpage
\section{Global symmetry}
\label{sec:sym}

\subsection{Group theory notations}
\label{sec:group}
Let $G$ be the global symmetry group of the CFT, and we assume that all irreps of $G$ is finite dimensional and can be unitarized.
This assumption is true for all compact groups.
All elements in $G$ commute with the conformal algebra, and the maximal torus we selected is enhanced by that of $G$.
Now a primary operator is classified by scaling dimension $\Delta$, spin $l$ (generally, irrep under $SO(d)$) and irrep $r$ under $G$.
Before talking about a CFT with global symmetry, we introduce some notations for group theoretic data\cite{autoboot}.

Let $\Irr{G}$ be the set of unitary representatives of equivalent (finite-dimensional) irreps.
In other words, for each irrep $r$ of $G$, there exists just one irrep $r'\in \Irr{G}$ such that $r\cong r'$ and $r'$ is a unitary irrep.
$r\in\Irr{G}$ has its representation space $V(r)$ and $r\colon G\to U(V(r))$ maps $g\in G$ to a unitary matrix $r(g)_{ab}$
, for $a,b=1,\ldots,\dim(r)$.
The dual representation $r^*$ is defined by
\begin{equation}
	r^*(g)_{ab}=\overline{r(g)_{ab}}=r\left(g^{-1}\right)_{ba}
\end{equation}
for $g\in G$ and $a,b=1,\ldots,\dim(r)$. Even if $r\in\Irr{G}$ or $r\cong r^*$, $r^*$ does not always belong to $\Irr{G}$.
We denote by $\bar{r}$ the irrep in $\Irr{G}$ that is isomorphic to $r^*$.

We denote the $G$-invariant subspace of $r_1\otimes\cdots\otimes r_n$ by $\inv{r_1,\ldots,r_n}$
and
\begin{equation}
	\inv{r_1,\ldots,r_n|s_1,\ldots, s_m} \colonequals \inv{r^*_1,\ldots,r^*_n,s_1,\ldots,s_m}.
\end{equation}
Let
\begin{equation}
	\Braket{\frac{a_1,\ldots,a_n}{r_1,\ldots,r_n}|\frac{b_1,\ldots,b_m}{s_1,\ldots,s_m}}_k,\quad k=1,\ldots,\dim\inv{r_1,\ldots,r_n|s_1,\ldots,s_m}
\end{equation}
be the orthonormal basis of $\inv{r_1,\ldots,r_n|s_1,\ldots, s_m}$, where $a_i$ (resp. $b_i$) is an index for irrep $r_i$ (resp. $s_i$).
We show that all of the orthonormal bases for $r_i,s_i\in\Irr{G}$ can be explicitly constructed by
the generalized Clebsch-Gordan coefficients $\Set{\frac{c}{t}|\frac{a,b}{r,s}}_n$ for $r,s,t\in\Irr{G}$,
defined by
\begin{equation}
	\Set{\frac{c}{t}|\frac{a,b}{r,s}}_n = \sqrt{\dim{t}}\Braket{\frac{c}{t}|\frac{a,b}{r,s}}_n.
\end{equation}
It is easy to see that
\begin{align}
	\sum_{ab}\overline{\Set{\frac{c}{t}|\frac{a,b}{r,s}}_n}\Set{\frac{c'}{t'}|\frac{a,b}{r,s}}_{n'}
	&= \delta_{tt'}\delta_{nn'}\delta_{cc'}, \\
	\sum_{t\in\Irr{G}}\sum_{c}\sum_{n}\overline{\Set{\frac{c}{t}|\frac{a,b}{r,s}}_n}\Set{\frac{c}{t}|\frac{a',b'}{r,s}}_{n}
	&= \delta_{aa'}\delta_{bb'}.
\end{align}
We symmetrize under $r\leftrightarrow s$ so that for $r\not\cong s$,
\begin{equation}
	\Set{\frac{c}{t}|\frac{a,b}{r,s}}_n = \Set{\frac{c}{t}|\frac{b,a}{s,r}}_n,
\end{equation}
and
\begin{equation}
	\Set{\frac{c}{t}|\frac{a_1,a_2}{r,r}}_n = \sigma_n(t|r,r)\Set{\frac{c}{t}|\frac{a_2,a_1}{r,r}}_n,
\end{equation}
where $\sigma_n(t|r,r)=\pm 1$.
An irrep $r$ is called complex if $r\not\cong\bar{r}$, real if $\sigma_1(\id|r,r)=1$ and pseudo-real otherwise.

When some of $r,s,t$ are the trivial representation $\id\in\Irr{G}$,
\begin{equation}
	\dim\inv{r|r,\id}=\dim\inv{\id|r,\bar{r}}=1,
\end{equation}
and we define
\begin{align}
	\Set{\frac{a}{r}|\frac{a',1}{r,\id}}_1 &= \delta_{aa'}, \\
	\Set{\frac{1}{\id}|\frac{a,\bar{a}}{r,\bar{r}}}_1 &\equalscolon \frac{1}{\sqrt{\dim(r)}}\set{a,\bar{a}}_r.
\end{align}
$\set{a,\bar{a}}_r$ is the intertwining matrix $r^*\to\bar{r}$, which might not be proportional to the identity matrix.

Using these building blocks, we can write down the orthonormal basis of $\inv{r,s,t}$ and $\inv{r_1,r_2,r_3,r_4}$:
\begin{align}
	\Braket{\frac{a,b,c}{r,s,t}}_n
	&= \frac{1}{\sqrt{\dim(t)}}\sum_{\bar{c}}\Set{\frac{\bar{c}}{\bar{t}}|\frac{a,b}{r,s}}_n \Set{\bar{c},c}_{\bar{t}},\\
	\Braket{\frac{a_1,a_2,a_3,a_4}{r_1,r_2,r_3,r_4}}_{s;nm}
	&= \sum_{\bar{b}}\Set{\frac{\bar{b}}{\bar{s}}|\frac{a_3,a_4}{r_3,r_4}}_m \Braket{\frac{a_1,a_2,\bar{b}}{r_1,r_2,\bar{s}}}_n\\
	&= \frac{1}{\sqrt{\dim(s)}}\sum_{b\bar{b}} \Set{\frac{b}{s}|\frac{a_1,a_2}{r_1,r_2}}_n
	\Set{\frac{\bar{b}}{\bar{s}}|\frac{a_3,a_4}{r_3,r_4}}_m\Set{b,\bar{b}}_{s}. \label{eq:4ptG}
\end{align}
The CG coefficients work like OPEs, and orthonormal bases corresponds to correlation functions.
This procedure is quite general to construct all bases of $\inv{r_1,\ldots,r_n}$ just as the calculation of $n$-point function using OPE.
The basis of $\inv{r_1,\ldots,r_n|s_1,\ldots,s_m}$ can be constructed from the basis of $\inv{r_1^*,\ldots,r_n^*,s_1,\ldots,s_m}$
using the intertwiner $r_i^*\to\overline{r_i}$.

Three-point and four-point functions have symmetries under swap of indices and complex conjugation.
First, as we symmetrized CG coefficients,
\begin{equation}
	\Braket{\frac{a,b,c}{r,s,t}}_n = \sigma_n(r,s,t) \Braket{\frac{b,a,c}{s,r,t}}_n
\end{equation}
where $\sigma_n(r,s,t) = \sigma_n(\bar{t}|r,s)$. The cyclic permutation brings $\tau$, which is defined by
\begin{equation}
	\sum_{n}\tau_{nm}(r,s,t)\Braket{\frac{a,b,c}{r,s,t}}_n = \Braket{\frac{b,c,a}{s,t,r}}_m.
\end{equation}
Similarly, the complex conjugation brings $\omega$, which is defined by
\begin{equation}
	\sum_{n}\omega_{nm}(r,s,t)\Braket{\frac{a,b,c}{r,s,t}}_n
	= \sum_{\bar{a}\bar{b}\bar{c}} \Set{\bar{a},a}_{\bar{r}} \Set{\bar{b},b}_{\bar{s}} \Set{\bar{c},c}_{\bar{t}}
	\overline{\Braket{\frac{\bar{a},\bar{b},\bar{c}}{\bar{r},\bar{s},\bar{t}}}_m}.
\end{equation}
For the four-point function, $\op{S}_4$ is generated by $(12)$, $(34)$ and $(24)$.
The actions of $(12)$ and $(34)$ are trivial:
\begin{align}
	\Braket{\frac{a_1,a_2,a_3,a_4}{r_1,r_2,r_3,r_4}}_{s;nm}
	&= \sigma_n(s|r_1,r_2)\Braket{\frac{a_1,a_2,a_3,a_4}{r_1,r_2,r_3,r_4}}_{s;nm} \\
	&= \sigma_m(\bar{s}|r_3,r_4)\Braket{\frac{a_1,a_2,a_3,a_4}{r_1,r_2,r_3,r_4}}_{s;nm}.
\end{align}
\def\six{
\begin{tikzpicture}[baseline=(0)]
	\node (n) at (-30pt,+30pt) {$n$};
	\node (m) at (+30pt,-30pt) {$m$};
	\node (0) at (0,0) {};
	\node (k) at (+30pt,+30pt) {$k$};
	\node (l) at (-30pt,-30pt) {$l$};
	\draw[thick,-] (n) -- node[pos=0.2,below] {$\vphantom{t}s$} (m);
	\draw[thick,-] (k) -- node[pos=0.2,below] {$t$} (l);
	\draw[thick,-] (n) -- node[pos=0.5,below] {$r_1$} (k);
	\draw[thick,-] (k) -- node[pos=0.5,right] {$r_4$} (m);
	\draw[thick,-] (m) -- node[pos=0.5,above] {$r_3$} (l);
	\draw[thick,-] (l) -- node[pos=0.5,left] {$r_2$} (n);
\end{tikzpicture}
}
The action of $(24)$ is nontrivial:
\begin{equation}
	\sum_{s;nm}\six
	\Braket{\frac{a_1,a_2,a_3,a_4}{r_1,r_2,r_3,r_4}}_{s;nm}
	=\Braket{\frac{a_1,a_4,a_3,a_2}{r_1,r_4,r_3,r_2}}_{t;kl},
\end{equation}
which can be solved as
\begin{equation}
	\six=
	\sum_{a_1a_2a_3a_4}\overline{\Braket{\frac{a_1,a_2,a_3,a_4}{r_1,r_2,r_3,r_4}}_{s;nm}}
	\Braket{\frac{a_1,a_4,a_3,a_2}{r_1,r_4,r_3,r_2}}_{t;kl}.
\end{equation}
The vertex $i$ with three edges $r,s,t$ (or $\bar{t}$) corresponds to a term like $\Braket{\frac{c}{t}|\frac{a,b}{r,s}}_n$,
which arises by using \cref{eq:4ptG},
and this tetrahedral object is known as the 6j symbol in the case $G=SU(2)$.

\subsection{CFT with global symmetry}
\label{sec:symcft}
A primary operator with spin $l$ in the irrep $r\in \Irr{G}$ has an index $a$, and we denote it by
\begin{equation}
	\op{O}_{r[a]}^i(x)\quad\ptext{$a=1,\ldots,\dim(r)$}.
\end{equation}
The number of independent OPE coefficients for $\phi_{1,r[a]}\phi_{2,s[b]}\to\op{O}^i_{t[c]}$ is
reduced from $\dim(r\otimes s\otimes t)$ to $\dim\inv{t|r,s}$ by the symmetry under $G$.
For two scalar primary fields $\phi_{1,2}$, we denote their OPE by
\begin{equation}
	\phi_{1,r[a]}(x)\phi_{2,s[b]}(y)
	=\sum_{\op{O}:t}\sum_{n=1}^{\dim\inv{t|r,s}}\lambda_{\phi_1\phi_2\op{O}}^n
	\Set{\frac{c}{t}|\frac{a,b}{r,s}}_n
	C_{\phi_1\phi_2\op{O},k}\left(x-y,\partial_y\right)\op{O}_{t[c]}^k(y),
	\label{eq:symope}
\end{equation}
where
$\op{O}:t$ means that the intermediate primary operator $\op{O}$ is in the representation $t$.
The superscript $k$ is an index of a spin-$\ell$ representation of the rotation group $SO(d)$,
and $C_{\phi_1\phi_2\op{O},k}(x-y,\partial_y)$ describes the contribution of the descendants of $\op{O}$.
A primary operator $\op{O}:r$ with spin-$\ell$ has its dual $\bar{\op{O}}:\bar{r}$
with the same scaling dimension and spin (note that the spin-$\ell$ representation is real).
We normalize the primary operators so that the two-point function is written as
\begin{equation}
	\Braket{\op{O}^i_{r[a]}(x)\op{O}'^{j}_{s[b]}(0)}=\delta_{\op{O}\bar{\op{O}}'}
	\Set{a,b}_r\frac{\sigma(\op{O})I^{ij}(x)}{\abs{x}^{2\Delta}}.
	\label{eq:sym2pt}
\end{equation}
We introduced signs of operators $\sigma(\op{O})=\pm 1$ to preserve the commutativity
even in case that $\Set{a,b}_r$ is anti-symmetric, i.e., $r$ is pseudo-real.
We take $\sigma(\op{O})=1$ when $r$ is complex or strictly real,
and choose signs of $\op{O}$, $\bar{\op{O}}$ so that $\sigma(\op{O})\sigma(\bar{\op{O}})=-1$ when $r$ is pseudo-real.
The conjugate of $\op{O}$ under inversion which is consistent with the two-point functions \eqref{eq:sym2pt} is
\begin{equation}
	\op{O}_{r[a]}^{\mu_1\cdots\mu_l}(x)^\dagger
	= \sum_{\bar{a}}\abs{x}^{-2\Delta}I^{\mu_1}_{\nu_1}(x)\cdots I^{\mu_l}_{\nu_l}(x)
	\sigma\left(\bar{\op{O}}\right)\overline{\Set{\bar{a},a}_{\bar{r}}}\bar{\op{O}}_{\bar{r}[\bar{a}]}^{\nu_1\cdots\nu_l}(x'),
	\label{eq:conj}
\end{equation}
where $x'=x/\abs{x}^2$.

The three-point functions are calculated using \cref{eq:symope,eq:sym2pt} as
\begin{align}
	& \Braket{\phi_{1,r[a]}(x_1)\phi_{2,s[b]}(x_2)\op{O}^i_{t[c]}(x_3)} \nonumber\\
	&= \sum_n\alpha_{\phi_1\phi_2\op{O}}^n\Braket{\frac{a,b,c}{r,s,t}}_n
	\frac{\sigma(\bar{\op{O}})Z^i(x_{13},x_{23})}{\abs{x_{12}}^{\Delta_1+\Delta_2-\Delta}
	\abs{x_{23}}^{\Delta_2+\Delta-\Delta_1}\abs{x_{31}}^{\Delta+\Delta_1-\Delta_2}},
\end{align}
where
\begin{equation}
	\alpha_{\phi_1\phi_2\op{O}}^n=\lambda_{\phi_1\phi_2\bar{\op{O}}}^n\sqrt{\dim(t)}.
\end{equation}

The three-point functions have the following symmetries:
\begin{align}
	\Braket{\phi_{1,r[a]}(x_1)\phi_{2,s[b]}(x_2)\op{O}^i_{t[c]}(x_3)}
	&= \Braket{\phi_{2,s[b]}(x_2)\phi_{1,r[a]}(x_1)\op{O}^i_{t[c]}(x_3)}, \\
	\overline{\Braket{\phi_{1,r[a]}(x_1)\phi_{2,s[b]}(x_2)\op{O}^i_{t[c]}(x_3)}}
	&= \Braket{\phi_{1,r[a]}(x_1)^\dagger\phi_{2,s[b]}(x_2)^\dagger\op{O}^i_{t[c]}(x_3)^\dagger},
\end{align}
and if $\ell=0$,
\begin{equation}
	\Braket{\phi_{1,r[a]}(x_1)\phi_{2,s[b]}(x_2)\phi_{3,t[c]}(x_3)}
	= \Braket{\phi_{2,s[b]}(x_2)\phi_{3,t[c]}(x_3)\phi_{1,r[a]}(x_1)}.
\end{equation}
These symmetries put constraints on OPE coefficients as follows:
\begin{align}
	\alpha_{\phi_1\phi_2\op{O}}^n
	&= \sigma_n(r,s,t)(-1)^l\alpha_{\phi_2\phi_1\op{O}}^n, \label{eq:sigmaope}\\
	\alpha_{\phi_1\phi_2\op{O}}^n
	&= \sigma(\bar{\phi}_1)\sigma(\bar{\phi}_2)\sigma(\op{O})
	\sum_m\omega_{nm}(r,s,t)\overline{\alpha_{\bar{\phi}_1\bar{\phi}_2\bar{\op{O}}}^{m}}, \label{eq:conjope}\\
	\alpha_{\phi_1\phi_2\phi_3}^n
	&= \sigma(\bar{\phi}_1)\sigma(\bar{\phi}_3)\sum_m\tau_{nm}(r_1,r_2,r_3)\alpha_{\phi_2\phi_3\phi_1}^m. \label{eq:tauope}
\end{align}
In case that $G$ is the trivial group, these relations are reduced to well-known facts:
\begin{align}
	\lambda_{\phi_1\phi_2\op{O}} &= (-1)^l\lambda_{\phi_2\phi_1\op{O}}, \\
	\lambda_{\phi_1\phi_2\op{O}} &\in \mathbb{R}, \\
	\lambda_{\phi_1\phi_2\phi_3} &= \lambda_{\phi_2\phi_3\phi_1}.
\end{align}

When $\op{O}$ is an unknown intermediate operator, \cref{eq:sigmaope,eq:conjope,eq:tauope} gives relations among
\begin{equation}
	\alpha_{12\op{O}}^n, \alpha_{21\op{O}}^n, \alpha_{\bar{1}\bar{2}\bar{\op{O}}}^n, \alpha_{\bar{2}\bar{1}\bar{\op{O}}}^n
\end{equation}
for $n=1,\ldots,\dim\inv{r,s,t}$, and when $\op{O}$ is an external scalar $\phi_3$, we have relations among
\begin{equation}
	\alpha_{123}^n, \alpha_{231}^n, \alpha_{312}^n, \alpha_{213}^n, \alpha_{321}^n, \alpha_{132}^n,
	\alpha_{\bar{1}\bar{2}\bar{3}}^n, \alpha_{\bar{2}\bar{3}\bar{1}}^n, \alpha_{\bar{3}\bar{1}\bar{2}}^n,
	\alpha_{\bar{2}\bar{1}\bar{3}}^n, \alpha_{\bar{3}\bar{2}\bar{1}}^n, \alpha_{\bar{1}\bar{3}\bar{2}}^n.
\end{equation}
Since these relations are $\mathbb{R}$-linear, there is a set of independent real parameters $\beta_{12\op{O}}^m$ that generates the solution space.

\subsection{Bootstrap equations}
\label{sec:boot}

The four-point functions takes the form
\begin{align}
	& \Braket{\phi_{1,r_1[a_1]}(x_1)\phi_{2,r_2[a_2]}(x_2)\phi_{3,r_3[a_3]}(x_3)\phi_{4,r_4[a_4]}(x_4)} \nonumber\\
	&\qquad= \frac{1}{\abs{x_{12}}^{\Delta_1+\Delta_2}\abs{x_{34}}^{\Delta_3+\Delta_4}}
	\left(\frac{\abs{x_{24}}}{\abs{x_{14}}}\right)^{\Delta_{12}}
	\left(\frac{\abs{x_{14}}}{\abs{x_{13}}}\right)^{\Delta_{34}} \nonumber\\
	&\qquad \times\sum_{\op{O}:s}\sum_{nm}\sigma(\op{O})\alpha_{\phi_1\phi_2\phi_3\phi_4\op{O}}^{nm}
	\Braket{\frac{a_1,a_2,a_3,a_4}{r_1,r_2,r_3,r_4}}_{s;nm}
	g^{\Delta_{12},\Delta_{34}}_{\op{O}}(u,v),
\end{align}
where
\begin{equation}
	\alpha_{\phi_1\phi_2\phi_3\phi_4\op{O}}^{nm}
	= \sqrt{\dim(s)}\lambda_{12\op{O}}^n\lambda_{34\bar{\op{O}}}^m
	= \lambda_{12\op{O}}^n\alpha_{34\op{O}}^m.
\end{equation}
The bootstrap equation $\Braket{1234}=\Braket{1432}$ symmetrized under $z\swap 1-z$ yields
\begin{align}
	0 &= \sum_s\sum_{\op{O}:s}\sum_{nm}\sigma(\op{O})\alpha_{1234\op{O}}^{nm}
	\Braket{\frac{a_1,a_2,a_3,a_4}{r_1,r_2,r_3,r_4}}_{s;nm}F_{\mp,\op{O}}^{12,34}(u,v) \nonumber\\
	&\pm \sum_t\sum_{\op{O}:t}\sum_{kl}\sigma(\op{O})\alpha_{1432\op{O}}^{kl}
	\Braket{\frac{a_1,a_4,a_3,a_2}{r_1,r_4,r_3,r_2}}_{t;kl}F_{\mp,\op{O}}^{14,32}(u,v)
\end{align}
where
\begin{equation}
	F_{\mp,\op{O}}^{ij,kl}(u,v)
	= v^{\frac{\Delta_k+\Delta_j}{2}}g_{\op{O}}^{\Delta_{ij},\Delta{kl}}(u,v)
	\mp u^{\frac{\Delta_k+\Delta_j}{2}}g_{\op{O}}^{\Delta_{ij},\Delta{kl}}(v,u).
\end{equation}
The inner product with $\Braket{\frac{a_1,a_2,a_3,a_4}{r_1,r_2,r_3,r_4}}_{s;nm}$ gives
\begin{equation}
	0 = F_{\mp,s;nm}^{1234}(u,v) \pm \sum_{t;kl}\six F_{\mp,t;kl}^{14,32}(u,v)
	\label{eq:bootsym}
\end{equation}
where
\begin{equation}
	F_{\mp,s;nm}^{1234}(u,v) = \sum_{\op{O}:s}\sigma(\op{O})\alpha_{1234\op{O}}^{nm}F_{\mp,\op{O}}^{12,34}(u,v).
	\label{eq:symblock}
\end{equation}
In case that $\abs{G}=1$, we have
\begin{equation}
	0 = \sum_{\op{O}}\left[\lambda_{12\op{O}}\lambda_{34\op{O}}F_{\mp,\op{O}}^{1234}(u,v)
	\pm \lambda_{14\op{O}}\lambda_{32\op{O}}F_{\mp,\op{O}}^{14,32}(u,v)\right].
\end{equation}

For a given collection of external primary scalars $\phi_i$ with scaling dimension $\Delta_i$ in the representation $r_i\in\Irr{G}$,
\code{autoboot} generates bootstrap equations which consist of equations in the form \eqref{eq:bootsym}
where externals $\phi_1,\ldots,\phi_4$ are in the given collection.
The summation over intermediates is decomposed into two types: `continuous' and `discrete' sectors.
A `continuous sector' is written in summation of $F_{\mp,\op{O}}^{12,34}(u,v)$ over unknown intermediates $\op{O}$,
and a `discrete sector' has concrete $\op{O}$ (typically, $\op{O}$ itself belongs to given collection of externals).
With this classification and taking real and imaginary parts, the bootstrap equations are collected as one equation:
\begin{equation}
	0 = \sum_{j: \text{disc.}} \boldsymbol{\beta}_j^t \vec{V}_{j} \boldsymbol{\beta}_j
	+ \sum_{j: \text{cont.}}\sum_{\ell}\int \boldsymbol{\beta}_j^t \vec{V}_{j,\Delta} \boldsymbol{\beta}_j d\Delta,
	\label{eq:bootscheme}
\end{equation}
where the subscript $j$ runs over sectors, $\boldsymbol{\beta}_j\in\mathbb{R}^{n_j}$ is a vector of OPE coefficients in the sector $j$
and $\vec{V}_j$ is a vector of symmetric matrices which elements are ($\mathbb{R}$-)linear combinations of $F_{\mp,\op{O}}^{12,34}(u,v)$.

To ensure that the four-point functions are invariant under the permutations of $1234$,
we have other two types of bootstrap equations $\Braket{1234}=\Braket{2134}$ and $\Braket{1234}=\Braket{1243}$.
These equations does not involve OPE coefficients
and can be satisfied by the symmetry between two OPE structures $C_{12\op{O},k}(x,P)$ and $C_{34\op{O},k}(x,P)$,
which read as the properties of the conformal blocks\cite{Poland:2018epd,Dolan:2011dv,Dolan:2000ut}:
\begin{align}
	g_{\Delta,l}^{\Delta_{12},\Delta_{34}}(u/v,1/v)
	&= (-1)^lv^{\Delta_{34}/2}g_{\Delta,l}^{-\Delta_{12},\Delta_{34}}(u,v) \nonumber\\
	&= (-1)^lv^{-\Delta_{12}/2}g_{\Delta,l}^{\Delta_{12},-\Delta_{34}}(u,v),\\
	F_{\mp,\op{O}}^{ij,kl}(u,v) &= F_{\mp,\op{O}}^{ji,lk}(u,v). \label{eq:f_jilk}
\end{align}
These properties can also be shown by the recursion relation in \autoref{eq:rat_rec}, and we also have\footnote{
	\cref{eq:f_jilk,eq:f_klij} are automatically used in \code{autoboot} to reduce bootstrap equations.
}
\begin{equation}
	F_{\mp,\op{O}}^{ij,kl}(u,v) = F_{\mp,\op{O}}^{kl,ij}(u,v). \label{eq:f_klij}
\end{equation}
Only the bootstrap equation from $\Braket{1234}=\Braket{1432}$ is the nontrivial constraints on OPE coefficients
and cannot be satisfied block by block. The $z\to 0$ limit shows that this bootstrap equation cannot be consistent
in case that the CFT has only finite number of primaries.
Indeed, the papers \cite{Fitzpatrick:2012yx,Komargodski:2012ek} showed that the existence of double-twist operators:
if the CFT has two primary operators $\op{O}_1,\op{O}_2$, then we must have a family of operators
$[\op{O}_1\op{O}_2]_n$ for $n=0,1,2,\ldots$, called the double-twist operators, with increasing spins and
twists $\tau=\Delta-l$ approaching $\tau\to\tau_{\op{O}_1}+\tau_{\op{O}_2}+2n$ as $l\to\infty$.
The paper \cite{Simmons-Duffin:2016wlq} analyzes the double-twist operators in the Ising model.

We symmetrized the bootstrap equations under $z\swap 1-z$ and it is natural to expand functions of $z$ such as
$g_{\op{O}}^{\Delta_{ij},\Delta{kl}}(u,v)$, $F_{\mp,\op{O}}^{12,34}(u,v)$
around the crossing symmetric point $z=z_*=1/2$ as
\begin{equation}
	f\left(z=x+\sqrt{y}\right)
	=\sum_{m+2n\leq\Lambda}f_{m,n}(x-1/2)^m y^n+o(\abs{z-1/2}^{\Lambda}),
	\label{eq:taylor}
\end{equation}
with a cutoff parameter $\Lambda$, where $x=\Re z$ and $\sqrt{y}=i\Im z$ (terms with odd power in $y$ vanish from the symmetry
of the conformal blocks under $z\swap\bar{z}$).
In \autoref{sec:numerical}, we discuss the methods to calculate the coefficients $f_{n,m}$ for the conformal blocks.
$z=1/2$ is also the center of the overlapping region of convergence $\abs{z},\abs{1-z}<1$.
At this point, we have $u_*=v_*=1/4$, $(x_*,y_*)=(1/2,0)$, $\rho_*=r_*=3-2\sqrt{2}\approx 0.1716$ and $\theta_*=0$ in \autoref{tb:coord}.

Now consider a linear functional $\alpha$ which acts on a function of $z$ in \eqref{eq:taylor} as
\begin{equation}
	\alpha(f)
	=\sum_{m+2n\leq\Lambda}\alpha_{m,n}f_{m,n}\in\mathbb{R}.
\end{equation}
From \eqref{eq:bootscheme}, a vector $\vec{\alpha}$ of such functionals gives
\begin{equation}
	0 = \vec{\alpha}\cdot\left(\vec{V}_{\texttt{unit}}\right)
	+ \sum_{j: \text{disc.}} \boldsymbol{\beta}_j^t \vec{\alpha}\cdot\left(\vec{V}_{j}\right) \boldsymbol{\beta}_j
	+ \sum_{j: \text{cont.}}\sum_{\ell}\int \boldsymbol{\beta}_j^t \vec{\alpha}\cdot\left(\vec{V}_{j,\Delta}\right) \boldsymbol{\beta}_j d\Delta,
	\label{eq:boot_with_alpha}
\end{equation}
where we separated the contribution from the unit operator $1(x)=1$, which has trivial OPE coefficients
\begin{equation}
	\alpha_{1\op{O}\op{O}'}=\delta_{\op{O}\op{O}'}.
\end{equation}
If there is a vector of functionals such that
\begin{itemize}
	\item $\vec{\alpha}\cdot\left(\vec{V}_{\texttt{unit}}\right)=1$,
	\item for any discrete sector $j$, $\vec{\alpha}\cdot\left(\vec{V}_{j}\right)\succeq 0$,
	\item for any continuous sector $j$, and for any $\Delta$ allowed, $\vec{\alpha}\cdot\left(\vec{V}_{j,\Delta}\right)\succeq 0$,
\end{itemize}
then Eq.~\eqref{eq:boot_with_alpha} contradicts from the reality of $\beta$.
Finding $\alpha$ (using rational approximation discussed in \autoref{sec:rational})
is nothing but a semidefinite problem (SDP), which we discuss further in \autoref{sec:hot}.
The functional vector $\vec{\alpha}$ works as a proof of inconsistency in bootstrap equations
and, as we will see later in \autoref{sec:examples}, this condition is strong enough to put nontrivial constraints on the CFT data.
In our \code{qboot}, \code{find_contradiction} constructs a semidefinite problem to find such a $\vec{\alpha}$.

Next, let us consider a method to extract information about the OPE coefficient in sector $j_0$ from \eqref{eq:boot_with_alpha},
following \cite{Caracciolo:2009bx,Poland:2011ey}.
We fix the direction of OPE coefficients as $\boldsymbol{\beta}_{j_0}=\abs{\beta_0}\boldsymbol{\lambda_0}$
where $\abs{\beta_0}\geq 0$ is an unknown norm. Rewrite \eqref{eq:boot_with_alpha} as
\begin{equation}
	\vec{\alpha}\cdot\left(\vec{V}_{\texttt{unit}}\right)
	=- \abs{\beta_0}^2 \boldsymbol{\lambda}_{0}^t \vec{\alpha}\cdot\left(\vec{V}_{j_0}\right) \boldsymbol{\lambda}_0
	- \sum_{j: \text{disc.}} \boldsymbol{\beta}_j^t \vec{\alpha}\cdot\left(\vec{V}_{j}\right) \boldsymbol{\beta}_j
	- \sum_{j: \text{cont.}}\sum_{\ell}\int \boldsymbol{\beta}_j^t \vec{\alpha}\cdot\left(\vec{V}_{j,\Delta}\right) \boldsymbol{\beta}_j d\Delta,
\end{equation}
and find $\vec{\alpha}$ such that
\begin{itemize}
	\item maximize $\vec{\alpha}\cdot\left(\vec{V}_{\texttt{unit}}\right)$,
	\item $\boldsymbol{\lambda}_{0}^t \vec{\alpha}\cdot\left(\vec{V}_{j_0}\right) \boldsymbol{\lambda}_0=1$,
	\item for any discrete sector $j\neq j_0$, $\vec{\alpha}\cdot\left(\vec{V}_{j}\right)\succeq 0$,
	\item for any continuous sector $j$, and for any $\Delta$ allowed, $\vec{\alpha}\cdot\left(\vec{V}_{j,\Delta}\right)\succeq 0$,
\end{itemize}
then such $\vec{\alpha}$ gives a constraint on $\abs{\beta_0}$ as
\begin{align}
	\vec{\alpha}\cdot\left(\vec{V}_{\texttt{unit}}\right)
	&= -\abs{\beta_0}^2 - \cdots \leq -\abs{\beta_0}^2, \\
	\abs{\beta_0}^2 &\leq -\vec{\alpha}\cdot\left(\vec{V}_{\texttt{unit}}\right). \label{eq:ope_max}
\end{align}
We obtain the tightest upper bound by the $\vec{\alpha}$ which maximizes $\vec{\alpha}\cdot\left(\vec{V}_{\texttt{unit}}\right)$,
which corresponds to a semidefinite programming with a objective function.
The lower bound
\begin{equation}
	\vec{\alpha}'\cdot\left(\vec{V}_{\texttt{unit}}\right) \leq \abs{\beta_0}^2 \label{eq:ope_min}
\end{equation}
also can be obtained from $\vec{\alpha}'$ which minimizes $\vec{\alpha}'\cdot\left(\vec{V}_{\texttt{unit}}\right)$
and satisfies the norm condition
\begin{equation}
	\boldsymbol{\lambda}_{0}^t \vec{\alpha}'\cdot\left(\vec{V}_{j_0}\right) \boldsymbol{\lambda}_0=-1
\end{equation}
and the same semidefiniteness conditions $\vec{\alpha}'\cdot\left(\vec{V}_{j}\right)\succeq 0$.
In our \code{qboot}, \code{ope_maximize} (resp. \code{ope_minimize}) constructs a semidefinite problem to find
the optimal $\vec{\alpha}$ (resp. $\vec{\alpha}'$).

\begin{figure}[htpb]
	\centering
	\begin{tabular}{c}
		\begin{minipage}{0.48\hsize}
			\centering
			\includegraphics[width=\textwidth]{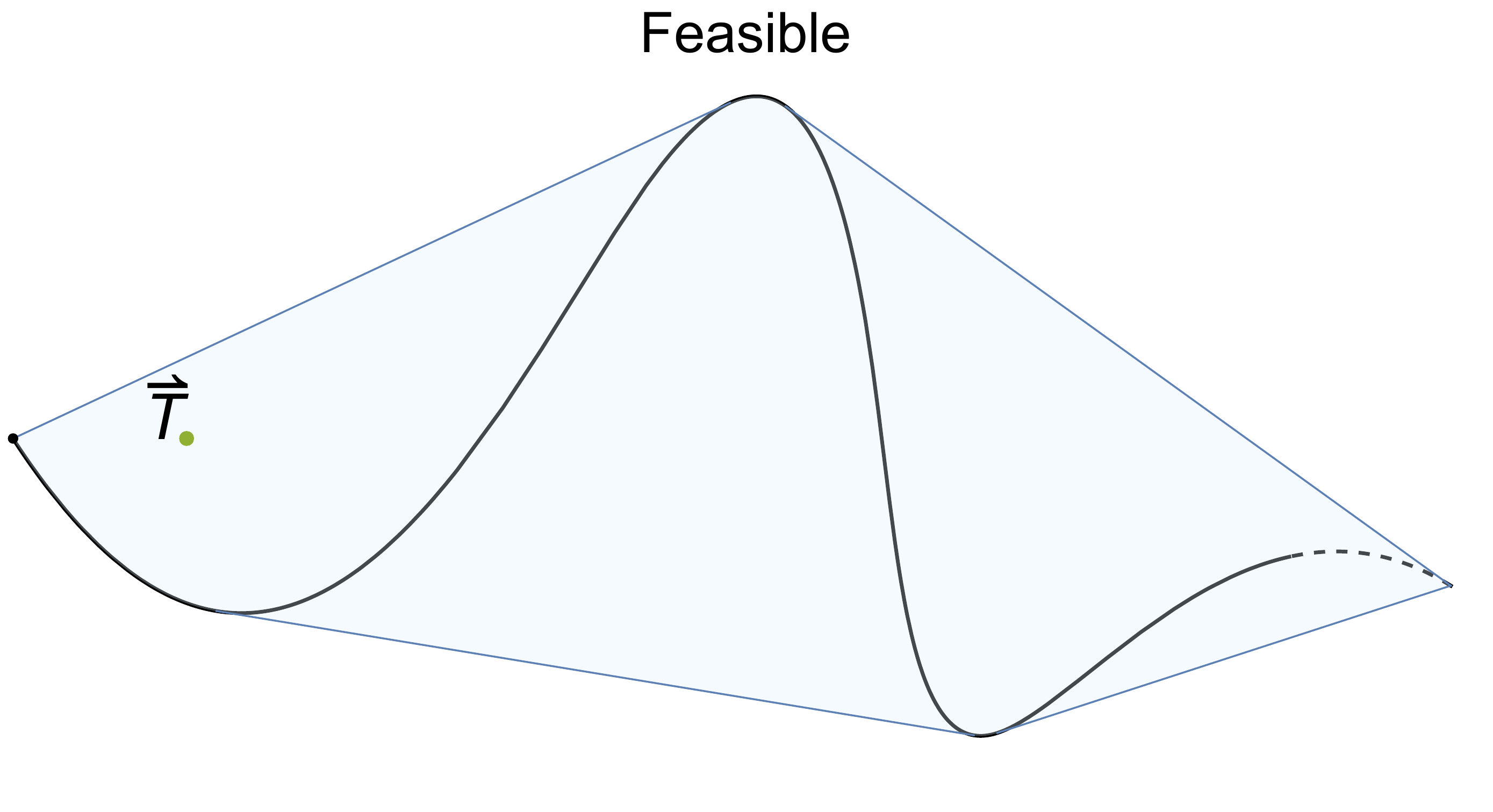}
		\end{minipage}
		\begin{minipage}{0.48\hsize}
			\centering
			\includegraphics[width=\textwidth]{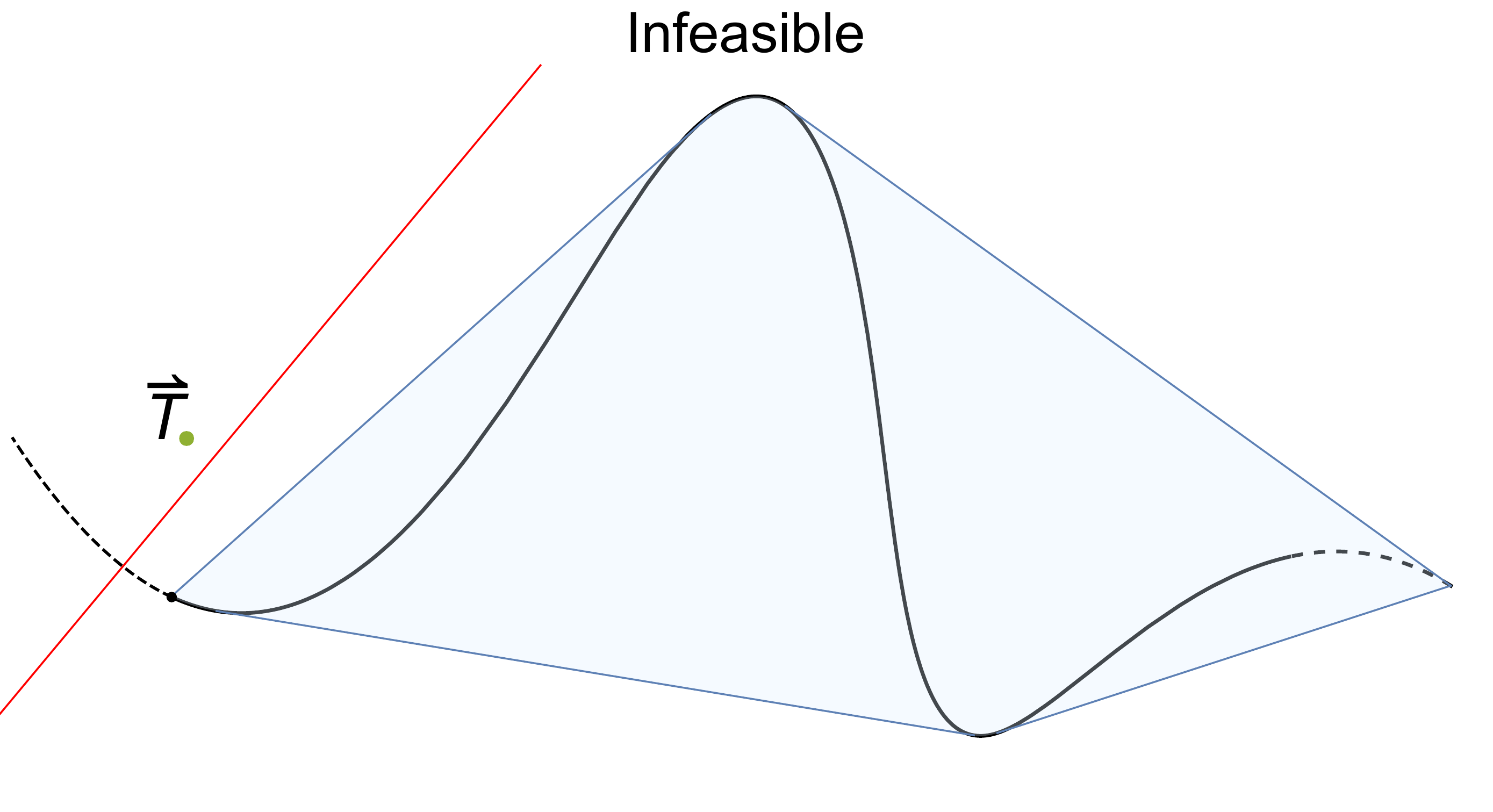}
		\end{minipage}
	\end{tabular}
	\includegraphics[width=0.48\textwidth]{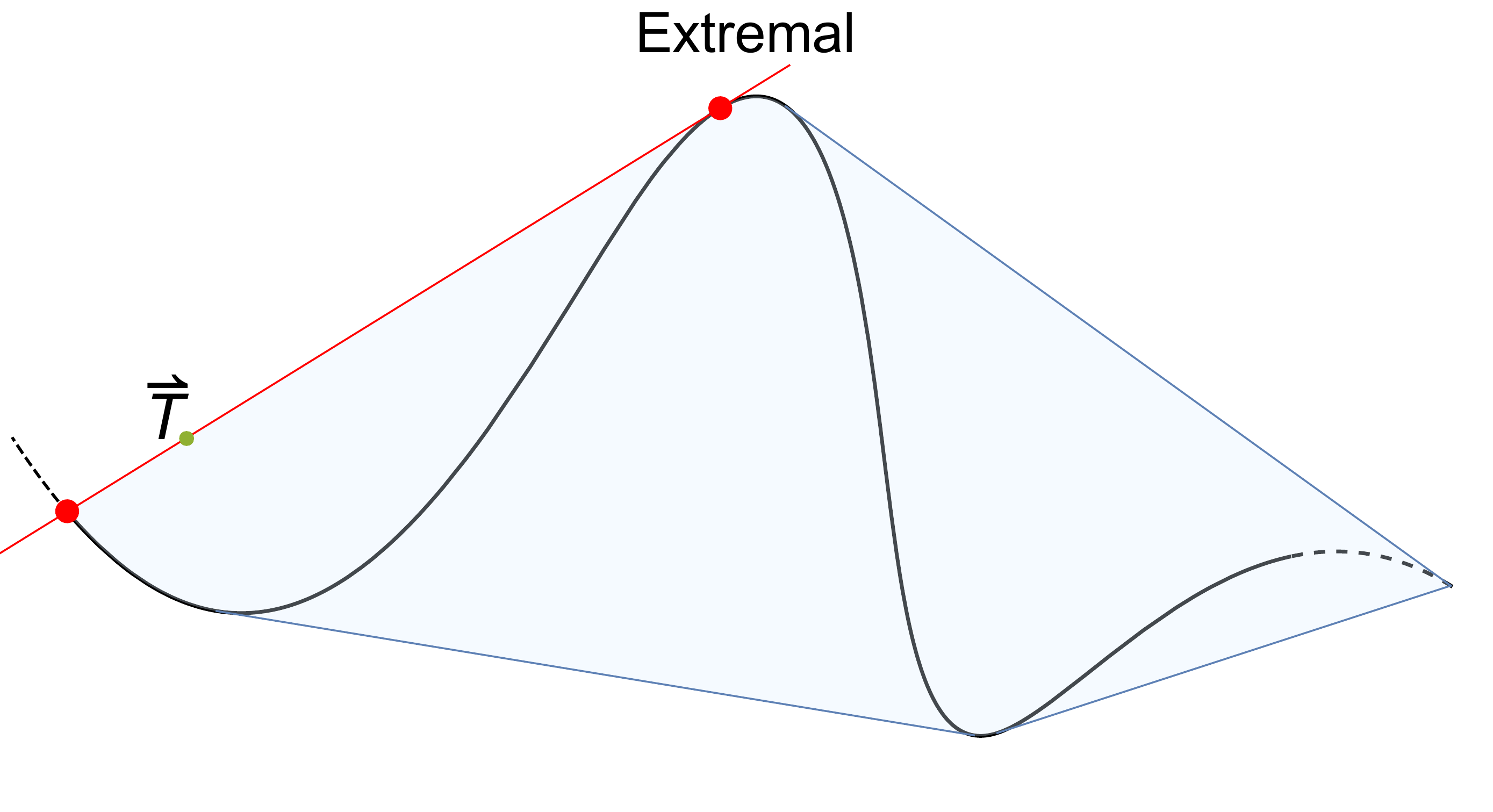}
	\caption{
		Feasible and infeasible cases in conical programming that the target point $\vec{T}$ is included in the conical hull of $S$.
		We consider the case with $N=3$ and that $S$ has one curve (the black line), and the figures are projected on some plane away from the origin.
		A conical hull becomes a convex hull in this projection, and shown as blue colored region.
		Varying $S$ continuously, we have the extremal case as a boundary between feasible and infeasible cases.
		The red line separates $\vec{T}$ from $S$ and is a proof of infeasibilily.
		We have infinitely many separating hyper-planes, and at the extremal case, we have only one line, which defines the extremal functional.
		$\vec{T}$ is expressed as a unique linear combination of two red points on the extremal hyper-plane.
	}
	\label{fig:efm}
\end{figure}

A technique to extract information from the bootstrap equation,
called the extremal functional method (EFM), was introduced in \cite{ElShowk:2012hu,El-Showk:2016mxr}
and used in \cite{Simmons-Duffin:2016wlq} to study the spectrum of the Ising model.
For simplicity, we assume that each sector has only one OPE coefficient: $\boldsymbol{\beta}_j=(\beta_j)$.
Separate the unit sector in \eqref{eq:bootscheme}
\begin{equation}
	-\vec{V}_\texttt{unit} = \sum_{j: \text{disc.}} \beta_j^2 \vec{V}_{j}
	+ \sum_{j: \text{cont.}}\sum_{\ell}\int \beta_j^2 \vec{V}_{j,\Delta} d\Delta,
	\label{eq:target}
\end{equation}
and consider a problem: ``Is there any solution $\Set{\beta_j|j}$ to \eqref{eq:target} upto first $N$ terms in Taylor series around $z_*$?''
Each block $\vec{V}_j$ is regarded as one point and $\set{\vec{V}_{j,\Delta}|\Delta}$ is a curve in $\mathbb{R}^N$.
Let $S=\set{\vec{V}_j|j: \text{disc.}}\cup\set{\vec{V}_{j,\Delta}|j: \text{cont.}}$.
Then the nonnegativity of $\beta_j^2$ implies that the right hand of \eqref{eq:target} defines
a convex cone including all points and curves in all sectors,
and the problem can be rephrased as ``Does the conical hull of $S$ contain $\vec{T}\colonequals-\vec{V}_\texttt{unit}$?'',
and by the duality, also equivalent to a non-existence of a hyper-plane $\alpha\cdot x=0$, $\alpha\in\mathbb{R}^N$ such that
all points and curves in $S$ are on one side $\alpha\cdot x\geq 0$ and $\vec{T}$ is on the other side $\alpha\cdot x<0$.
This is just the condition we discussed above, but gives us a geometric picture of the conformal bootstrap.
A small variation of parameters such as scaling dimensions of externals or the gap in the spectrum
causes variations of $S$ and $\vec{T}$. At the boundary where the target vector $\vec{T}$ is just on the surface of the cone,
we can expect that there is a unique separating hyper-plane and $\alpha$, which is called the extremal functional (\autoref{fig:efm}),
and only terms on the hyper-plane, $\alpha\cdot\vec{V}_{j,\Delta}=0$, can appear in \eqref{eq:target}.
Assuming the uniqueness, the number of zero points is $N$ and we have a spectrum with $N$ primaries.
$\vec{T}$ is also uniquely decomposed as a linear combination of $\vec{V}_{j,\Delta}$ in the spectrum,
and we can calculate OPE coefficients which solve the truncated bootstrap equations.
This is called the EFM and gives us information about CFT data on the boundary of the exclusion plot.
The EFM is also applicable to the case with multiple OPE coefficients by changing the zero point condition to
$\det \left(\alpha\cdot\vec{V}_{j,\Delta}\right)=0$.
Generally, the island shrinks as we take $N\to\infty$, so the truly consistent point resides not on the boundary but inside the island,
and we have to execute the EFM with sufficiently small island to get reasonable results\cite{Simmons-Duffin:2016wlq}.

\section{Numerical methods}
\label{sec:numerical}
In \autoref{sec:series_real}, \autoref{sec:series_trans} and \autoref{sec:rational},
we discuss numerical methods to compute conformal blocks numerically.
In \autoref{sec:hot}, we explain how SDP is used to solve bootstrap equations and also introduce a new technique which we call hot-starting.

\subsection{Diagonal limit for conformal blocks}
\label{sec:series_real}
Let a conformal block $g_{\Delta,l}^{\Delta_{12},\Delta_{34}}$ on real line $\rho=\bar{\rho}$ be expanded around $\rho=0$ as
\begin{align}
	g_{\Delta,l}^{\Delta_{12},\Delta_{34}}(\rho,\rho) &= (4\rho)^{\Delta}\sum_{n=0}^{\infty}b_n\rho^n, \label{eq:rho_series}\\
	b_0 &= \frac{(2\epsilon)_l}{(-2)^l(\epsilon)_l},
\end{align}
where $b_0$ follows from the normalization condition \cref{eq:gnorm}.
We selected $\rho$ coordinate instead of $z$ coordinate according to \cite{Hogervorst:2013sma}.
One can derive ordinary differential equations
\begin{align}
	D_4 g_{\Delta,l}^{\Delta_{12},\Delta_{34}}(\rho,\rho) &= 0 \quad\ptext{$\ell>0$}, \label{eq:d_rho_nonzero}\\
	D_3 g_{\Delta,0}^{\Delta_{12},\Delta_{34}}(\rho,\rho) &= 0, \label{eq:d_rho_zero}
\end{align}
from the Casimir equations \cref{eq:cas2,eq:cas4}, where $D_4$ is some order-4 differential operator of $\rho$ and $D_3$ is order-3 \cite{hor}.
These differential equations imply recursion relations of $b_n$:
\begin{align}
	\sum_{i=0}^7 p_{i,n} b_{n-i} &= 0 \quad\ptext{$l>0$}, \label{eq:rec_nonzero}\\
	\sum_{i=0}^5 p_{i,n} b_{n-i} &= 0 \quad\ptext{$l=0$}, \label{eq:rec_zero}
\end{align}
where $b_n=0$ for $n<0$.
$p_{i,n}$ is a polynomial of $n$ and
\begin{equation}
	p_{0,n} =
	\begin{cases}
		-n(n+2\Delta-2\epsilon-2)(n+\Delta-2\epsilon-l-1)(n+\Delta+l-1) & \ptext{$l>0$}, \\
		n(n+2\Delta-2\epsilon-2)(n+\Delta-1) & \ptext{$l=0$}.
	\end{cases}
\end{equation}
If $\Delta$ is a root of some $p_{0,n}$,
we calculate at $\Delta\to\Delta\pm x$ for some small $x$ and just take their average by continuity of conformal blocks in $\Delta$.
We get approximation of conformal blocks around the crossing symmetric point $\rho=\rho_*=3-2\sqrt{2}$ (and their derivatives on $\rho=\bar{\rho}$)
by taking sufficiently many terms:
\begin{equation}
	\left.\partial_\rho^k g_{\Delta,l}^{\Delta_{12},\Delta_{34}}(\rho,\rho)\right|_{\rho=\rho_*}
	\approx \sum_{n=0}^{n_\texttt{max}}4^{\Delta}b_n(\Delta+n)(\Delta+n-1)\cdots(\Delta+n-k+1)\rho_*^{\Delta+n-k}.
	\label{eq:series}
\end{equation}
To obtain results in $P$ binary digits, the sufficient number for $n_\texttt{max}$ is roughly $n_\texttt{max}>0.4P$.
The explicit form of the polynomials $p_{i,n}$ used in \code{qboot} was given in \code{cboot}\footnote{
	\url{https://github.com/tohtsky/cboot/blob/master/scalar/hor_formula.c}
}.
To get approximation in $z$ coordinate, we can use
\begin{equation}
	z-\frac{1}{2}=\left(2\sqrt{2}r'-\frac{1}{2}r'^2\right)\left(4-2\sqrt{2}+r'\right)^{-2},
\end{equation}
where $r'=\rho-\rho_*$, and (inversed) substitution of variables in
\begin{equation}
	g_{\Delta,l}^{\Delta_{12},\Delta_{34}}(\rho,\rho)
	= \sum_{n=0}^{\Lambda} \left.\partial_\rho^k g_{\Delta,l}^{\Delta_{12},\Delta_{34}}(\rho,\rho)\right|_{\rho=\rho_*} r'^n + o(r'^\Lambda)
\end{equation}
gives
\begin{equation}
	g_{\Delta,l}^{\Delta_{12},\Delta_{34}}(z,z)
	= \sum_{n=0}^{\Lambda} \left.\partial_z^k g_{\Delta,l}^{\Delta_{12},\Delta_{34}}(z,z)\right|_{z=1/2} (z-1/2)^n + o((z-1/2)^\Lambda).
\end{equation}

\subsection{Transverse derivatives of conformal blocks}
\label{sec:series_trans}
The conformal block expanded at $z=\bar{z}$
\begin{equation}
	g_{\Delta,l}^{\Delta_{12},\Delta_{34}}(z,z)=\sum_{k=0}^{\Lambda}g_k(z-1/2)^k+o\left((z-1/2)^{\Lambda}\right).
\end{equation}
is sufficient to recover the transverse components
\begin{equation}
	g_{\Delta,l}^{\Delta_{12},\Delta_{34}}(z,\bar{z})
	= \sum_{n=0}^{\Lambda/2}\sum_{m=0}^{\Lambda-2n}g_{m,n}\left(x-1/2\right)^m y^n+o\left(\abs{z-1/2}^{\Lambda}\right).
\end{equation}
The technique to achieve this was introduced in \cite{ElShowk:2012ht} and generalized to the case with unequal externals in \cite{pycftboot}.
It was derived by rewriting the Casimir equation \cref{eq:cas2} in $(x,y)$ coordinate,
and with our notation, the formula is:
\begin{align}
	& 2 n (2n + 2 \epsilon - 1) \left[g_{m, n} - \left(8 g_{m - 3, n} + 4 g_{m - 2, n} - 2 g_{m - 1, n}\right)\right] \nonumber\\
	&= -(m + 1) (m + 2) g_{m + 2, n - 1} \nonumber\\
	&\quad + 2 (m + 1) \left(2 (\epsilon + S) - m - 4 n + 6)\right) g_{m + 1, n - 1} \nonumber\\
	&\quad + 4 \left[4 c_2 + 4 (m + n - 1) \epsilon + m ^ 2 + 8 m n + 4 n ^ 2\right. \nonumber\\
	&\quad\qquad \left.- 5 m - 2 n - 2 + 2 P + 4 (m + 2 n - 2) S\right] g_{m, n - 1} \nonumber\\
	&\quad + 8 \left[2 (m - 2 n + 1) \epsilon + m ^ 2 + 12 m n + 12 n ^ 2\right. \nonumber\\
	&\quad\qquad \left.- 13 m - 34 n + 22 + 2 P + 2 (m + 4 n - 5) S\right] g_{m - 1, n - 1} \nonumber\\
	&\quad + 8 (m + 1) \left(-2 \epsilon + 3 m + 4 n - 6 + 2 S\right) g_{m + 1, n - 2} \nonumber\\
	&\quad + 4 (m + 1) (m + 2) g_{m + 2, n - 2},
\end{align}
and the boundary conditions are
\begin{align}
	g_{m,0} &= g_m, \\
	g_{m,n} &= 0 \quad \ptext{$m<0$ or $n<0$}.
\end{align}
This technique, combined with the diagonal limit in \autoref{sec:series_real} gives us a method
to compute an arbitrary conformal block $g_{\Delta,l}^{\Delta_{12},\Delta_{34}}$ with fixed $\Delta,l,\Delta_{12},\Delta_{34}$
as a Taylor series around $z=1/2$ upto $m+2n\leq\Lambda$.
Now, the symmetrized object $F_{\mp,\op{O}}^{ij,kl}(u,v)$ defined as
\begin{equation}
	F_{\mp,\op{O}}^{ij,kl}(u,v)
	= v^{\frac{\Delta_k+\Delta_j}{2}}g_{\op{O}}^{\Delta_{ij},\Delta{kl}}(u,v)
	\mp u^{\frac{\Delta_k+\Delta_j}{2}}g_{\op{O}}^{\Delta_{ij},\Delta{kl}}(v,u),
\end{equation}
is also computable, because we can compute the Taylor series of
\begin{equation}
	v^{\frac{\Delta_k+\Delta_j}{2}} = \left((x-1/2)^2-y\right)^{\frac{\Delta_k+\Delta_j}{2}}
\end{equation}
easily and $F_{\mp,\op{O}}^{ij,kl}(u,v)/2$ is just the projection of
$v^{\frac{\Delta_k+\Delta_j}{2}}g_{\op{O}}^{\Delta_{ij},\Delta{kl}}(u,v)$
onto even (resp. odd) sector spanned by $(x-1/2)^m y^n$ with even (resp. odd) $y$.

\subsection{Rational Approximation}
\label{sec:rational}
A conformal block represented by the projection operator
\begin{equation}
	\abs{\op{O}} = \sum_{\Ket{\alpha},\Ket{\beta}=\Ket{\op{O}},P\Ket{\op{O}},PP\Ket{\op{O}},\ldots}
	\left(G^{-1}\right)^{\alpha\beta}\Ket{\alpha}\Bra{\beta},
\end{equation}
in \eqref{eq:proj} shows that $g_{\Delta,l}^{\Delta_{12},\Delta_{34}}$ has poles in case that the Gram matrix
$G^{\alpha,\beta}=\Braket{\alpha|\beta}$ becomes singular.
The singular descendant states which arise at a pole $\Delta=\Delta_*$
are called the null states, and the null states are descendants of a state $\Ket{\chi}$,
which is a descendant of $\op{O}$ and also a primary $K_\mu\Ket{\chi}=0$,
and thus can be gathered into conformal towers included in the tower of $\op{O}$.
Then the residue of $g_{\Delta,l}^{\Delta_{12},\Delta_{34}}$ at $\Delta=\Delta_*$ are described by another conformal block,
and the paper \cite{Penedones:2015aga} proved a recursion relation for conformal blocks in odd $d$:
\begin{align}
	g_{\Delta,l}^{\Delta_{12},\Delta_{34}}(r,\eta) &= (4r)^{\Delta}h_{\Delta,l}^{\Delta_{12},\Delta_{34}}(r,\eta), \\
	h_{\Delta,l}^{\Delta_{12},\Delta_{34}}(r,\eta)
	&= \widetilde{h}_l^{\Delta_{12},\Delta_{34}}(r,\eta)
	+\sum_{i,k}\frac{c_i^{\Delta_{12},\Delta_{34}}(k)r^{n_i(k)}}{\Delta-\Delta_i(k)}
	h_{\Delta_i(k)+n_i(k),l_i(k)}^{\Delta_{12},\Delta_{34}}(r,\eta), \label{eq:rat_rec} \\
	\widetilde{h}_l^{\Delta_{12},\Delta_{34}}(r,\eta)
	&= \frac{(-1/2)^ll!C_l^{(\epsilon)}(\eta)}
	{\left(\epsilon\right)_l(1-r^2)^{\epsilon}(1-2\eta r+r^2)^{1-\Delta_{12}+\Delta_{34}}(1+2\eta r+r^2)^{1+\Delta_{12}-\Delta_{34}}},
\end{align}
which was the main key in the seminal work \cite{Kos:2014bka}.
Here $i$ runs over $\set{1,2,3}$ and $k$ runs over all positive integer for $i=1,2$ and $1\leq k\leq l$ for $i=3$.
\begin{table}[htbp]
	\centering
	\begin{tabular}{c|c|c|c|l}
		$i$ & $\Delta_i(k)$ & $n_i(k)$ & $l_i(k)$ & $k$ runs over \\ \hline
		1 & $1-l-k$ & $k$ & $l+k$ & $1,2,\ldots$ \\
		2 & $1+\epsilon-k$ & $2k$ & $l$ & $1,2,\ldots$ \\
		3 & $1+l+2\epsilon-k$ & $k$ & $l-k$ & $1,2,\ldots,l$
	\end{tabular}
	\caption{Ingredients in the recursion relation of conformal blocks.}
	\label{tb:recursion}
\end{table}
$\Delta_i(k)$, $n_i(k)$ and $l_i(k)$ is defined in \autoref{tb:recursion}, and\footnote{
	In case that $\Delta_{12}=0$ or $\Delta_{34}=0$, we have $c_1(k)=c_3(k)=0$ for odd $k$,
	and the poles $\Delta_1(k)$, $\Delta_3(k)$ vanishes from the conformal blocks.
	This property is incorporated in \code{cboot} and \code{qboot} to reduce the number of poles.
}
\begin{align}
	c_1^{\Delta_{12},\Delta_{34}}(k)
	&= -\frac{(-8)^kk\left(\frac{1-k+\Delta_{12}}{2}\right)_k\left(\frac{1-k+\Delta_{34}}{2}\right)_k}{(k!)^2}, \\
	c_2^{\Delta_{12},\Delta_{34}}(k)
	&= -\frac{(-16)^kk(\epsilon-k)_{2k}\prod_{\pm,j=12,34}\left(\frac{l+\epsilon+1-k\pm\Delta_j}{2}\right)_k}
	{(k!)^2(l+\epsilon-k)_{2k}(l+\epsilon+1-k)_{2k}}, \\
	c_3^{\Delta_{12},\Delta_{34}}(k)
	&= -\frac{(-2)^kk(2\epsilon+l-k)_k\left(\frac{1-k+\Delta_{12}}{2}\right)_k\left(\frac{1-k+\Delta_{34}}{2}\right)_k}
	{(k!)^2(l-k)!(l+\epsilon-k)_k(l+\epsilon+1-k)_k}.
\end{align}
The poles $\Delta_i(k)$ are below the unitarity bound (and the unitarity bound is the maximum pole by definition).
This formula cannot be applied in case that $d$ is even, because $\Delta_1(k)=\Delta_2(k+l+d-2)$ and poles of order-two emerge,
but we can take a fractional dimension around $d$ and the limit to
the even integer reproduces consistent results with the closed-form known for even $d$.
From this recursion relation, a conformal block as a function of $\Delta$ can be approximated in the following form
\begin{equation}
	g_{\Delta,l}^{\Delta_{12},\Delta_{34}}(r,\eta)
	\approx \frac{(4r)^{\Delta}}{\prod_j(\Delta-\Delta_j)}P(\Delta), \label{eq:rat_approx}
\end{equation}
where $\set{\Delta_j}$ is a finite set of poles and $P(\Delta)$ is a polynomial of $\Delta$,
by just taking finite terms in \eqref{eq:rat_rec}.
Note that
the prefactor $(4r)^{\Delta}/\prod_j(\Delta-\Delta_j)$ is automatically positive in case that $(\Delta, l)$ satisfies the unitarity condition.

Eq.~\eqref{eq:rat_rec} depends on $\Delta$ only through the denominator $1/(\Delta-\Delta_i(k))$,
and we have an approximation around $z=1/2$:
\begin{align}
	h_{\Delta,l}^{\Delta_{12},\Delta_{34}}(x,y)
	&\approx \frac{1}{\prod_j(\Delta-\Delta_j)}\sum_{m+2n\leq \Lambda}h_{m,n}(\Delta)(x-1/2)^m y^n + o(\abs{z-1/2}^{\Lambda}), \\
	\deg h_{m,n} &= \nu,
\end{align}
where $\nu$ is the number of poles $\Delta_j$.
$(4r)^\Delta$ can be expanded around the crossing symmetric point $(r_*,\theta_*)=(3-2\sqrt{2},0)$ as
\begin{equation}
	(r/r_*)^\Delta=\sum_{m+2n\leq \Lambda}p_{m,n}(\Delta)(x-1/2)^m y^n + o(\abs{z-1/2}^{\Lambda}),
\end{equation}
with some polynomials $p_{m,n}(\Delta)$ of degree $m+n$, and we obtain
\begin{equation}
	g_{\Delta,l}^{\Delta_{12},\Delta_{34}}(x,y)
	\approx \frac{(4r_*)^\Delta}{\prod_j(\Delta-\Delta_j)}\sum_{m+2n\leq \Lambda}g_{m,n}(\Delta)(x-1/2)^m y^n + o(\abs{z-1/2}^{\Lambda}),
\end{equation}
where the maximum degree of $g_{m,n}(\Delta)$ is $\Lambda+\nu$.
More accurate approximation requires larger value of $\nu$,
and increases the running time of \code{SDPB}, which depends on the degree of polynomials $g_{m,n}(\Delta)$.
We review the technique to suppress the degree \emph{and} improve the accuracy, which was introduced in \cite{Kos:2013tga}
and implemented in \code{PyCFTBoot} and \code{cboot}.
The idea is to take two collection of poles $P_1\subset P_2$ and calculate approximation with $P_2$
and then rewrite it only with poles in $P_1$. The second step is done by replacing contributions from $P_2\setminus P_1$
with those from $P_1$:
\begin{equation}
	\frac{1}{\Delta-\Delta_i} \approx \sum_{j\in P_1}\frac{a_{j}}{\Delta-\Delta_j}. \label{eq:pole_replace}
\end{equation}
The coefficients $a_1,\ldots,a_n$ are chosen by matching the first $n/2$ derivatives at the unitary bound and $\infty$.

\subsection{Semidefinite programming solver and hot-starting}
\label{sec:hot}
We briefly describe a simple method to often significantly reduce the running time of the SDP solver during the numerical bootstrap.

The semidefinite programming solver \code{SDPB} \cite{sdpb} can solve a polynomial matrix program (PMP), which is defined from
$b_n\in\mathbb{R}$ and symmetric matrices $M_j^n(x)$ of size $m_j\times m_j$, whose elements are polynomials of $x$,
as follows:
\begin{align}
	\text{maximize}&\quad \sum_{n=0}^{N-1} b_n y_n+b_N& \text{over}&\quad y\in\mathbb{R}^N,\nonumber\\
	\text{such that}&\quad \sum_{n=0}^{N-1} y_n M_j^n(x)\succeq M_j^N(x)& \text{for all}&\quad x\geq 0, 0\leq j<J.
\end{align}
A SDP can easily be converted into a PMP just by taking constant polynomials,
and as discussed in \cite{sdpb}, a PMP is also convertible to a SDP:
\begin{align}
	\text{maximize}&\quad b\cdot y\colonequals\sum_{n=0}^{N-1} b_n y_n& \text{over}&\quad y\in\mathbb{R}^N, Y\in\mathcal{S}^K,\nonumber\\
	\text{such that}&\quad Y\succeq 0 \nonumber\\
	\text{and}&\quad\operatorname{Tr}(A_p Y)+(By)_p=c_p& \text{for all}&\quad 0\leq p<P.
	\label{eq:dual}
\end{align}
where $\mathcal{S}^K$ is a set of real symmetric matrices of size $K$,
and $A_0,\ldots,A_{P-1}\in\mathcal{S}^K$, $B\in\mathbb{R}^{P\times N}$ and $c\in\mathbb{R}^P$
are defined by $M_j^n(x)$. The key idea in this transformation is a generalization of the following fact:
a polynomial $p(x)$ with real coefficients is nonnegative in $x\geq 0$ if and only if
$p(x)$ is a sum of squares, i.e., there exist positive semidefinite matrices $Y_1,Y_2$ such that
\begin{equation}
	p(x) = \sum_{ij} Y_{1;ij}q_i(x)q_j(x) + x \sum_{ij} Y_{2;ij}q_i(x)q_j(x),
\end{equation}
where $q_i(x)=x^i$, and equivalent to the existence of a symmetric matrices
$Y_1\in\mathcal{S}^{d/2}$, $Y_2\in\mathcal{S}^{(d-1)/2}$ such that
\begin{align}
	\begin{pmatrix}
		Y_1 & 0 \\
		0 & Y_2
	\end{pmatrix} &\succeq 0, \\
	p(x_k) &= \sum_{ij} Y_{1;ij}q_i(x_k)q_j(x_k) + x_k \sum_{ij} Y_{2;ij}q_i(x_k)q_j(x_k),
\end{align}
where $d=\deg p$ and $(x_0,\ldots,x_d)$ are arbitrary but distinct sample points.
The generalization of this theorem to the nonnegativity of a polynomial matrix
allows us to rewrite the $J$ constraints $\sum_{n=0}^{N-1} y_n M_j^n(x)\succeq M_j^N(x)$ for $x\geq 0$
into constraints on the positive-semidefiniteness of $Y_1,\ldots,Y_{2J}$ and sufficient number of linear equations.
\code{SDPB} utilizes the sparsity of a coefficient matrix $A$ and
a combined block-diagonal matrix $Y=\operatorname{diag}(Y_1,\ldots,Y_{2J})$.
The previous study before the appearance of \code{SDPB} relied on older, more generic SDP solver \code{SDPA} \cite{SDPA6,SDPA7,SDPA-GMP},
and the specialized solver \code{SDPB} significantly improved the performance in the study of the conformal bootstrap.

Eq.~\eqref{eq:dual} is related to its dual optimization problem:
\begin{align}
	\text{minimize}&\quad c\cdot x& \text{over}&\quad x\in\mathbb{R}^P, X\in\mathcal{S}^K,\nonumber\\
	\text{such that}&\quad X\succeq 0, \nonumber\\
	&\quad X=\sum_{p} A_p x_p, \nonumber\\
	\text{and}&\quad B^t x=b.
	\label{eq:primal}
\end{align}
We call \eqref{eq:dual} the dual problem and \eqref{eq:primal} the primal problem.

When $(x,X)$ or $(y,Y)$ satisfies the respective constraints in \eqref{eq:primal} or \eqref{eq:dual}, they are called primal or dual feasible.
A primal (resp.~dual) feasible point is called optimal if it has the minimum (resp.~maximum) objective.
The duality gap defined as $c\cdot x-b\cdot y$ is guaranteed to be non-negative for a primal feasible $(x,X)$ and a dual feasible $(y,Y)$.
When the duality gap vanishes, both $(x,X)$ and $(y,Y)$ are the optimal point, and $XY=0$.
A SDP solver starts from an initial point $(x,X,y,Y)$,
which is allowed not to satisfy the equality constraints in \eqref{eq:dual} and \eqref{eq:primal},
and updates the values of $(x,X,y,Y)$ approximately along the central path $XY=\mu I$, $\mu\to +0$
via a generalized Newton search so that they become feasible up to an allowed numerical error we specify.

In the application to the numerical bootstrap, the bootstrap constraints are turned into
a maximization problem of the dual form discussed above.
The aim is to construct an exclusion plot of the scaling dimensions $\Delta_{1,\ldots,n}$ of external operators $\phi_{1,\ldots,n}$.
Depending on the precision we want to impose, we pick a fixed value of $K,N,P$,
and we construct $c$, $B$, $A_p$ as a function of $\Delta_{1,\ldots,n}$.
We often simply set $b=0$ and look for a dual feasible solution.
If one is found, the chosen set of values $\Delta_{1,\ldots,n}$ is excluded.
To construct an exclusion plot, we repeat this operation for many sets of values $\Delta_{1,\ldots,n}$.

In the existing literature before \cite{autoboot}, and in the sample implementations available in the community,
the SDP solver was often repeatedly run with the fixed initial value
$(x,X,y,Y)=(0,\Omega_P I_{K\times K},0,\Omega_D I_{K\times K})$ where $I_{K\times K}$ is the unit matrix and $\Omega_{P,D}$ are real constants.
Our improvement is simple and straightforward: for two sets of nearby input values $\Delta_{1,\ldots,n}$ and $\Delta'_{1,\ldots,n}$,
we reuse the final value $(x_*,X_*,y_*,Y_*)$ for the previous run as the initial value for the next run.
For nearby values of $\Delta_{1,\ldots,n}$, the updates of the values $(x,X,y,Y)$
via the generalized Newton search are expected to follow a similar path.
Therefore, we can expect that reusing the values of $(x,X,y,Y)$ might speed up the running time, possibly significantly.
We call this simple technique the hot-starting of the semidefinite solver.
For this purpose, we contributed a new option \code{--initialCheckpointFile} to \code{SDPB}, so that the initial value of $(x,X,y,Y)$
can be specified at the launch of \code{SDPB}. This technique was later used in \cite{Chester:2019ifh} very effectively.

We have not performed any extensive, scientific measurement of the actual speedup by this technique.
But in our experience, the \code{SDPB} finds the dual feasible solutions about 10 to 20 times faster than starting from the default initial value.

There are a couple of points to watch out in using this technique:
\begin{itemize}
	\item In the original description of \code{SDPB} in \cite{sdpb}, it is written in Sec.~3.4 that
	\begin{quotation}
		In practice, if \code{SDPB} finds a primal feasible solution $(x,X)$ after some number of iterations,
		then it will never eventually find a dual feasible one. Thus, we additionally include the option \code{--findPrimalFeasible}
	\end{quotation}
	and that finding a primal feasible solution corresponds to the chosen set of values $\Delta_{1,\ldots,n}$ is considered allowed.
	This observation does not hold, however, once the hot-start technique is applied.
	We indeed found that often a primal feasible solution is quickly found, and then a dual feasible solution is found later.
	Therefore, finding a primal feasible solution should not be taken as a substitute for never finding a dual feasible solution.
	Instead, we need to turn on options \code{--findDualFeasible} and \code{--detectPrimalFeasibleJump} and turn off \code{--findPrimalFeasible}
	\footnote{
		Walter Landry pointed out that our observation here seems to be related to the bug in \code{SDPB},
		where the primal error was not correctly evaluated.
		This bug is corrected in the \code{SDPB} version 2 \cite{sdpb2}, released in early March 2019.
	}.
	\item
	From our experiences, it is useful to \emph{prepare} the tuple $(x,X,y,Y)$
	by running the \code{SDPB} for two values of $\Delta_{1,\ldots,n}$,
	such that one is known to belong to the rejected region
	and another is known to belong to the accepted region, so that the tuple $(x,X,y,Y)$
	\emph{experiences} both finding of a dual feasible solution
	and detecting of a primal feasible jump.
	Somehow this significantly speeds up the running time of the subsequent runs\footnote{
		In one example we solved using a 14-core CPU, the first two points took about $10$ days,
		and the average runtime in the subsequent runs was about $9.5$ hours per point.
	}.
	\item
	When one reuses the tuple $(x,X,y,Y)$ too many times,
	the control value $\mu$ which is supposed to decrease sometimes mysteriously starts to increase.
	At the same time, one observes that the primal and dual step lengths $\alpha_P$ and $\alpha_D$
	(in the notation of \cite{sdpb}) become very small.
	This effectively stops the updating of the tuple $(x,X,y,Y)$.
	When this happens, it is better to start afresh,
	or to reuse the tuple $(x,X,y,Y)$ from some time ago which did not show this pathological behavior.
\end{itemize}
\newpage

\section{Implementation}
\label{sec:implementation}
In this section, we discuss the actual implementation of the ideas for \code{autoboot} and \code{qboot} explained in previous sections.
In \autoref{sec:qboot_impl}, all symbols provided by \code{qboot} are defined in the namespace \code{qboot}
and we omit \code{qboot::} for simplicity.

\subsection{Group theory data}
\label{sec:group_impl}
In \code{autoboot} we provide a proof-of-concept implementation of the strategy described in the previous section.
For each compact group $G$ to be supported in \code{autoboot},
one needs to provide the following information:
\begin{itemize}
	\item Labels $r$ of irreducible representations together with their dimensions,
	\item The complex conjugation map $r \mapsto \bar r$,
	\item Abstract tensor product decompositions of $r_i\otimes r_j$ into irreducible representations,
	\item Explicit unitary representation matrices of the generators of $G$ for each irreducible representation $r$.
\end{itemize}

Currently we support
small finite groups in the \code{SmallGrp} library \cite{SmallGrp} of the computer algebra system \code{GAP} \cite{GAP}
and small classical groups $SO(2),O(2),SO(3),O(3),U(1),SU(2)$.
For classical groups, these data can in principle be generated automatically,
but at present we implement by hand only a few representations we actually support.

For small finite groups, we use a separate script to extract these data from \code{GAP}
and convert them using a \code{C#} program into a form easily usable from \code{autoboot}.
Currently the script uses \code{IrreducibleRepresentationsDixon} in the \code{GAP} library \code{ctbllib},
which is based on the algorithm described in \cite{Dixon}.
Due to the slowness of this algorithm, the distribution of \code{autoboot} as of December 2019
does not contain the converted data for all the small groups in the \code{SmallGrp} library.

We have in fact implemented two variants, one where matrix elements are computed as rigorous algebraic numbers,
and another where matrix elements are numerically evaluated with arbitrary precision.
The line
\begin{lstlisting}
	<<"group.m"
\end{lstlisting}
or
\begin{lstlisting}
	<<"ngroup.m"
\end{lstlisting}
loads the algebraic or numerical version, respectively.

The invariant tensor
$f(a,b,c)=\text{\code{ope[r,s,t][n][a,b,c]}}$ in $\inv{t|r,s}$ needs to satisfy
\begin{equation}
	\sum_{a'b'}r(g)_{aa'}s(g)_{bb'}f(a',b',c)
	=\sum_{c'}f(a,b,c')t(g)_{c'c}
\end{equation} for the discrete part $g\in G$
and
\begin{equation}
	\sum_{a'}r(x)_{aa'}f(a',b,c)
	+\sum_{b'}s(x)_{bb'}f(a,b',c)
	=\sum_{c'}f(a,b,c')t(x)_{c'c}
\end{equation}
for infinitesimal generators $x\in \mathfrak{g}$,
where we use $r_{aa'}$ for the representation matrices for a representation $r$, etc.
Our \code{autoboot} enumerates these equations from the given explicit representation matrices,
and solves them using \code{NullSpace} and \code{Orthogonalize} of \code{Mathematica}.
We also make sure that for $r=s$ these coefficients are either even or odd under $a\swap b$,
and for $r\neq s$, $\text{\code{ope[r,s,t][n][a,b,c]}}=\text{\code{ope[s,r,t][n][b,a,c]}}$.

The notations in this thesis and in the code are mapped as follows:
\begin{align}
	\text{\code{inv[r,s,t]}} &= \dim\inv{r,s,t}, \\
	\text{\code{ope[r,s,t][n][a,b,c]}} &= \Set{\frac{c}{t}|\frac{a,b}{r,s}}_n, \\
	\text{\code{ope[r][a,b]}} &= \Set{a,b}_r, \\
	\text{\code{cor[r,s,t][n][a,b,c]}} &= \Braket{\frac{a,b,c}{r,s,t}}_n, \\
	\text{\code{cor[}$r_1,r_2,r_3,r_4$\code{][s,n,m][}$a_1,a_2,a_3,a_4$\code{]}}
	&= \Braket{\frac{a_1,a_2,a_3,a_4}{r_1,r_2,r_3,r_4}}_{s;nm}.
\end{align}
Various isomorphisms among the invariant tensors are given by the following:
\begin{align}
	\bm{\textcolor[rgb]{.4,.6,.56}{\sigma}}\text{\code{[r,s,t][n]}} &= \sigma_n(r,s,t),\\
	\bm{\textcolor[rgb]{.4,.6,.56}{\tau}}\text{\code{[r,s,t][n,m]}} &= \tau_{nm}(r,s,t),\\
	\bm{\textcolor[rgb]{.4,.6,.56}{\omega}}\text{\code{[r,s,t][n,m]}} &= \omega_{nm}(r,s,t),\\
	\text{\code{six[}$r_1,r_2,r_3,r_4$\code{][s,n,m,t,k,l]}} &= \six.
\end{align}
These are memoized and (except $\sigma$) can be computed using the inner product of the invariant tensors as explained already.
Since the matrix elements of invariant tensors are often very sparse, and that the dimension of the space of invariant tensors is often simply 1,
our \code{autoboot} uses a quicker method in computing them,
by using only the first $\dim\inv{r,s,t}$ linearly-independent entries of the invariant tensors and actually solving the linear equations.

\subsection{CFT data}
\label{sec:cft_impl}
A primary operator $\op{O}$ in the representation $r$ in $G$,
with the sign $\sigma(\op{O})=p$ and the spin $l$ such that $(-1)^l=q$
is represented by
\begin{equation}
	\op{O}=\text{\code{op[}$\op{O}$\code{,r,p,q]}}.
\end{equation}
The complex conjugate operator is then \code{dualOp[}$\op{O}$\code{]}.
The first argument is the name of the operator;
all intermediate operators share the name \code{op}.
The unit operator is given by $1=\text{\code{op[0,id,1,1]}}$.
To register an external primary scalar operator, call
\begin{equation}
	\text{\code{setOps[\{op[x,r,p,q],...\}]}}.
\end{equation}
As a shorthand, we can use \code{op[x,r]} for \code{op[x,r,1,1]}.

The OPE coefficients are denoted by
\begin{align}
	\bm{\textcolor[rgb]{.4,.6,.56}{\lambda}}\text{\code{[}$\phi_1,\phi_2,\op{O}$\code{][n]}} &= \lambda_{\phi_1\phi_2\op{O}}^n, &
	\bm{\textcolor[rgb]{.4,.6,.56}{\alpha}}\text{\code{[}$\phi_1,\phi_2,\op{O}$\code{][n]}} &= \alpha_{\phi_1\phi_2\op{O}}^n
\end{align}
for intermediate operators $\op{O}$, and by
\begin{align}
	\bm{\textcolor[rgb]{.4,.6,.56}{\mu}}\text{\code{[}$\phi_1,\phi_2,\op{O}$\code{][n]}} &= \lambda_{\phi_1\phi_2\phi_3}^n, &
	\bm{\textcolor[rgb]{.4,.6,.56}{\nu}}\text{\code{[}$\phi_1,\phi_2,\op{O}$\code{][n]}} &= \alpha_{\phi_1\phi_2\phi_3}^n
\end{align}
for external operators $\phi_3=\op{O}$.
Internally, we solve the constraints \eqref{eq:sigmaope}, \eqref{eq:tauope}, \eqref{eq:conjope}
as explained at the end of \autoref{sec:symcft}
and represent them all by linear combinations of real constants
$\bm{\textcolor[rgb]{.4,.6,.56}{\beta}}\text{\code{[}$\phi_1,\phi_2,\op{O}$\code{][m]}}$.

\subsection{Bootstrap equations}
\label{sec:booteq_impl}
The bootstrap equations are obtained by calling \code{bootAll[]}.
When the bootstrap equations are given by $a=0\land b=0\land\cdots$,
the return value of \code{bootAll[]} is $\text{\code{eqn[}$\Set{a,b,\ldots}$\code{]}}$.
Here, $a,b,\ldots$ are given by real linear combinations of \code{sum} and \code{single},
where \code{sum[f,op[x,r,p,q]]} represents
\begin{equation}
	\sum_{\substack{\op{O}:r\\\sigma(\op{O})=p\\(-1)^l=q}}f
\end{equation}
and \code{single[f]} corresponds to just $f$.

Inside the code, the conformal blocks are represented by
\begin{align}
	\text{\code{Fp[a,b,c,d,o]}} &= F_{-,o}^{ab,cd},&
	\text{\code{Hp[a,b,c,d,o]}} &= F_{+,o}^{ab,cd},\\
	\text{\code{F[a,b,c,d]}} &= F_{-,\op{O}}^{ab,cd},&
	\text{\code{H[a,b,c,d]}} &= F_{+,\op{O}}^{ab,cd}.
\end{align}
Then the function $f$ inside \code{single} is a product of two $\bm{\textcolor[rgb]{.4,.6,.56}{\beta}}$'s
and \code{Fp} or \code{Hp},
and the function $f$ inside \code{sum} is a product of two $\bm{\textcolor[rgb]{.4,.6,.56}{\beta}}$'s
and \code{F} or \code{H}.
The function \code{format} gives a more readable representation of the equations.

We convert the bootstrap equations into an SDP following the standard method.
To do this, \code{makeSDP[...]} first finds all the sectors $\op{O}$ in the intermediate channel,
and for each sector $j$, we list all the OPE coefficients $\beta^I_j$ involved in that sector.
We then extract the vector of matrices $\vec{V}^{IJ}_j$ as described in \eqref{eq:bootscheme}.
\code{autoboot} splits this matrix $\vec{V}^{IJ}_j$ into block-diagonal components.

At this point, it is straightforward to convert it into a form understandable by an SDP solver \code{spdb}.
We implemented a function \code{toCboot[...]} which constructs a \code{Python} program which uses \code{cboot} \cite{Nakayama:2016jhq}.
After saving the \code{Python} program to a file, some minor edits will be necessary
to set up the gaps in the assumed spectrum etc.
Then the \code{Python} program will output the XML file which can be fed to \code{SDPB}.
We also implemented a function \code{toQboot[...]} which constructs a \code{C++} program which uses \code{qboot}.

\subsection{Semidefinite programming}
\label{sec:sdp}

\code{SDPB} \cite{sdpb} can solve a polynomial matrix program (PMP), which is defined from
$b_n\in\mathbb{R}$ and symmetric matrices $M_j^n(x)$ of size $m_j\times m_j$, whose elements are polynomials of $x$,
as follows:
\begin{quotation}
	maximize $\sum_{n=0}^{N-1} b_n y_n+b_N$ over $y\in\mathbb{R}^N$,\\
	such that $\sum_{n=0}^{N-1} y_n M_j^n(x)\succeq M_j^N(x)$ for all $x\geq 0$ and $0\leq j< J$.
\end{quotation}
There is an obvious generalization of a PMP, which we call a function matrix program (FMP),
which allows an element of $M_j^n(x)$ at $(r,c)$ to be in the form $\chi_j(x)Q_{j;r,c}^n(x)$,
where $\chi_j(x)$ is a positive function in $x\geq 0$ and $Q_{j;r,c}^n(x)$ is a polynomial of $x$.
An FMP can also be solved by \code{SDPB} by cancelling a positive common factor $\chi_j(x)$ from the inequalities, but as noted in \cite{sdpb},
keeping the natural scale in $\chi_j(x)$ improves performance.
In \code{qboot}, $\chi_j(x)$ is represented by an instance of \code{ScaleFactor}.
A scale factor $\chi_j(x)$ should know the maximum degree $d_j=\max_{n,r,c}\deg Q_{j;r,c}^n$,
because \code{SDPB} requires
\begin{itemize}
	\item bilinear bases $q_m^j(x)$ ($0\leq m\leq d_j/2$),
	\item sample points $x_k^j$ ($0\leq k\leq d_j$),
	\item sample scalings $s_k^j$ ($0\leq k\leq d_j$),
\end{itemize}
and it is natural to take $s_k^j=\chi(x_k^j)$ and $q_m^j(x)$ to be an orthogonal polynomials:
\begin{align}
	\deg q_m^j &= m, \\
	\int_{0}^{\infty} q_m^j(x)q_{m'}^j(x)\chi_j(x)dx &= \delta_{mm'}.
	\label{eq:bil_orth}
\end{align}

In the conformal bootstrap, the typical form of scale factors for a sector with spectrum $[\Delta_*,\infty)$ is
\begin{equation}
	\chi_j(x) = \frac{(4\rho_*)^{x+\Delta_*}}{\prod_i (x+\Delta_*-\Delta_i)},
\end{equation}
where $\Delta_i$ are the poles, discussed in \autoref{sec:rational}.
The integral
\begin{equation}
	\int_{0}^{\infty}\frac{e^{-kx}x^n}{x+p}dx
	= (-p)^n e^{pk} \Gamma(0, pk)+\frac{1}{k^n}\sum_{i=0}^{n-1} (n-i-1)! (-pk)^i
\end{equation}
can be evaluated exactly, therefore we can easily calculate the orthogonal polynomials $q_m^j(x)$.
Sample points can be any sequence of distinct points, and we followed \cite{sdpb} in choosing\footnote{
	This sequence was in \code{Mathematica} script in \code{SDPB}, \url{https://github.com/davidsd/sdpb/blob/master/mathematica/SDPB.m},
	and used in \code{cboot}.
}:
\begin{equation}
	x_k^j=\frac{\pi^2}{-4\log r_*}\cdot\frac{(k-1/4)^2}{d_j+1}.
\end{equation}

Consider a generalized inequality
\begin{equation}
	\forall x\in\left[a,b\right), \chi(x)\sum_{n=0}^{d} M_n x^n\geq 0, \label{eq:ineq_a_b}
\end{equation}
where $\chi(x)$ is a positive function in $a\leq x<b$ and we assume $b>0$, $M_n$ is a symmetric matrix.
A M\"{o}bius transformation\footnote{
	The assumption $b>0$ is needed to ensure $y\geq 0$ in $a\leq x<b$.
	$y=\frac{x-a}{b-x}$ is valid even if $b\leq 0$, and we chose the normalization as in \eqref{eq:mobius}
	to get $y=x-a$ in the limit $b\to\infty$.
}
\begin{equation}
	x = \frac{b(y+a)}{y+b},\quad y = \frac{b(x-a)}{b-x} \label{eq:mobius}
\end{equation}
converts \eqref{eq:ineq_a_b} into
\begin{equation}
	\forall y\in\left[0,\infty\right), \frac{1}{(y+b)^d}\chi\left(\frac{b(y+a)}{y+b}\right)\sum_{n=0}^{d} M_n b^n(y+a)^n(y+b)^{d-n}\geq 0.
\end{equation}
This is a FMP with a new scale factor $\tilde{\chi}(y)=\frac{1}{(y+b)^d}\chi\left(\frac{b(y+a)}{y+b}\right)$ and
we can solve it by \code{SDPB}.
$\tilde{\chi}(y)$, however, does not have a natural orthogonal polynomial associated to it,
because an element of the Gram matrix in \eqref{eq:bil_orth} associated with $\tilde{\chi}$
\begin{equation}
	\int_{0}^{\infty}\tilde{\chi}(y)y^n dy
	= \frac{b^n}{(b(b-a))^{d-1}}\int_{a}^b(x-a)^n(b-x)^{d-n-2}\chi(x) dx
\end{equation}
does not converge for $n\geq d-1$ if $\chi(x)=O(1)$ around $x=b$, and we just take $q_m^j(y)=y^m$ if $b<\infty$.

A (unitary) conformal bootstrap using semidefiniteness has been studied with gap assumptions on the spectrum:
\begin{quote}
	Any operator $(\Delta,l)$ in some sector must satisfy $\Delta_*\leq \Delta$.
\end{quote}
With \code{qboot}, \cref{eq:mobius} allows more complicated assumptions such as:
\begin{quote}
	Any operator $(\Delta,l)$ in some sector must satisfy $\Delta_1\leq \Delta\leq \Delta_2$ or $\Delta_3\leq \Delta$.
\end{quote}
We describe how to write spectrums in \code{qboot} in \autoref{sec:qboot_impl}.

The elements of matrices $M_n$ appearing in an inequality \eqref{eq:ineq_a_b} is a linear combination of
$F_{\mp,\op{O}}^{ab,cd}$. If all primaries $\op{O},a,b,c,d$ are fixed, it is straightforward to
evaluate the element using the methods described in \autoref{sec:series_real} and \autoref{sec:series_trans}.
We also have to evaluate with fixed $a,b,c,d$ but with unknown primary $\op{O}$, as a function of its scaling dimension $\Delta$.
Previously, this has been done by the recursion relation in \autoref{sec:rational}, but we do not use it explicitly.
We use the rational approximation \eqref{eq:rat_rec} only to determine the scaling factor $\chi(x)$ (and the maximum degree),
and evaluate conformal blocks only at sample points $x_k^j$.
This method depends on the structure of the input files for \code{SDPB}:
\code{SDPB} has two types of input. One is the XML file which needs full coefficients of polynomials $\sum_n M_nx^n$,
and the other is a directory with some files\footnote{
	The directory format for \code{SDPB} is not documented in its manual, while the XML format is fully supported.
	We checked that we understand the directory format correctly by asking the developers of \code{SDPB}
	at \url{https://github.com/davidsd/sdpb/issues/37}.
} which can be constructed only by the evaluations $\chi(x_k^j)\sum_n M_n(x_k^j)^n$.
In \code{qboot}, a conformal block is only evaluated with a fixed internal operator and fixed externals
and \code{qboot} does not have to calculate polynomials explicitly and is greatly simplified.
Furthermore, the accuracy of conformal blocks does not depend on the number of poles, which determines the accuracy in rational approximation,
and can be improved, even with less poles, just by taking large cutoff \code{n_Max} in the series expansion of the diagonal limit \eqref{eq:series}.

\subsection{QBoot}
\label{sec:qboot_impl}
% \code{qboot} is implemented with \code{mpfr} \cite{MPFR}, a \code{C} library for multiple-precision floating-point computations
% based on \code{GMP} \cite{GMP}, a \code{C} library for arbitrary precision arithmetic,
% operating on signed integers, rational numbers, and floating-point numbers.
% With the help of these libraries, \code{qboot} works in arbitrary precision.

In \code{qboot}, bootstrap equations are represented by an instance of \code{BootstrapEquation},
which internally has three kinds of information; a list of sectors, a list of bootstrap equations and a reference to \code{Context}.

Each sector is a instance of \code{Sector} and corresponds to a label $j$ in the schematic bootstrap equation in \autoref{sec:boot},
\begin{equation}
	0 = \sum_{j: \text{disc.}} \boldsymbol{\beta}_j^t \vec{V}_{j} \boldsymbol{\beta}_j
	+ \sum_{j: \text{cont.}}\sum_{\ell}\int \boldsymbol{\beta}_j^t \vec{V}_{j,\Delta} \boldsymbol{\beta}_j d\Delta.
	\tag{\ref{eq:bootscheme}}
\end{equation}
A sector has its name of type \code{std::string} and the number $n_j$ of OPE coefficients
($\boldsymbol{\beta}_j\in\mathbb{R}^{n_j}$), and a continuous sector has the associated spectrum.
A discrete sector is then obtained as following.
\begin{lstlisting}
	Sector sec1("name1",2);
	Sector sec2("name2",3,{real("5.6"),real("1.2"),real("-3.4")});
\end{lstlisting}
In this example, \code{sec1} represents a sector with two unknown OPE coefficients
and \code{sec2} represents one with three fixed (upto overall factor) OPE coefficients.
Fixed coefficients arises for the unit sector in any CFT, which has only the trivial OPE coefficients.
Another important example is the stress tensor $T_{\mu\nu}$; the Ward identity fixes the OPE coefficients with one $T_{\mu\nu}$
as \cite{Osborn:1993cr,Poland:2018epd}
\begin{equation}
	\lambda_{\phi\phi T} = -\frac{d}{(d-1)S_d}\Delta_{\phi},\quad
	S_d = \frac{2\pi^{d/2}}{\Gamma(d/2)}.
\end{equation}
In a continuous sector, the spectrum is a list of \code{GeneralPrimaryOperator}.
An instance of \code{GeneralPrimaryOperator} stands for an intermediate operator of spin $l$ and scaling dimension $\Delta$ running over
$\Delta\in[\Delta_0,\Delta_1)$ ($\Delta_1$ is allowed to be $\infty$) and added into a \code{Section} variable \code{sec}
with \code{nj} OPE coefficients by:
\begin{lstlisting}
	Sector sec("name",nj,ContinuousType);
	sec.add_op(l_0,d_0,d_1);
	sec.add_op(l_0,d_2);
	sec.add_op(l_1,d_3);
	sec.add_op(l_2);
\end{lstlisting}
In this example, the spectrum defined by \code{sec} is
\begin{equation}
	\Set{(\Delta,l_0)|d_0\leq\Delta<d_1\lor d_2\leq\Delta}\cup\Set{(\Delta,l_1)|d_3\leq\Delta}\cup\Set{(\Delta,l_2)|\Delta_\texttt{unitary}\leq\Delta},
\end{equation}
where $\Delta_\texttt{unitary}$ is the unitarity bound.

After all sectors are added, bootstrap equations are given by a list of \code{Equation}.
An instance of \code{Equation} corresponds to one line in bootstrap equations,
\begin{equation}
	0 = \sum_{j: \text{disc.}}\sum_{1\leq I,J\leq n_j} \beta^I_j\beta^J_j V^{IJ}_{j}
	+ \sum_{j: \text{cont.}}\sum_{\ell}\int \sum_{1\leq I,J\leq n_j}\beta^I_j\beta^J_j V_{j,\Delta} d\Delta.
\end{equation}
Matrix elements $V^{IJ}_j$ are written as a linear combination of $F_{\mp,\op{O}}^{ab,cd}$,
and $F_-,F_+$ do not appear in one line at the same time.
From this property, an \code{Equation} is assigned a parity, \code{Even} for $F_+$ and \code{Odd} for $F_-$.
One term $p \beta^I_j\beta^J_j F_{\mp,\op{O}}^{ab,cd}$ in a continuous sector $j$ can be added by
\begin{lstlisting}
	eq.add("cont. j", I, J, p, std::array{a, b, c, d});
\end{lstlisting}
and a term $p \beta^I_j\beta^J_j F_{\mp,o}^{ab,cd}$ in a discrete sector $j$ can be added by
\begin{lstlisting}
	eq.add("disc. j", I, J, p, o, std::array{a, b, c, d});
\end{lstlisting}
Here, $a,b,c,d,o$ are instances of \code{PrimaryOperator} and a primary operator with scaling dimension \code{delta}
and spin \code{spin} can be created by
\begin{lstlisting}
	PrimaryOperator op(delta, spin, c);
\end{lstlisting}
in which \code{c} is an instance of \code{Context}.

An instance of \code{Context} is a basic object which provides a context in which conformal blocks are evaluated;
the spacetime dimension \code{dim}, the cutoff \code{n_Max} of the series expansion \eqref{eq:series},
\code{lambda} which controls the number of terms to represent a complex function \eqref{eq:taylor}
and the number of threads \code{p} invoked by \code{Context}.
When the method \code{F_block} in \code{Context} is called,
one of the threads calculates the related conformal block using the methods described in \autoref{sec:series_real} and \autoref{sec:series_trans},
and the result is memoized during the lifetime of \code{Context}.

We show a simple example for two equations
\begin{align}
	0 &= F_{-,1}^{\sigma\sigma,\sigma\sigma}(u,v)
	+ \lambda_{\sigma\sigma\epsilon}^2 F_{-,\epsilon}^{\sigma\sigma,\sigma\sigma}(u,v)
	+ \sum_{l:\text{even}}\lambda_{\sigma\sigma\op{O}}^2F_{-,\op{O}}^{\sigma\sigma,\sigma\sigma}(u,v) \label{eq:sample_ising_1}\\
	0 &= F_{+,1}^{\epsilon\epsilon,\sigma\sigma}(u,v)
	- \lambda_{\sigma\sigma\epsilon}^2 F_{+,\sigma}^{\epsilon\sigma,\sigma\epsilon}(u,v)
	+ \lambda_{\sigma\sigma\epsilon}\lambda_{\epsilon\epsilon\epsilon} F_{+,\epsilon}^{\epsilon\epsilon,\sigma\sigma}(u,v) \nonumber\\
	&\quad- \sum_{\op{O}:\text{odd+}}\sum_{l:\text{even}}\lambda_{\sigma\sigma\op{O}}^2F_{+,\op{O}}^{\epsilon\sigma,\sigma\epsilon}(u,v)
	+ \sum_{\op{O}:\text{odd-}}\sum_{l:\text{odd}}\lambda_{\sigma\sigma\op{O}}^2F_{+,\op{O}}^{\epsilon\sigma,\sigma\epsilon}(u,v) \nonumber\\
	&\quad+ \sum_{\op{O}:\text{even}}\sum_{l:\text{even}}\lambda_{\sigma\sigma\op{O}}\lambda_{\epsilon\epsilon\op{O}}
	F_{+,\op{O}}^{\epsilon\epsilon,\sigma\sigma}(u,v) \label{eq:sample_ising_2}
\end{align}
of the Ising model in \code{qboot}\footnote{
	See \url{https://github.com/selpoG/qboot/tree/master/sample} for a working example.
}, which is discussed in \autoref{sec:example_ising} in detail. The code is as follows:
\begin{lstlisting}
	qboot::mp::global_prec=1000; //L0
	uint32_t n_Max=450,lambda=10,dim=3,numax=5,maxspin=20,parallel=8; //L1
	Context c(n_Max,lambda,dim,parallel); //L2
	Op one{real(0),0,c},s{real("0.5181489"),0,c},e{real("1.412625"),0,c}; //L3
	vector<Sector> secs{{"unit",1,{real(1)}},{"scalar",2}}; //L4
	{
		Sector even("even",2,ContinuousType); //L5
		even.add_op(0,real("3.8"),real("3.83")); //L6
		even.add_op(0,real("6.8")); //L7
		for(uint32_t spin=2;spin<=maxspin;spin+=2)even.add_op(spin);
		secs.push_back(even);
		Sector oddp("odd+",1,ContinuousType); //L8
		oddp.add_op(0,real(3)); //L9
		for(uint32_t spin=2;spin<=maxspin;spin+=2)oddp.add_op(spin);
		secs.push_back(oddp);
		Sector oddm("odd-",1,ContinuousType); //L10
		for(uint32_t spin=1;spin<=maxspin;spin+=2)oddm.add_op(spin);
		secs.push_back(oddm);
	}
	BootstrapEquation boot(c,secs,numax); //L11
	Eq eq1(boot,Odd); //L12
	eq1.add(  "unit",0,0, real(1),one,array{s,s,s,s});
	eq1.add("scalar",0,0, real(1),  e,array{s,s,s,s});
	eq1.add(  "even",0,0, real(1),array{s,s,s,s});
	boot.add_equation(eq1);
	Eq eq2(boot,Even); //L13
	eq2.add(  "unit",0,0, real(1),one,array{e,e,s,s});
	eq2.add("scalar",0,0,real(-1),  s,array{e,s,s,e});
	eq2.add("scalar",0,1, real(1),  e,array{e,e,s,s});
	eq2.add(  "odd+",0,0,real(-1),array{e,s,s,e});
	eq2.add(  "odd-",0,0, real(1),array{e,s,s,e});
	eq2.add(  "even",0,1, real(1),array{e,e,s,s});
	boot.add_equation(eq2);
	boot.finish(); //L14
	auto pmp=boot.find_contradiction("unit",parallel); //L15
	auto input=move(pmp).create_input(parallel); //L16
	input.write("sdp",parallel); //L17
\end{lstlisting}
Let us go through the example line by line:
\begin{itemize}
	\item The line \code{L0} tells \code{MPFR} to use binary $1000$ digits in the significand of floating-point number.
	\item The lines \code{L1} and \code{L2} set parameters and create a \code{Context} \code{c}. We use $8$ threads to calculate conformal blocks.
	\item The line \code{L3} creates three primary scalars, the unit operator, $\sigma$ and $\epsilon$.
	\item The line \code{L4} initializes \code{secs} with two discrete sectors \code{"unit"} and \code{"scalar"}.
	\item The lines \code{L5}, \code{L8} and \code{L10} defines three continuous sectors \code{"even"}, \code{"odd+"} and \code{"odd-"}.
	The scalar part of the \code{"even"} sector is set to $[3.8,3.83)\cup [6.8,\infty)$ as in \code{L6} and \code{L7}, which is a stronger assumption
	than the unitarity $[1/2,\infty)$. The scalar part of the \code{"odd+"} is set to $[3,\infty)$ as in \code{L9}, which corresponds to
	the uniqueness of relevant $\mathbb{Z}_2$-odd scalars in the Ising model.
	\item The line \code{L11} initializes a \code{BootstrapEquation}.
	\item In \code{qboot}, bootstrap equations are described line by line and all the transposition process required to build an input to \code{SDPB}
	are executed internally.
	The line \code{L12} describes \eqref{eq:sample_ising_1} and \code{L13} describes \eqref{eq:sample_ising_2}.
	The OPE coefficients are $1$ in \code{"unit"},
	$(\lambda_{\sigma\sigma\epsilon},\lambda_{\epsilon\epsilon\epsilon})$ in \code{"scalar"},
	$\lambda_{\sigma\sigma\op{O}}$ in \code{"odd+"} and \code{"odd-"}
	and $(\lambda_{\sigma\sigma\op{O}},\lambda_{\epsilon\epsilon\op{O}})$ in \code{"even"}.
	\item The line \code{L14} tells \code{boot} that all lines of bootstrap equations has been added.
	\item The line \code{L15} creates a polynomial matrix program from the bootstrap equations in $8$ threads.
	For now, we can use \code{find_contradiction}, \code{ope_maximize} and \code{ope_minimize} discussed in
	\cref{eq:boot_with_alpha,eq:ope_max,eq:ope_min}.
	The calculation of all conformal blocks are done only in this line.
	\item The line \code{L16} computes data to be fed to \code{SDPB}.
	\item The line \code{L17} writes \code{input} to a directory \code{"sdp"}.
\end{itemize}
Most of computations are executed in \code{parallel} threads, so the runtime is roughly proportional to $1/\texttt{parallel}$.
\code{qboot} can create the output $14-19$ times faster than \code{cboot} on a 14-core 28-thread CPU.
\newpage

\section{Examples}
\label{sec:examples}
\subsection{3d Ising model}
\label{sec:example_ising}
\begin{figure}[htpb]
	\centering
	\includegraphics[width=\textwidth]{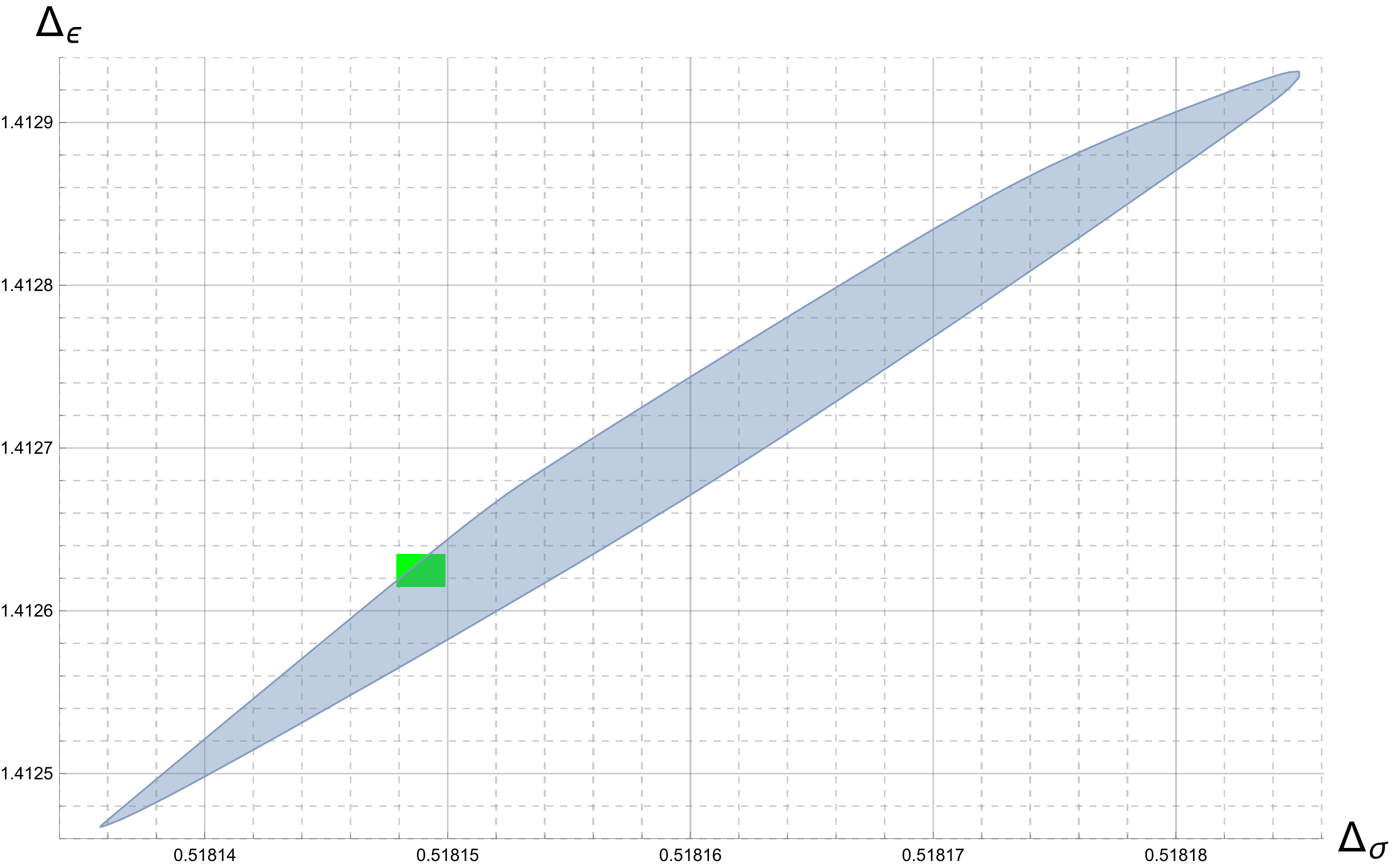}
	\caption{
		The island of the Ising model.
		The green rectangle is the bounding box of most precise result obtained in \cite{Kos:2016ysd},
		and the blue region is our result. Our result is weaker but illustrate the use of \code{qboot}.
	}
	\label{fig:ising-island}
\end{figure}

As discussed in \autoref{sec:intro}, the three dimensional Ising model is $\mathbb{Z}_2$ symmetric
and has one relevant primary scalar $\epsilon$ in the $\mathbb{Z}_2$-even sector
and one relevant scalar $\sigma$ in the odd sector.
Bootstrap equations for $\epsilon,\sigma$ are
\begin{align}
	0 &= F^{\sigma\sigma,\sigma\sigma}_{-,1}
	+ \lambda_{\sigma\sigma\epsilon}^2 F^{\sigma\sigma,\sigma\sigma}_{-,\epsilon}
	+ \sum_{\op{O}^+:\text{even}} \lambda_{\sigma\sigma\op{O}}^2 F^{\sigma\sigma,\sigma\sigma}_{-,\op{O}}, \\
	0 &= F^{\epsilon\epsilon,\epsilon\epsilon}_{-,1}
	+ \lambda_{\epsilon\epsilon\epsilon}^2 F^{\epsilon\epsilon,\epsilon\epsilon}_{-,\epsilon}
	+ \sum_{\op{O}^+:\text{even}} \lambda_{\epsilon\epsilon\op{O}}^2 F^{\epsilon\epsilon,\epsilon\epsilon}_{-,\op{O}}, \\
	0 &= \lambda_{\sigma\sigma\epsilon}^2 F^{\epsilon\sigma,\epsilon\sigma}_{-,\sigma}
	+ \sum_{\op{O}^-:\text{odd}} \lambda_{\epsilon\sigma\op{O}}^2 F^{\epsilon\sigma,\epsilon\sigma}_{-,\op{O}}
	+ \sum_{\op{O}^+:\text{odd}} \lambda_{\epsilon\sigma\op{O}}^2 F^{\epsilon\sigma,\epsilon\sigma}_{-,\op{O}}, \\
	0 &= F^{\epsilon\epsilon,\sigma\sigma}_{-,1}
	+ \lambda_{\sigma\sigma\epsilon}^2 F^{\epsilon\sigma,\sigma\epsilon}_{-,\sigma}
	+ \lambda_{\sigma\sigma\epsilon}\lambda_{\epsilon\epsilon\epsilon} F^{\epsilon\epsilon,\sigma\sigma}_{-,\epsilon} \nonumber\\
	&- \sum_{\op{O}^-:\text{odd}} \lambda_{\epsilon\sigma\op{O}}^2 F^{\epsilon\sigma,\sigma\epsilon}_{-,\op{O}}
	+ \sum_{\op{O}^+:\text{odd}} \lambda_{\epsilon\sigma\op{O}}^2 F^{\epsilon\sigma,\sigma\epsilon}_{-,\op{O}}
	+ \sum_{\op{O}^+:\text{even}} \lambda_{\sigma\sigma\op{O}}\lambda_{\epsilon\epsilon\op{O}} F^{\epsilon\epsilon,\sigma\sigma}_{-,\op{O}}, \\
	0 &= F^{\epsilon\epsilon,\sigma\sigma}_{+,1}
	- \lambda_{\sigma\sigma\epsilon}^2 F^{\epsilon\sigma,\sigma\epsilon}_{+,\sigma}
	+ \lambda_{\sigma\sigma\epsilon}\lambda_{\epsilon\epsilon\epsilon} F^{\epsilon\epsilon,\sigma\sigma}_{+,\epsilon} \nonumber\\
	&+ \sum_{\op{O}^-:\text{odd}} \lambda_{\epsilon\sigma\op{O}}^2 F^{\epsilon\sigma,\sigma\epsilon}_{+,\op{O}}
	- \sum_{\op{O}^+:\text{odd}} \lambda_{\epsilon\sigma\op{O}}^2 F^{\epsilon\sigma,\sigma\epsilon}_{+,\op{O}}
	+ \sum_{\op{O}^+:\text{even}} \lambda_{\sigma\sigma\op{O}}\lambda_{\epsilon\epsilon\op{O}} F^{\epsilon\epsilon,\sigma\sigma}_{+,\op{O}}.
\end{align}
Here, $\op{O}^{\pm}:\text{parity}$ means that $\op{O}$ has spin $l$ with $(-1)^l=\pm$ and belongs to $\text{parity}$ sector of $\mathbb{Z}_2$.
In the generated \code{C++} program for \code{qboot}, we used following parameters
\begin{itemize}
	\item \code{global_prec} $=1000$, $n_\texttt{max}=450$, $\Lambda=27$, \code{numax} $=10$,
	\item the spectrum of $\mathbb{Z}_2$-even, $l=0$ is $[3,\infty)$ and the spectrum of $\mathbb{Z}_2$-even, $l=0$ is $[3,\infty)$,
	\item $l=\Set{0,1,\ldots,26}\cup\Set{29,30,33,34,37,38,41,42,45,46,49,50}$.
\end{itemize}
We have to take a finite set of spins to get a finite size of SDP and our choice is the same as in \cite{sdpb}.
Solving problems generated by \code{find_contradiction} gives
the island in the $(\Delta_\sigma,\Delta_\epsilon)$ space shown in \autoref{fig:ising-island}.

\begin{figure}[htpb]
	\centering
	\includegraphics[width=\textwidth]{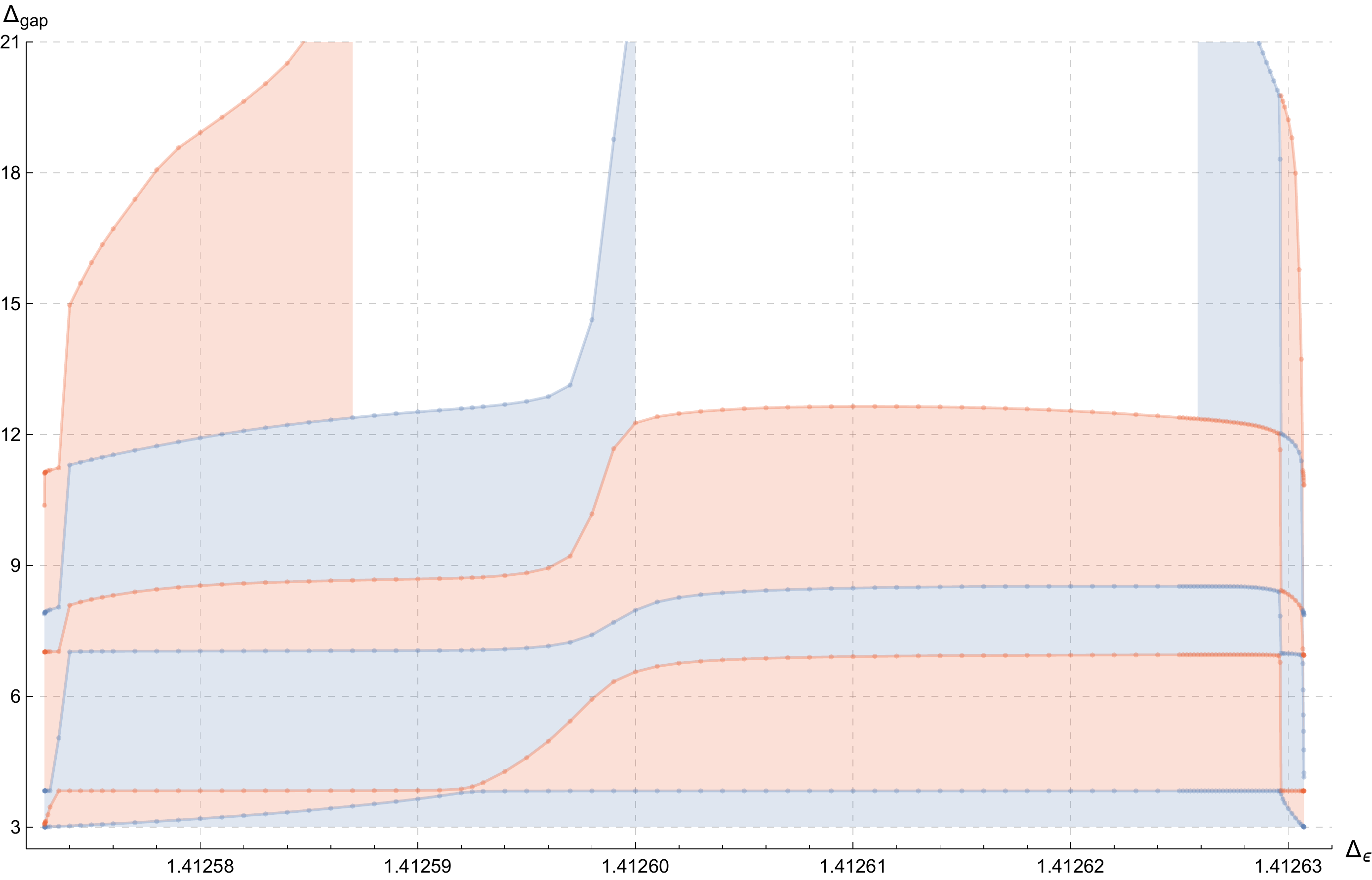}
	\caption{
		The Ising model must have at least one $\mathbb{Z}$-even spin-$0$ primary operator
		in each colored region (assuming that $\Delta_\sigma=0.518149$).
		We showed the first six regions surrounded by $3<f_1<\cdots<f_6$.
		$f_5$ and $f_6$ have divergences shown as white regions.
	}
	\label{fig:gap-e}
\end{figure}

\begin{figure}[htpb]
	\centering
	\begin{tabular}{cc}
		\begin{minipage}{0.47\hsize}
			\centering
			$\Delta_\texttt{gap}\approx 3.83$
			\includegraphics[width=\textwidth]{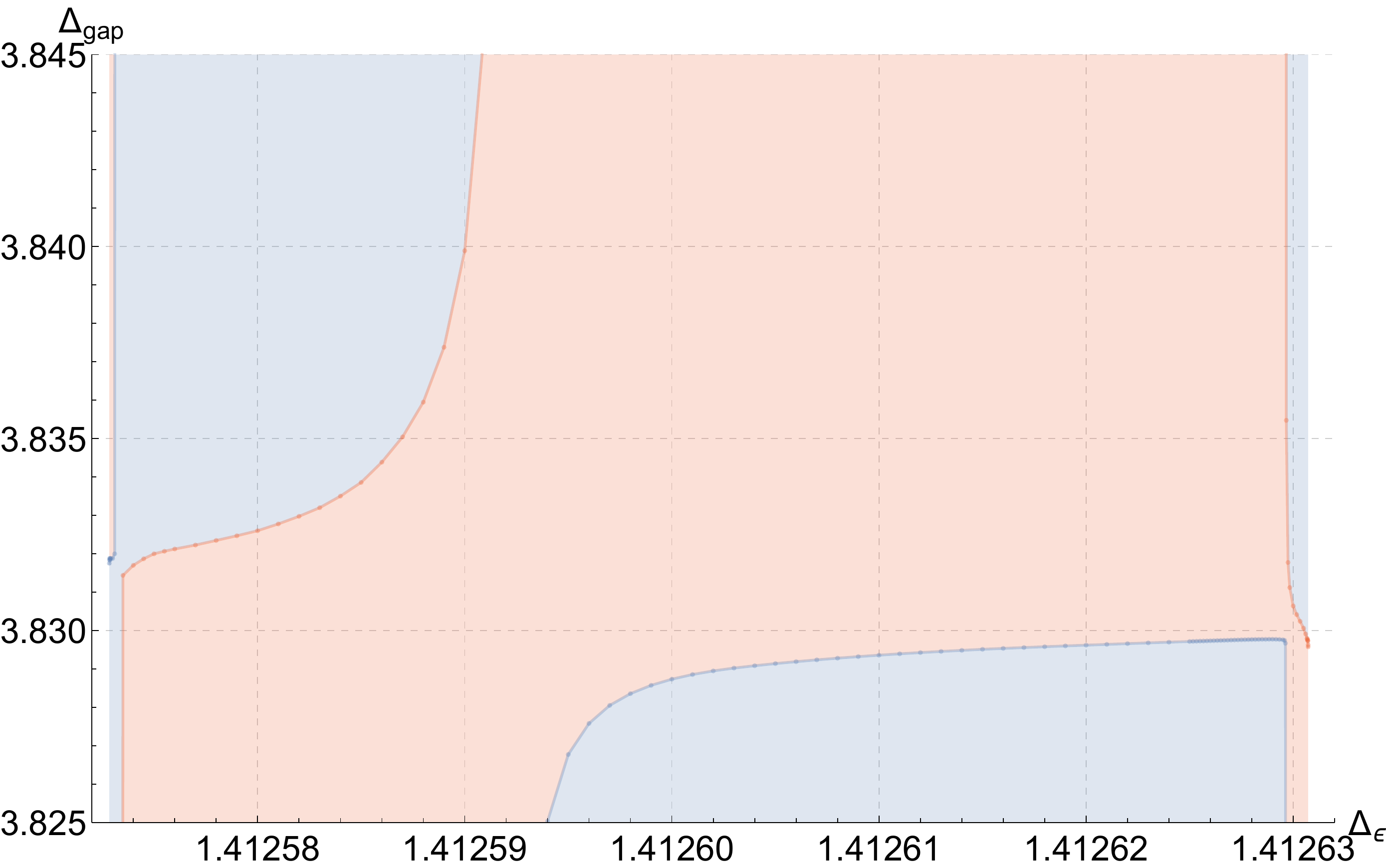}
		\end{minipage}&
		\begin{minipage}{0.47\hsize}
			\centering
			$\Delta_\texttt{gap}\approx 7.0$
			\includegraphics[width=\textwidth]{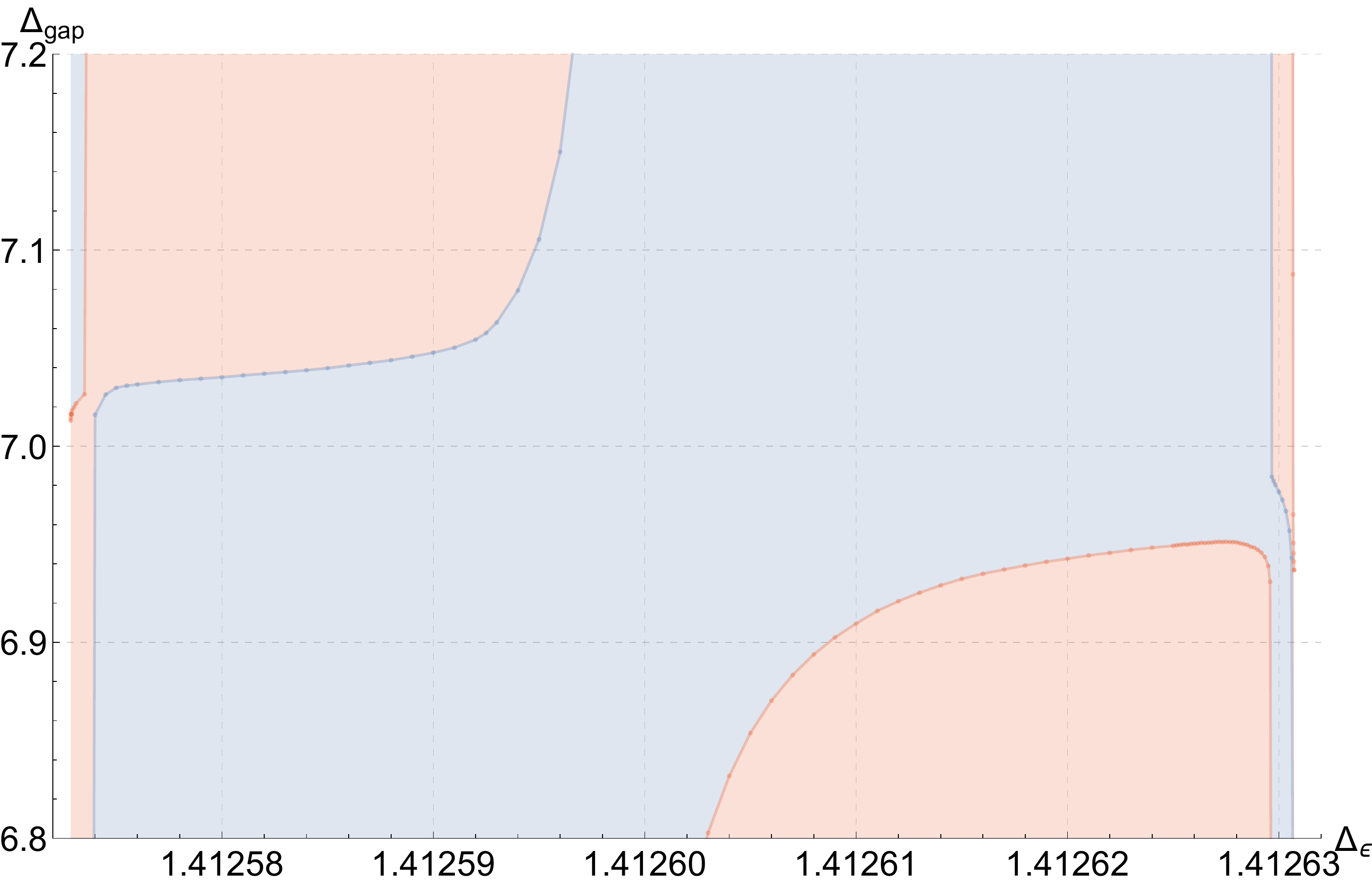}
		\end{minipage}\\\\
		\begin{minipage}{0.47\hsize}
			\centering
			$\Delta_\texttt{gap}\approx 8.5$
			\includegraphics[width=\textwidth]{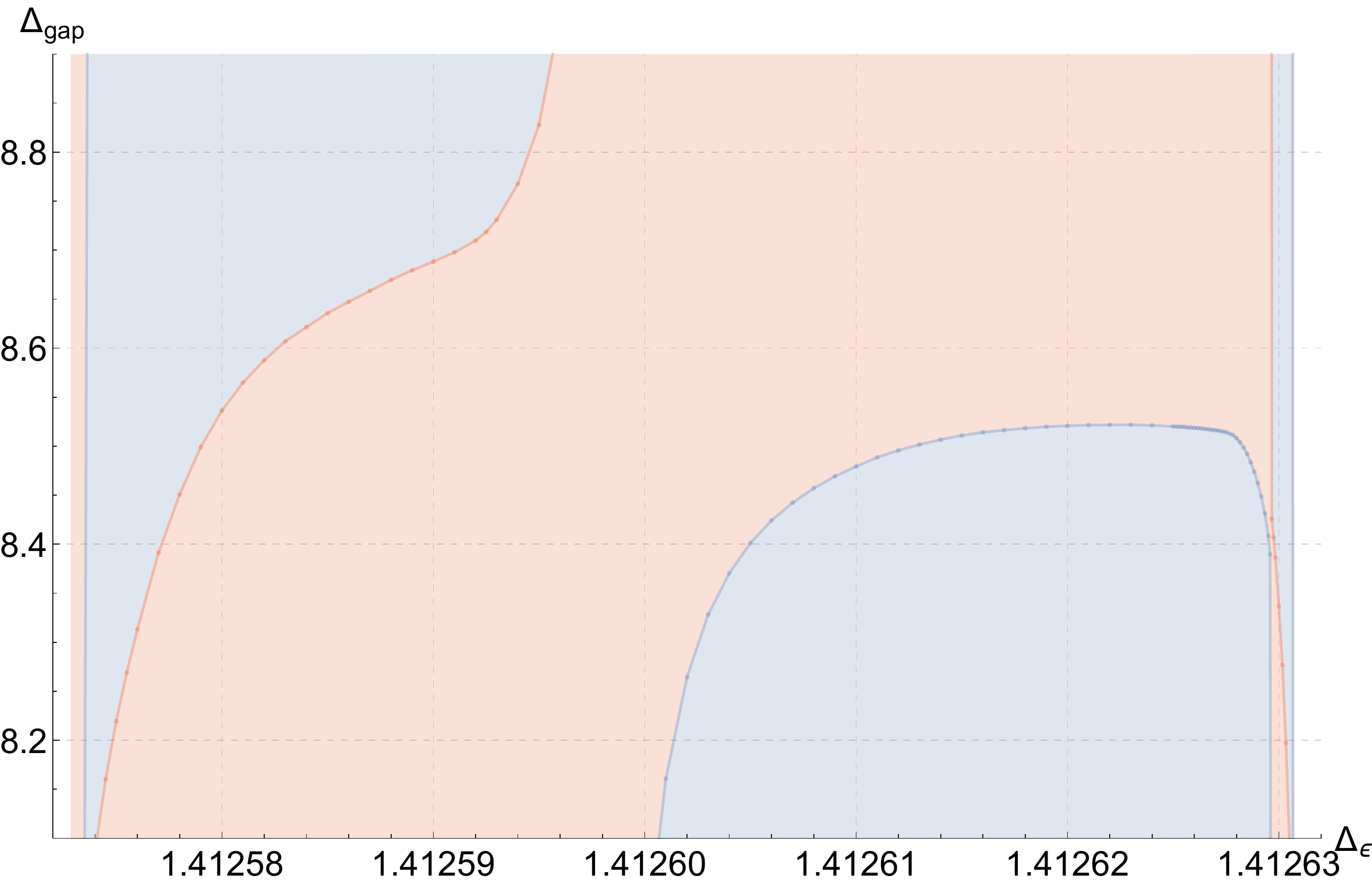}
		\end{minipage}&
		\begin{minipage}{0.47\hsize}
			\centering
			$\Delta_\texttt{gap}\approx 12$
			\includegraphics[width=\textwidth]{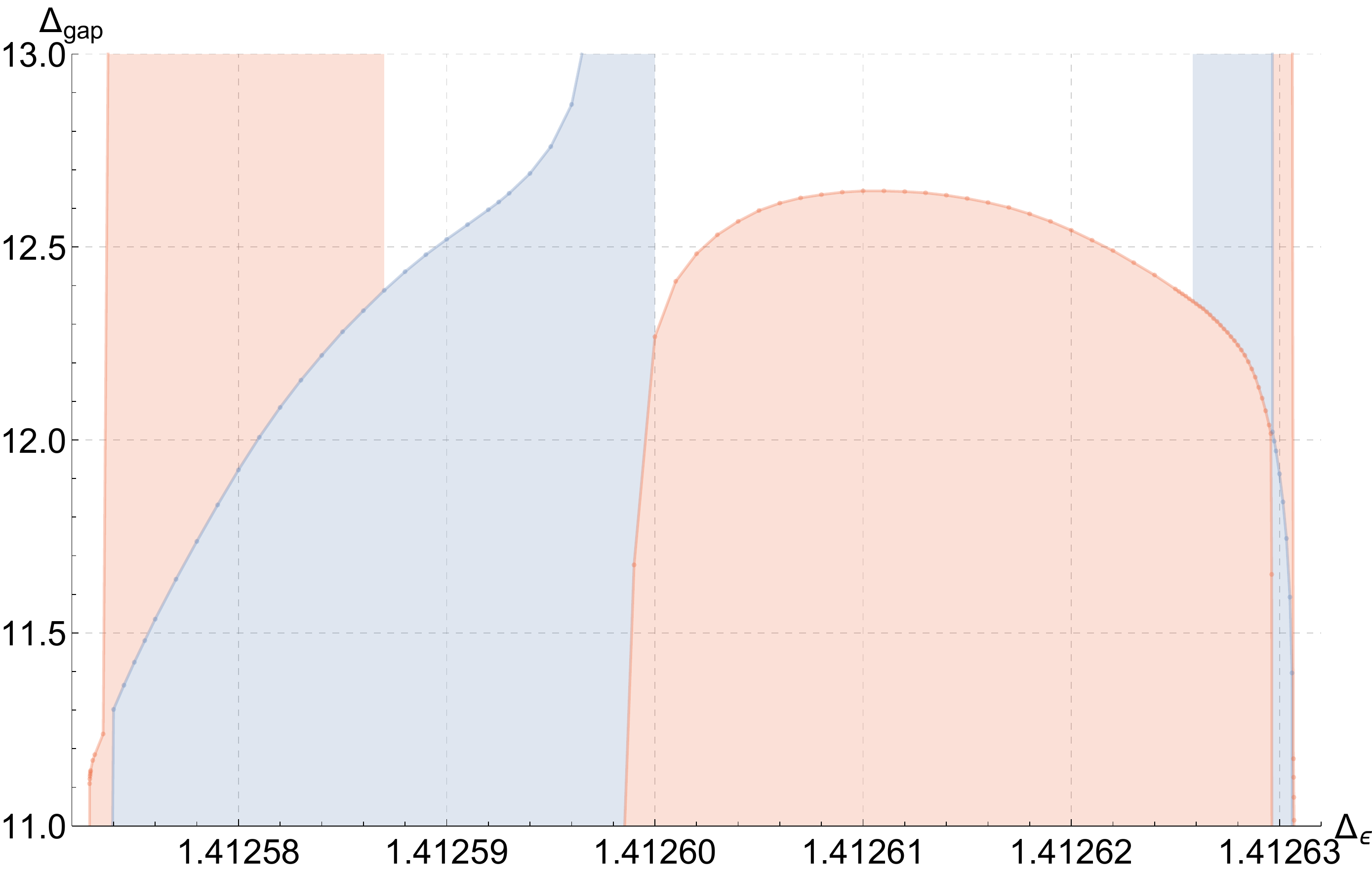}
		\end{minipage}
	\end{tabular}
	\caption{
		Enlarged plots of \autoref{fig:gap-e} (around $\Delta_\texttt{gap}\approx 3.83, 7.0, 8.5, 12$, respectively).
	}
	\label{fig:gap-e-zoom}
\end{figure}

\begin{figure}[htpb]
	\centering
	\includegraphics[width=\textwidth]{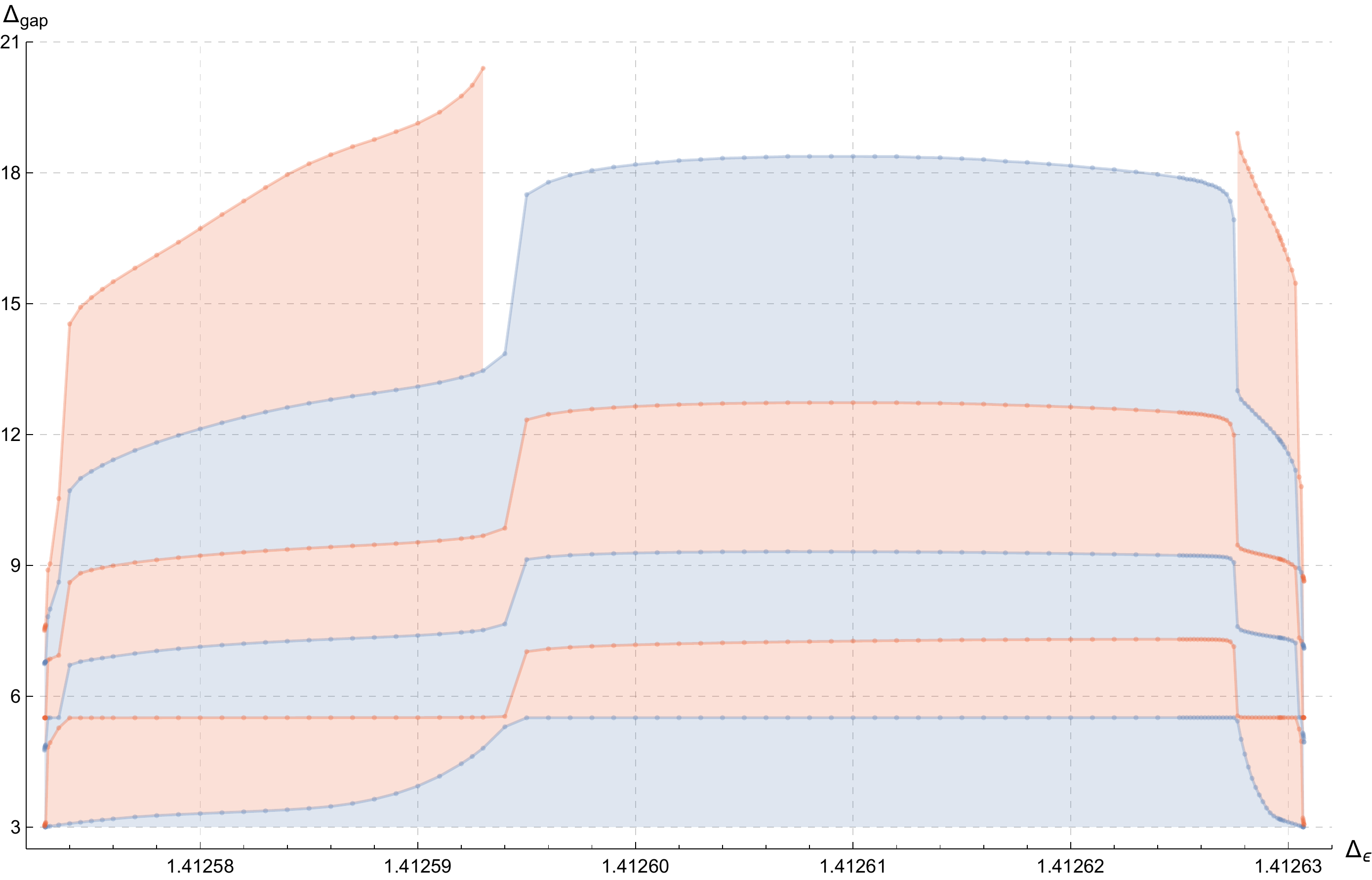}
	\caption{
		The Ising model must have at least one $\mathbb{Z}$-even spin-$2$ primary operator
		in each colored region (assuming that $\Delta_\sigma=0.518149$).
		We showed the first six regions surrounded by $3<g_1<\cdots<g_6$.
		$g_6$ has a divergence shown as a white region.
	}
	\label{fig:gap-t}
\end{figure}

\begin{figure}[htpb]
	\centering
	\begin{tabular}{cc}
		\begin{minipage}{0.47\hsize}
			\centering
			$\Delta_\texttt{gap}\approx 5.51$
			\includegraphics[width=\textwidth]{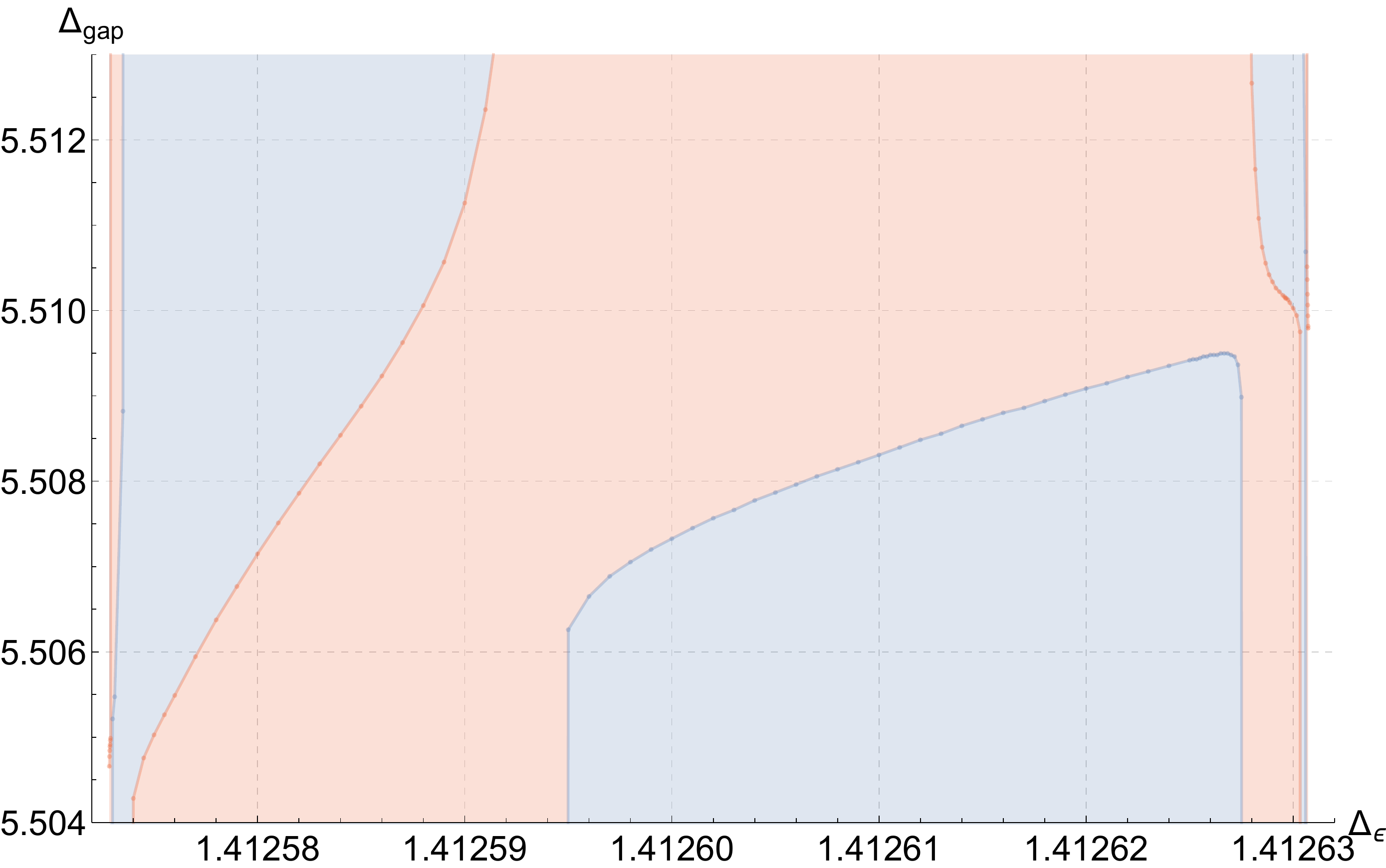}
		\end{minipage}&
		\begin{minipage}{0.47\hsize}
			\centering
			$\Delta_\texttt{gap}\approx 7.2$
			\includegraphics[width=\textwidth]{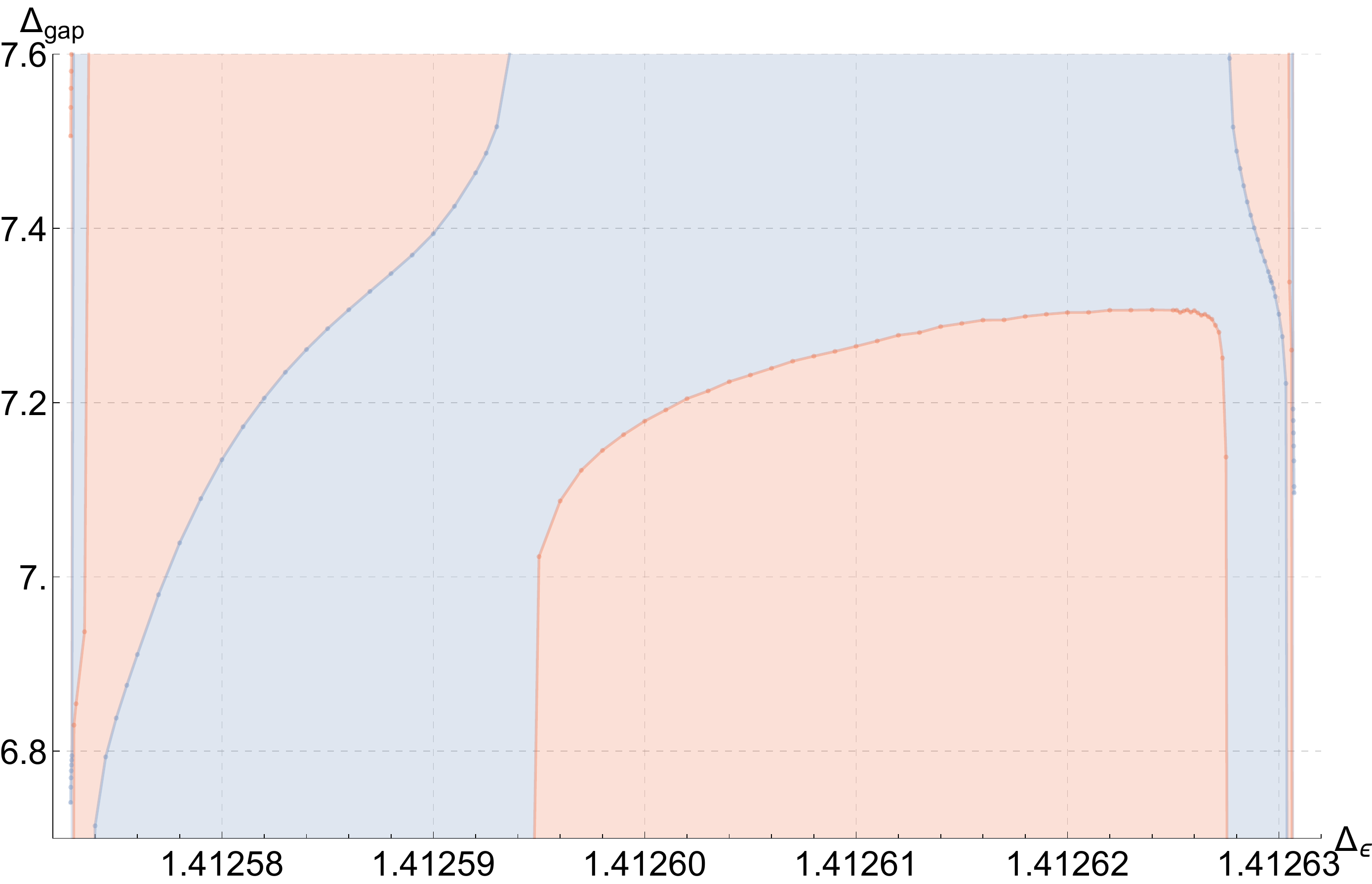}
		\end{minipage}\\\\
		\begin{minipage}{0.47\hsize}
			\centering
			$\Delta_\texttt{gap}\approx 9.3$
			\includegraphics[width=\textwidth]{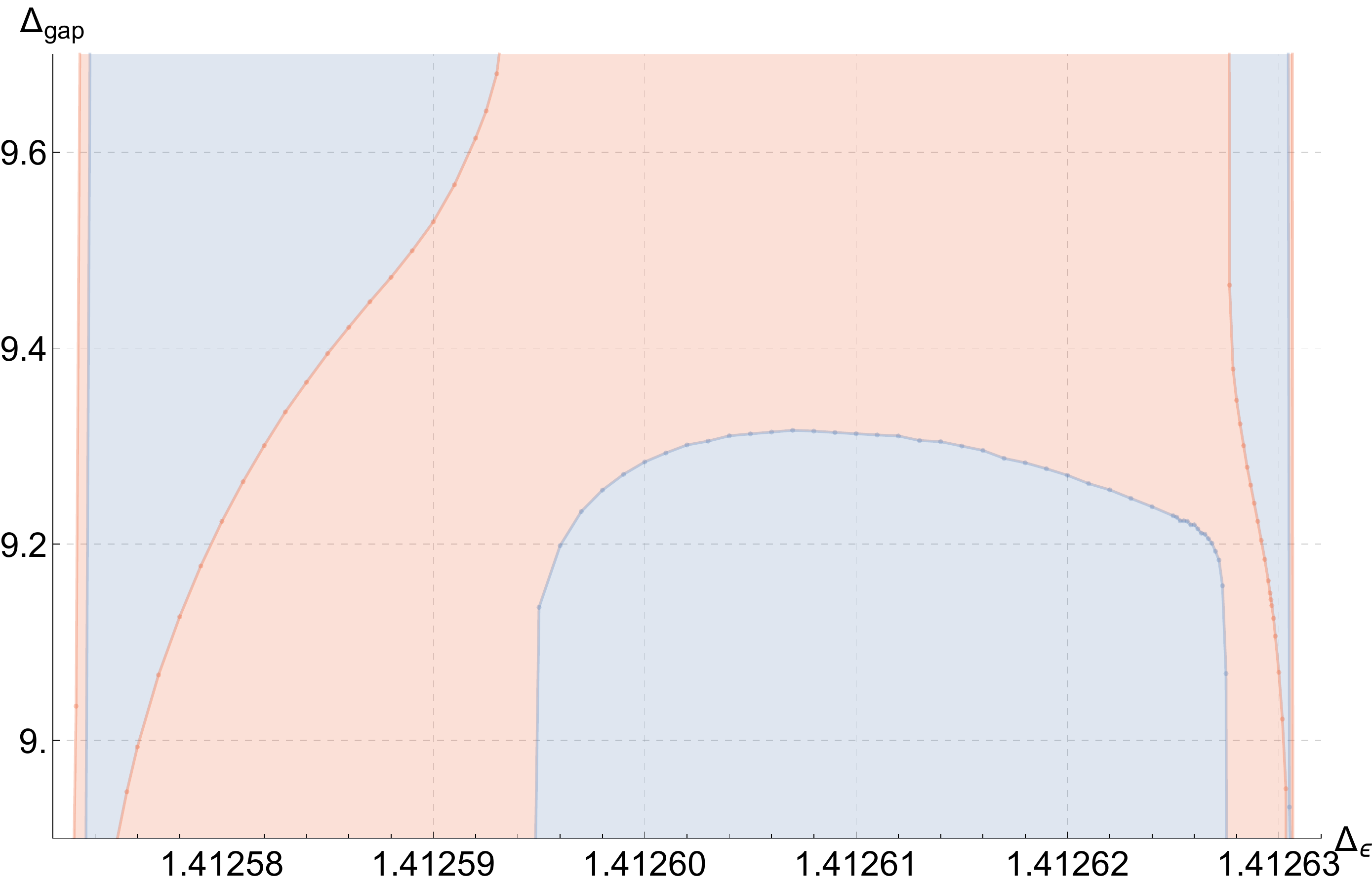}
		\end{minipage}&
		\begin{minipage}{0.47\hsize}
			\centering
			$\Delta_\texttt{gap}\approx 12.5$
			\includegraphics[width=\textwidth]{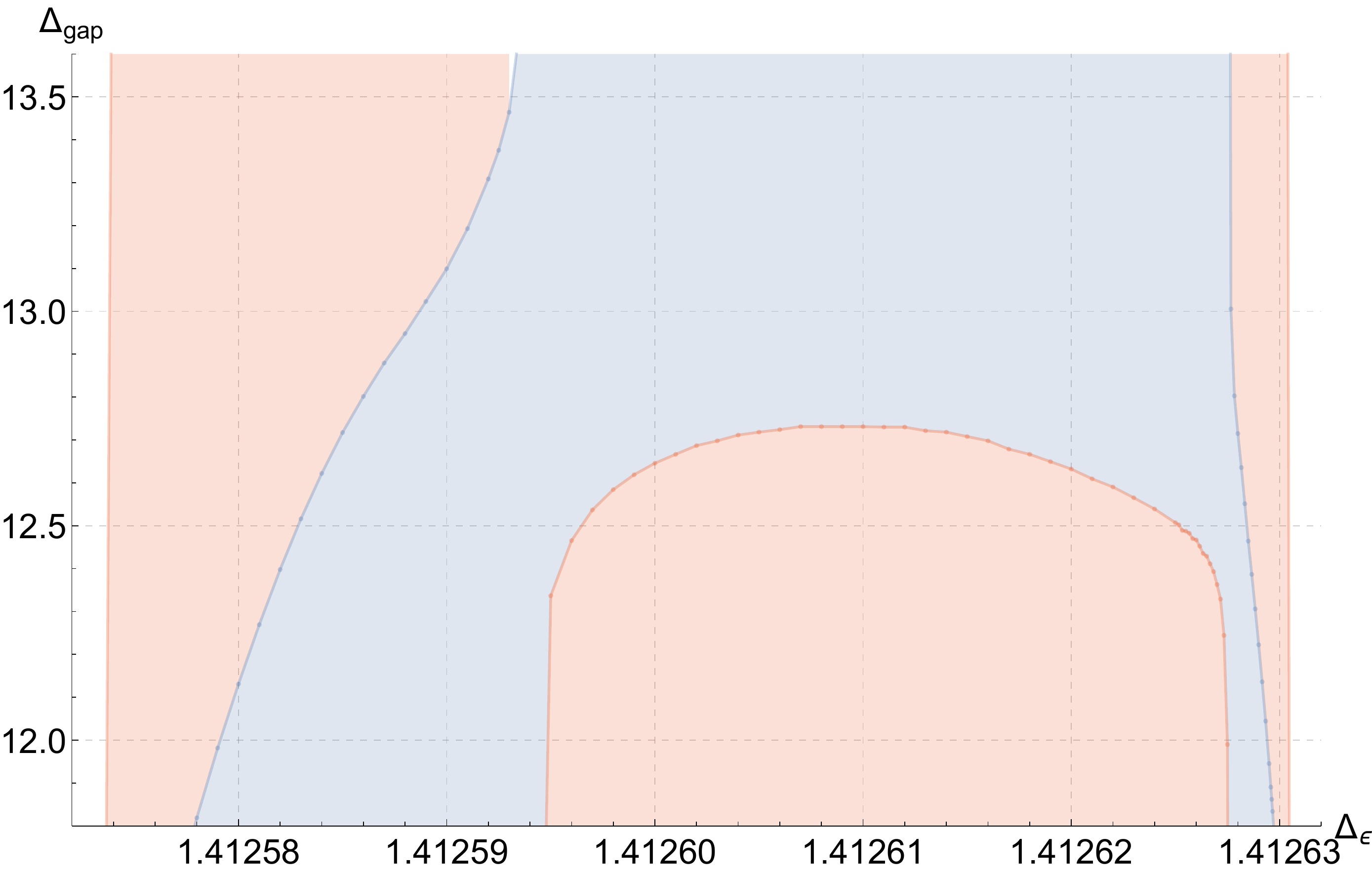}
		\end{minipage}
	\end{tabular}
	\caption{
		Enlarged plots of \autoref{fig:gap-t} (around $\Delta_\texttt{gap}\approx 5.51, 7.2, 9.3, 12.5$, respectively).
	}
	\label{fig:gap-t-zoom}
\end{figure}

\begin{figure}[htpb]
	\centering
	\includegraphics[width=\textwidth]{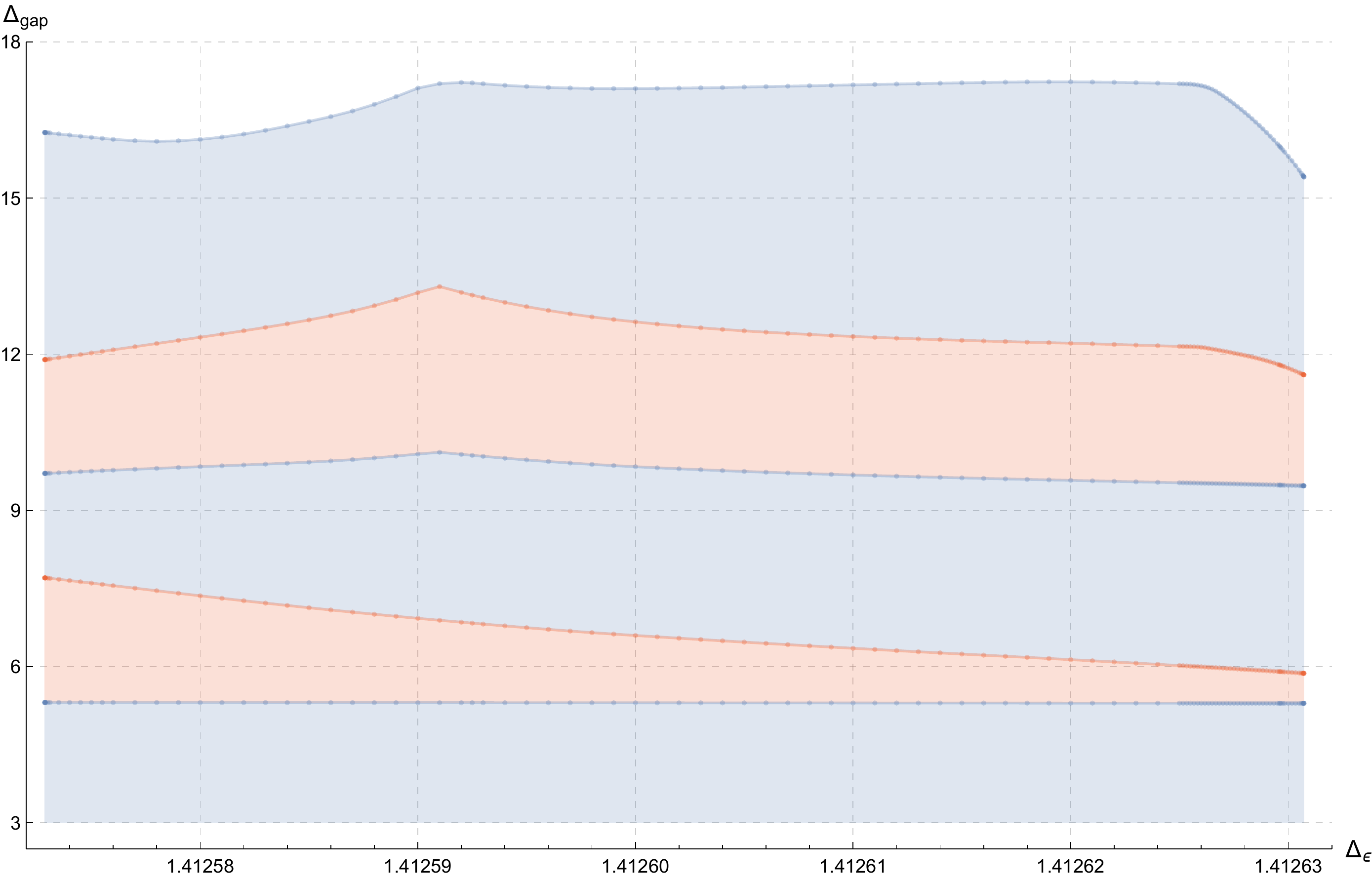}
	\caption{
		The Ising model must have at least one $\mathbb{Z}$-odd spin-$0$ primary operator
		in each colored region (assuming that $\Delta_\sigma=0.518149$).
		We showed the first five regions surrounded by $3<h_1<\cdots<h_5$.
	}
	\label{fig:gap-s}
\end{figure}

\begin{figure}[htpb]
	\centering
	\begin{tabular}{cc}
		\begin{minipage}{0.47\hsize}
			\centering
			$\Delta_\texttt{gap}\approx 5.31$
			\includegraphics[width=\textwidth]{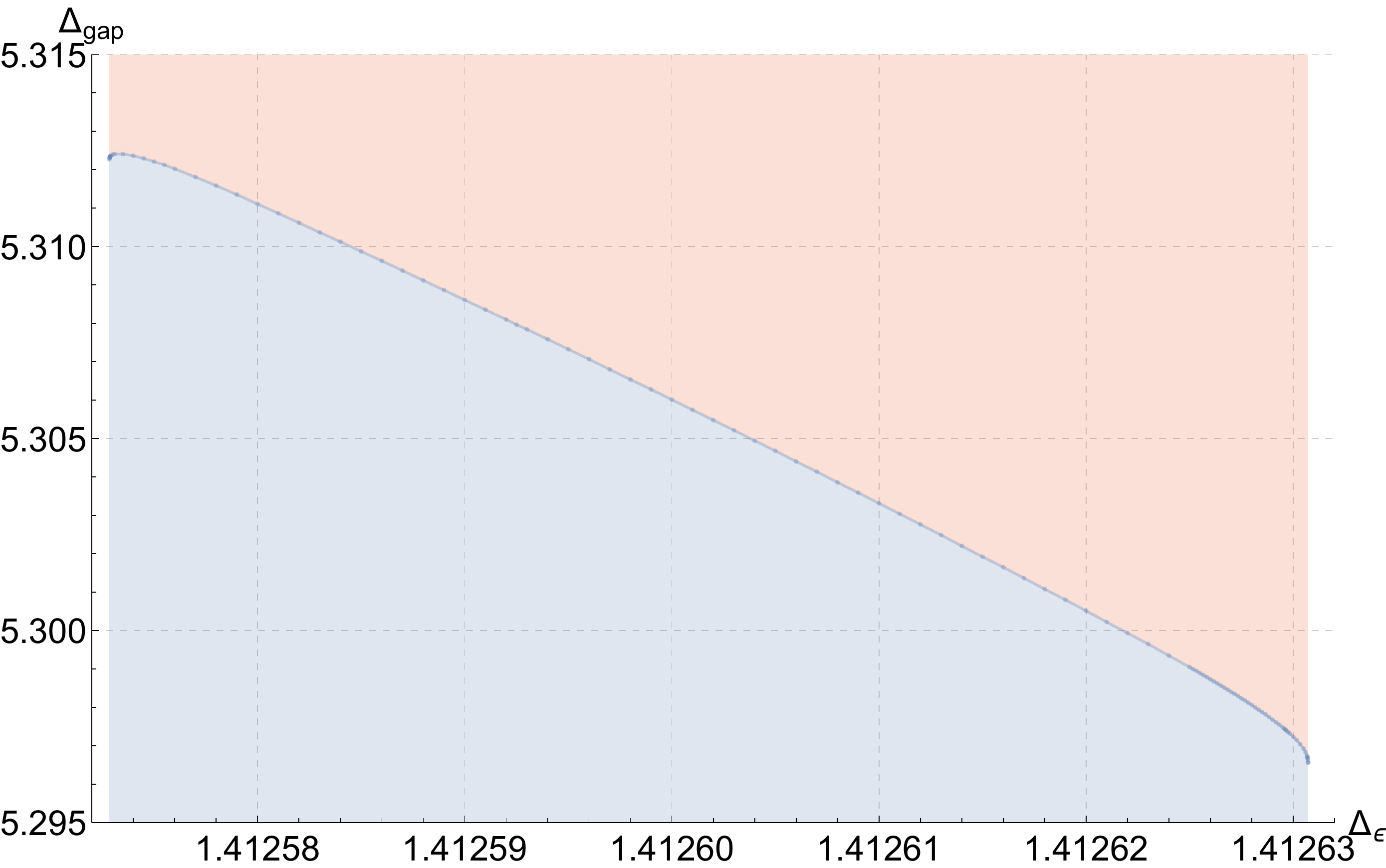}
		\end{minipage}&
		\begin{minipage}{0.47\hsize}
			\centering
			$\Delta_\texttt{gap}\approx 7.0$
			\includegraphics[width=\textwidth]{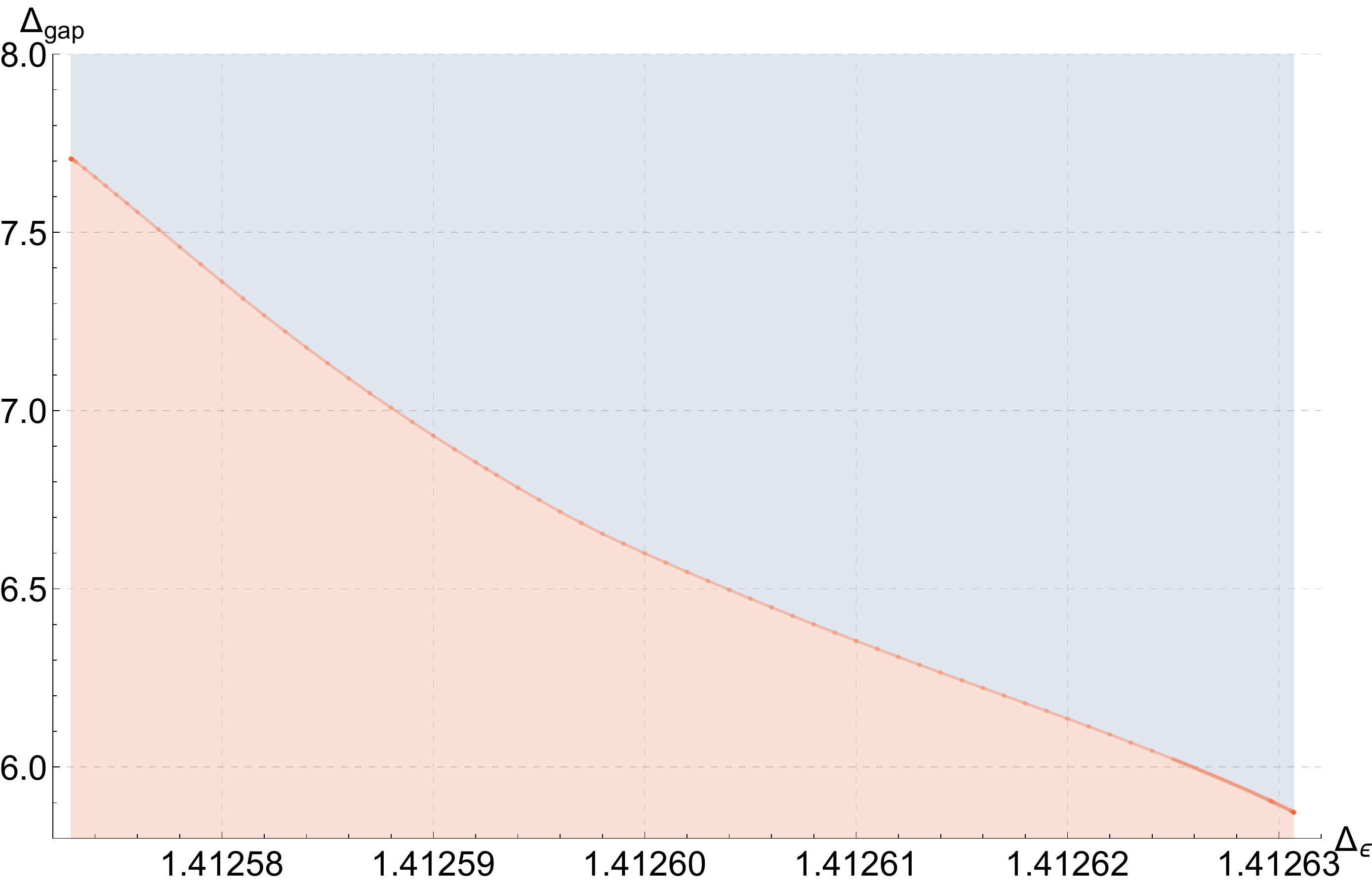}
		\end{minipage}\\\\
		\begin{minipage}{0.47\hsize}
			\centering
			$\Delta_\texttt{gap}\approx 9.8$
			\includegraphics[width=\textwidth]{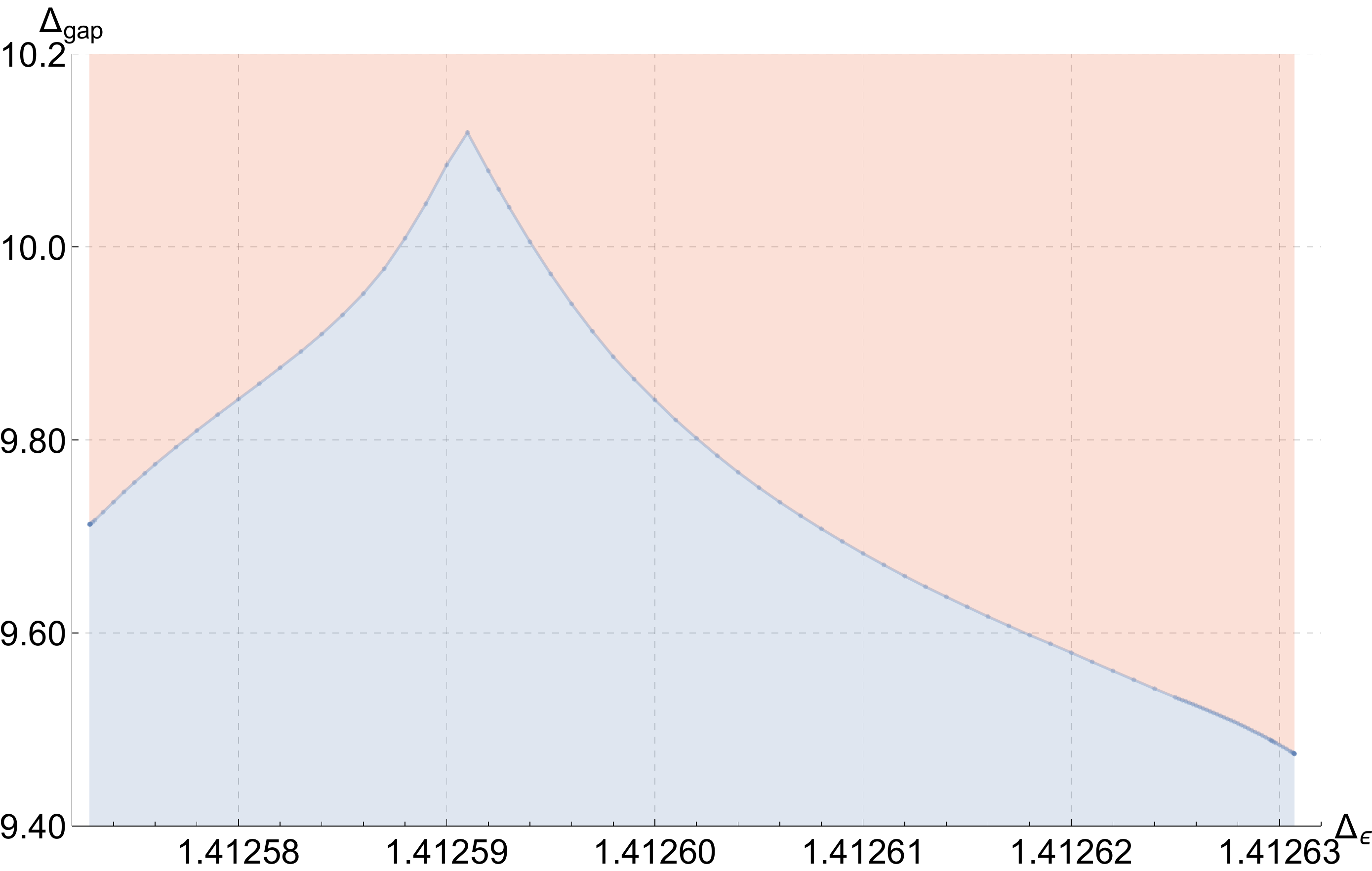}
		\end{minipage}&
		\begin{minipage}{0.47\hsize}
			\centering
			$\Delta_\texttt{gap}\approx 12.5$
			\includegraphics[width=\textwidth]{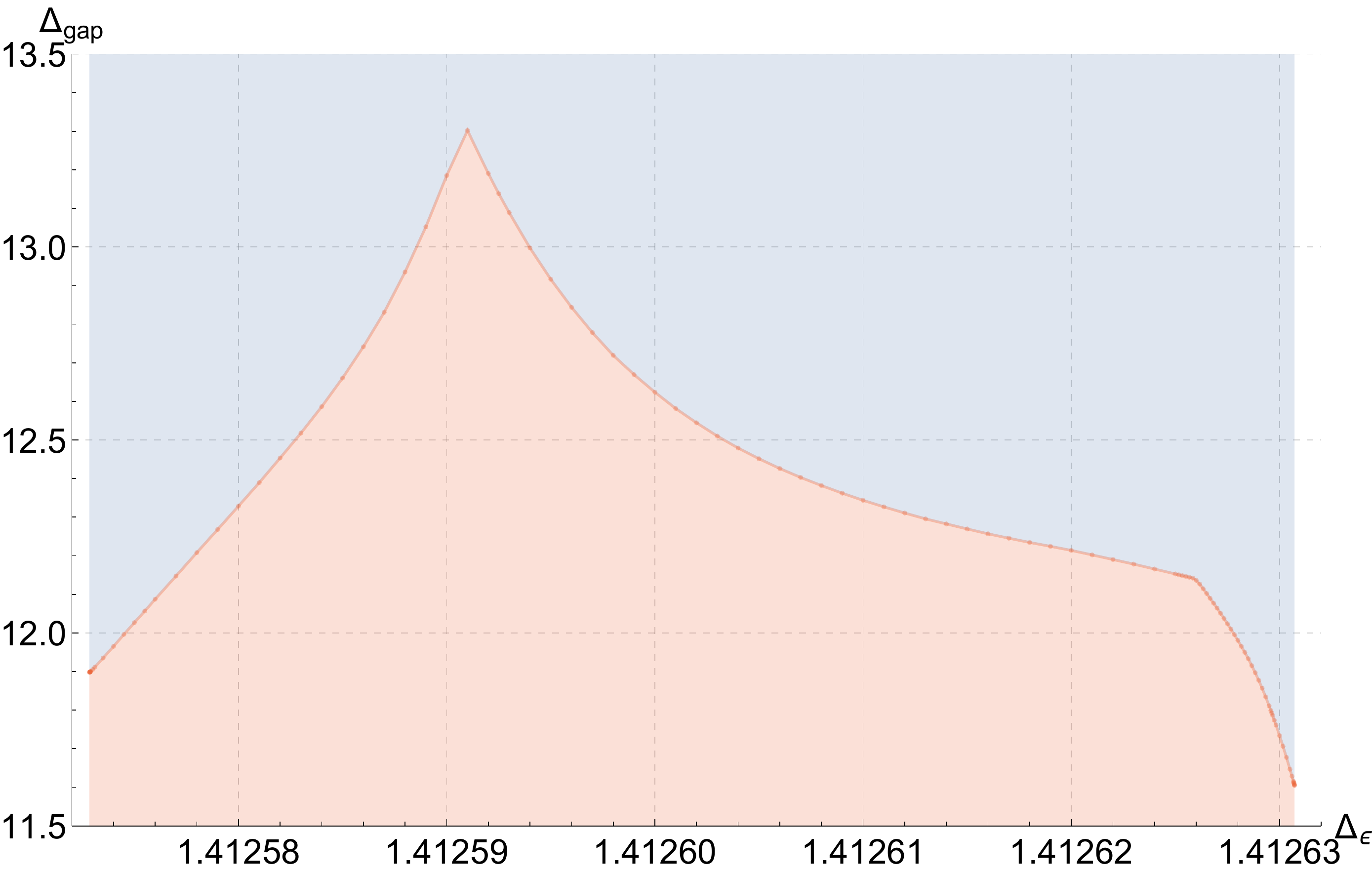}
		\end{minipage}
	\end{tabular}
	\caption{
		Enlarged plots of \autoref{fig:gap-s} (around $\Delta_\texttt{gap}\approx 5.31, 7.0, 9.8, 12.5$, respectively).
	}
	\label{fig:gap-s-zoom}
\end{figure}

Consider the case that the spectrum of $\mathbb{Z}_2$-even, $l=0$ is $[\Delta_\texttt{gap},\infty)$.
If $(\Delta_\sigma,\Delta_\epsilon,\Delta_\texttt{gap})$ is excluded and $\Delta_\texttt{gap}<\Delta'_\texttt{gap}$,
the assumption by $\Delta'_\texttt{gap}$ is stronger than $\Delta_\texttt{gap}$ and cannot be consistent.
Thus we get the upper bound by the bisection method: $(\Delta_\sigma,\Delta_\epsilon,\Delta_\texttt{gap})$ is not excluded
if and only if $\Delta_\texttt{gap}<f_1(\Delta_\sigma,\Delta_\epsilon)$.
$f_1$, which is defined in the island, has another implication;
we must have at least one $\mathbb{Z}_2$-even primary with scaling dimension
\begin{equation}
	3\leq \Delta\leq f_1(\Delta_\sigma,\Delta_\epsilon)
\end{equation}
in any consistent CFT, otherwise $f_1$ could be larger.
Next, we assume that the spectrum of $\mathbb{Z}_2$-even, $l=0$ is $[3,f_1(\Delta_\sigma,\Delta_\epsilon))\cup[\Delta_\texttt{gap},\infty)$,
which is computable by the M\"{o}bius transformation implemented in \code{qboot}.
By a similar discussion, we get the upper bound $f_2$ such that $(\Delta_\sigma,\Delta_\epsilon,\Delta_\texttt{gap})$ is not excluded
if and only if $\Delta_\texttt{gap}<f_2(\Delta_\sigma,\Delta_\epsilon)$.
$f_2$ implies that we must have at least one primary with scaling dimension
\begin{equation}
	f_1(\Delta_\sigma,\Delta_\epsilon)\leq \Delta\leq f_2(\Delta_\sigma,\Delta_\epsilon).
\end{equation}
We can continue this until $f_{N+1}$ diverges and there must be one primary in the $\mathbb{Z}_2$-even sector with scaling dimension
\begin{equation}
	\begin{cases}
		3\leq \Delta\leq f_1(\Delta_\sigma,\Delta_\epsilon), & \ptext{$n=1$} \\
		f_{n-1}(\Delta_\sigma,\Delta_\epsilon)\leq \Delta\leq f_{n}(\Delta_\sigma,\Delta_\epsilon), & \ptext{$n>1$}
	\end{cases}
\end{equation}
for $n=1,\ldots,N$. The divergence of $f_{N+1}$ can be detected by the feasibility of the finite spectrum $[3,f_N(\Delta_\sigma,\Delta_\epsilon))$.
This method also can be applied for any other sectors,
e.g., $\mathbb{Z}_2$-even spin-$2$ sector or $\mathbb{Z}_2$-odd scalar sector.
We did the computation only in the intersection of a line $\Delta_\sigma=0.518149$ and the island in \autoref{fig:ising-island},
and calculated $f_0,\ldots,f_6$ for $\mathbb{Z}_2$-even scalar sector in \autoref{fig:gap-e} and \autoref{fig:gap-e-zoom},
$g_0,\ldots,g_6$ for $\mathbb{Z}_2$-even spin-$2$ sector in \autoref{fig:gap-t} and \autoref{fig:gap-t-zoom},
and $h_0,\ldots,h_5$ for $\mathbb{Z}_2$-odd scalar sector in \autoref{fig:gap-s} and \autoref{fig:gap-s-zoom}.
Most of curves in \autoref{fig:gap-e} and \autoref{fig:gap-t} have
almost-flat segments in $1.412574\lesssim\Delta_\epsilon\lesssim 1.412594$ and $1.4126\lesssim\Delta_\epsilon\lesssim 1.412628$,
and jumping segments which connect almost-flat segments.
We call the values in almost-flat segments the `stable' values.
The plots imply that two successive curves have a common `stable' value, for example $f_0$ and $f_1$ share $\Delta_\texttt{gap}\approx 3.83$.

In the spectrum of the Ising model calculated in \cite{Simmons-Duffin:2016wlq}
using the EFM on the boundary of the island in $\Lambda=43$,
the existence of primaries in \autoref{tb:ising_efm} is estimated.
The errors shown are not rigorous because the EFM is only applicable to the boundary of the island.
Our results, however, is rigorous and implies that there exist primary operators with scaling dimension near `stable' values.
\begin{table}[htbp]
	\centering
	\begin{tabular}{l|r|c|l}
		$\op{O}$ & $\mathbb{Z}_2$ & $l$ & $\Delta$ \\ \hline
		$\epsilon$    & even & 0 & $1.412625(10)$  \\
		$\epsilon'$   & even & 0 & $3.82968(23)$   \\
		              & even & 0 & $6.8956(43)$    \\
		              & even & 0 & $7.2535(51)$    \\ \hline
		$T_{\mu\nu}$  & even & 2 & $3$             \\
		$T_{\mu\nu}'$ & even & 2 & $5.50915(44)$   \\ \hline
		$\sigma$      &  odd & 0 & $0.5181489(10)$ \\
		$\sigma'$     &  odd & 0 & $5.2906(11)$    \\
	\end{tabular}
	\caption{
		Operators with $\Delta\leq 8$ estimated in \cite{Simmons-Duffin:2016wlq}.
		All errors but $\epsilon,\sigma,T_{\mu\nu}$ are non-rigorous.
	}
	\label{tb:ising_efm}
\end{table}

\newpage

\subsection{3d \texorpdfstring{$O(2)$}{O(2)} model}
\label{sec:example_o2}
As the next example, we consider the 3d $O(2)$ model, defined by real scalar fields $(\phi_1,\phi_2)$ and the Hamiltonian
\begin{equation}
	\op{H}=\partial_\mu\phi_i\partial^\mu\phi^i+m^2\phi_i\phi^i+\lambda(\phi_i\phi^i)^2.
\end{equation}
The irreps of $O(2)$ are
\begin{itemize}
	\item the trivial representation $\mathbf{I}^+$,
	\item the sign representation $\mathbf{I}^-$,
	\item for $n=1,2,\ldots$, the $n$-th symmetric traceless tensor representation $\mathbf{S}^n$,
\end{itemize}
and $\dim \mathbf{I}^\pm=1$, $\dim\mathbf{S}^n=2$.
At the critical point, we have three relevant scalar operators: a singlet $s\sim \phi_i\phi^i$, a vector $\phi_i$ and a traceless-symmetric $t_{ij}$.
Generally \cite{Calabrese:2004ca}, $O(N)$ model has a harmonic operator $\op{O}_p$ with scaling dimension $\Delta_p$ for $p=2,3,\ldots$
in the spin-$l$ representation of the global symmetry $O(N)$ which can be schematically written as
\begin{equation}
	\op{O}_p = \phi_{i_1}(x)\cdots\phi_{i_p}(x)-\ptext{traces},
\end{equation}
and $t$ is the case for $p=2$.
Adding perturbation $\int d^d x h_p \op{O}_p$ to the Hamiltonian,
critical exponents $\beta_p,\phi_p$ are defined similarly in \autoref{sec:intro} as
\begin{align}
	\Braket{\op{O}_p}&\sim\abs{t}^{\beta_p},\\
	\chi_p=\int d^d x\Braket{\op{O}_p(x)\op{O}_p(0)}_c&\sim t^{-\gamma_p},
\end{align}
where these exponents are related with $\Delta_{p}$ as
\begin{align}
	\beta_p &= \frac{\Delta_p}{d-\Delta_s}, \\
	\gamma_p &= \frac{d-2\Delta_p}{d-\Delta_s}.
\end{align}

In \cite{Kos:2015mba}, the same model was analyzed using $s$ and $\phi$ as external operators,
and the information on $\Delta_t$ was obtained by specifying the condition in the intermediate channel.
We assume $O(2)$-symmetry and that each of $s,\phi,t$ is the unique relevant scalar primary in
$\mathbf{I}^+$, $\mathbf{V}\colonequals\mathbf{S}^1$, $\mathbf{T}\colonequals\mathbf{S}^2$, respectively.
The \code{Mathematica} code required is simply:
\begin{lstlisting}
	o2=getO[2];
	setGroup[o2];
	setOps[{op[s,o2[id]], op[v,v[1]], op[t,v[2]]}];
	eq=bootAll[];
	sdp=makeSDP[eq];
	py=toCboot[sdp];
	WriteString["O2.py",py];
\end{lstlisting}
Here \code{o2=getO[2]} creates the group $O(2)$ within \code{autoboot},
and \code{v[n]} stands for $\mathbf{S}^n$.
We use the symbol \code{v} to represent the operator $\phi$.
The bootstrap equations generated by \code{toTeX} are given below:\footnote{
	The author heard from the author of \cite{Chester:2019ifh} that they independently wrote these equations
	by hand or with the help of \code{Mathematica} and checked that their results agreed with each other after correction of many mistakes.
	Our results can be reproduced by just running the example file \url{https://github.com/selpoG/autoboot/blob/master/sample/O2.m}
	within 3 seconds in author's laptop.
}

{\tiny
\begin{align*}
	0&=F^{s s,s s}_{-,1}+{\lambda_{s s s}}^2 F^{s s,s s}_{-,s}+\sum_{\op{O}^+:{\mathbf{I}^+}}{\lambda_{s s \op{O}}}^2 F^{s s,s s}_{-,\op{O}},\\
	0&=\sum_{\op{O}^-:{\mathbf{I}^-}}{\lambda_{t t \op{O}}}^2 F^{t t,t t}_{-,\op{O}}+\frac{1}{2}\sum_{\op{O}^+:{\mathbf{S}^{4}}}{\lambda_{t t \op{O}}}^2 F^{t t,t t}_{-,\op{O}},\\
	0&={\lambda_{t t s}}^2 F^{s t,s t}_{-,t}+\sum_{\op{O}^-:{\mathbf{T}}}{\lambda_{s t \op{O}}}^2 F^{s t,s t}_{-,\op{O}}+\sum_{\op{O}^+:{\mathbf{T}}}{\lambda_{s t \op{O}}}^2 F^{s t,s t}_{-,\op{O}},\\
	0&={\lambda_{\phi \phi s}}^2 F^{s \phi,s \phi}_{-,\phi}+\sum_{\op{O}^-:{\mathbf{V}}}{\lambda_{s \phi \op{O}}}^2 F^{s \phi,s \phi}_{-,\op{O}}+\sum_{\op{O}^+:{\mathbf{V}}}{\lambda_{s \phi \op{O}}}^2 F^{s \phi,s \phi}_{-,\op{O}},\\
	0&=F^{t t,t t}_{-,1}+\frac{1}{2}{\lambda_{t t s}}^2 F^{t t,t t}_{-,s}+\frac{1}{2}\sum_{\op{O}^+:{\mathbf{I}^+}}{\lambda_{t t \op{O}}}^2 F^{t t,t t}_{-,\op{O}}+\frac{1}{4}\sum_{\op{O}^+:{\mathbf{S}^{4}}}{\lambda_{t t \op{O}}}^2 F^{t t,t t}_{-,\op{O}},\\
	0&={\lambda_{\phi t \phi}}^2 F^{\phi \phi,\phi \phi}_{-,t}+2\sum_{\op{O}^-:{\mathbf{I}^-}}{\lambda_{\phi \phi \op{O}}}^2 F^{\phi \phi,\phi \phi}_{-,\op{O}}+\sum_{\op{O}^+:{\mathbf{T}}}{\lambda_{\phi \phi \op{O}}}^2 F^{\phi \phi,\phi \phi}_{-,\op{O}},\\
	0&=\lambda_{\phi t \phi} \lambda_{\phi \phi s} F^{s \phi,t \phi}_{-,\phi}-\sum_{\op{O}^-:{\mathbf{V}}}\lambda_{s \phi \op{O}} \lambda_{\phi t \op{O}} F^{s \phi,t \phi}_{-,\op{O}}+\sum_{\op{O}^+:{\mathbf{V}}}\lambda_{s \phi \op{O}} \lambda_{\phi t \op{O}} F^{s \phi,t \phi}_{-,\op{O}},\\
	0&=F^{\phi \phi,\phi \phi}_{-,1}+\frac{1}{2}{\lambda_{\phi \phi s}}^2 F^{\phi \phi,\phi \phi}_{-,s}-\frac{1}{2}\sum_{\op{O}^-:{\mathbf{I}^-}}{\lambda_{\phi \phi \op{O}}}^2 F^{\phi \phi,\phi \phi}_{-,\op{O}}+\frac{1}{2}\sum_{\op{O}^+:{\mathbf{I}^+}}{\lambda_{\phi \phi \op{O}}}^2 F^{\phi \phi,\phi \phi}_{-,\op{O}},\\
	0&=F^{t t,t t}_{+,1}+\frac{1}{2}{\lambda_{t t s}}^2 F^{t t,t t}_{+,s}+\frac{1}{2}\sum_{\op{O}^-:{\mathbf{I}^-}}{\lambda_{t t \op{O}}}^2 F^{t t,t t}_{+,\op{O}}+\frac{1}{2}\sum_{\op{O}^+:{\mathbf{I}^+}}{\lambda_{t t \op{O}}}^2 F^{t t,t t}_{+,\op{O}}-\frac{1}{2}\sum_{\op{O}^+:{\mathbf{S}^{4}}}{\lambda_{t t \op{O}}}^2 F^{t t,t t}_{+,\op{O}},\\
	0&={\lambda_{\phi t \phi}}^2 F^{\phi t,\phi t}_{-,\phi}+\sum_{\op{O}^-:{\mathbf{V}}}{\lambda_{\phi t \op{O}}}^2 F^{\phi t,\phi t}_{-,\op{O}}+\sum_{\op{O}^+:{\mathbf{V}}}{\lambda_{\phi t \op{O}}}^2 F^{\phi t,\phi t}_{-,\op{O}}+\sum_{\op{O}^-:{\mathbf{S}^{3}}}{\lambda_{\phi t \op{O}}}^2 F^{\phi t,\phi t}_{-,\op{O}}+\sum_{\op{O}^+:{\mathbf{S}^{3}}}{\lambda_{\phi t \op{O}}}^2 F^{\phi t,\phi t}_{-,\op{O}},\\
	0&={\lambda_{\phi t \phi}}^2 F^{\phi t,\phi t}_{+,\phi}+\sum_{\op{O}^-:{\mathbf{V}}}{\lambda_{\phi t \op{O}}}^2 F^{\phi t,\phi t}_{+,\op{O}}+\sum_{\op{O}^+:{\mathbf{V}}}{\lambda_{\phi t \op{O}}}^2 F^{\phi t,\phi t}_{+,\op{O}}-\sum_{\op{O}^-:{\mathbf{S}^{3}}}{\lambda_{\phi t \op{O}}}^2 F^{\phi t,\phi t}_{+,\op{O}}-\sum_{\op{O}^+:{\mathbf{S}^{3}}}{\lambda_{\phi t \op{O}}}^2 F^{\phi t,\phi t}_{+,\op{O}},\\
	0&=F^{s s,t t}_{-,1}+\frac{\sqrt{2}}{2}\lambda_{s s s} \lambda_{t t s} F^{s s,t t}_{-,s}+\frac{1}{2}{\lambda_{t t s}}^2 F^{s t,t s}_{-,t}-\frac{1}{2}\sum_{\op{O}^-:{\mathbf{T}}}{\lambda_{s t \op{O}}}^2 F^{s t,t s}_{-,\op{O}}+\frac{1}{2}\sum_{\op{O}^+:{\mathbf{T}}}{\lambda_{s t \op{O}}}^2 F^{s t,t s}_{-,\op{O}}+\frac{\sqrt{2}}{2}\sum_{\op{O}^+:{\mathbf{I}^+}}\lambda_{s s \op{O}} \lambda_{t t \op{O}} F^{s s,t t}_{-,\op{O}},\\
	0&=F^{s s,t t}_{+,1}+\frac{\sqrt{2}}{2}\lambda_{s s s} \lambda_{t t s} F^{s s,t t}_{+,s}-\frac{1}{2}{\lambda_{t t s}}^2 F^{s t,t s}_{+,t}+\frac{1}{2}\sum_{\op{O}^-:{\mathbf{T}}}{\lambda_{s t \op{O}}}^2 F^{s t,t s}_{+,\op{O}}-\frac{1}{2}\sum_{\op{O}^+:{\mathbf{T}}}{\lambda_{s t \op{O}}}^2 F^{s t,t s}_{+,\op{O}}+\frac{\sqrt{2}}{2}\sum_{\op{O}^+:{\mathbf{I}^+}}\lambda_{s s \op{O}} \lambda_{t t \op{O}} F^{s s,t t}_{+,\op{O}},\\
	0&=\lambda_{t t s} \lambda_{\phi t \phi} F^{s t,\phi \phi}_{-,t}+\lambda_{\phi t \phi} \lambda_{\phi \phi s} F^{s \phi,\phi t}_{-,\phi}+\sum_{\op{O}^-:{\mathbf{V}}}\lambda_{s \phi \op{O}} \lambda_{\phi t \op{O}} F^{s \phi,\phi t}_{-,\op{O}}+\sum_{\op{O}^+:{\mathbf{V}}}\lambda_{s \phi \op{O}} \lambda_{\phi t \op{O}} F^{s \phi,\phi t}_{-,\op{O}}-\sum_{\op{O}^+:{\mathbf{T}}}\lambda_{s t \op{O}} \lambda_{\phi \phi \op{O}} F^{s t,\phi \phi}_{-,\op{O}},\\
	0&=\lambda_{t t s} \lambda_{\phi t \phi} F^{s t,\phi \phi}_{+,t}-\lambda_{\phi t \phi} \lambda_{\phi \phi s} F^{s \phi,\phi t}_{+,\phi}-\sum_{\op{O}^-:{\mathbf{V}}}\lambda_{s \phi \op{O}} \lambda_{\phi t \op{O}} F^{s \phi,\phi t}_{+,\op{O}}-\sum_{\op{O}^+:{\mathbf{V}}}\lambda_{s \phi \op{O}} \lambda_{\phi t \op{O}} F^{s \phi,\phi t}_{+,\op{O}}-\sum_{\op{O}^+:{\mathbf{T}}}\lambda_{s t \op{O}} \lambda_{\phi \phi \op{O}} F^{s t,\phi \phi}_{+,\op{O}},\\
	0&=F^{\phi \phi,\phi \phi}_{+,1}-\frac{1}{2}{\lambda_{\phi t \phi}}^2 F^{\phi \phi,\phi \phi}_{+,t}+\frac{1}{2}{\lambda_{\phi \phi s}}^2 F^{\phi \phi,\phi \phi}_{+,s}+\frac{1}{2}\sum_{\op{O}^-:{\mathbf{I}^-}}{\lambda_{\phi \phi \op{O}}}^2 F^{\phi \phi,\phi \phi}_{+,\op{O}}+\frac{1}{2}\sum_{\op{O}^+:{\mathbf{I}^+}}{\lambda_{\phi \phi \op{O}}}^2 F^{\phi \phi,\phi \phi}_{+,\op{O}}-\frac{1}{2}\sum_{\op{O}^+:{\mathbf{T}}}{\lambda_{\phi \phi \op{O}}}^2 F^{\phi \phi,\phi \phi}_{+,\op{O}},\\
	0&=F^{s s,\phi \phi}_{-,1}+\frac{\sqrt{2}}{2}\lambda_{s s s} \lambda_{\phi \phi s} F^{s s,\phi \phi}_{-,s}+\frac{1}{2}{\lambda_{\phi \phi s}}^2 F^{s \phi,\phi s}_{-,\phi}-\frac{1}{2}\sum_{\op{O}^-:{\mathbf{V}}}{\lambda_{s \phi \op{O}}}^2 F^{s \phi,\phi s}_{-,\op{O}}+\frac{1}{2}\sum_{\op{O}^+:{\mathbf{V}}}{\lambda_{s \phi \op{O}}}^2 F^{s \phi,\phi s}_{-,\op{O}}+\frac{\sqrt{2}}{2}\sum_{\op{O}^+:{\mathbf{I}^+}}\lambda_{s s \op{O}} \lambda_{\phi \phi \op{O}} F^{s s,\phi \phi}_{-,\op{O}},\\
	0&=F^{s s,\phi \phi}_{+,1}+\frac{\sqrt{2}}{2}\lambda_{s s s} \lambda_{\phi \phi s} F^{s s,\phi \phi}_{+,s}-\frac{1}{2}{\lambda_{\phi \phi s}}^2 F^{s \phi,\phi s}_{+,\phi}+\frac{1}{2}\sum_{\op{O}^-:{\mathbf{V}}}{\lambda_{s \phi \op{O}}}^2 F^{s \phi,\phi s}_{+,\op{O}}-\frac{1}{2}\sum_{\op{O}^+:{\mathbf{V}}}{\lambda_{s \phi \op{O}}}^2 F^{s \phi,\phi s}_{+,\op{O}}+\frac{\sqrt{2}}{2}\sum_{\op{O}^+:{\mathbf{I}^+}}\lambda_{s s \op{O}} \lambda_{\phi \phi \op{O}} F^{s s,\phi \phi}_{+,\op{O}},\\
	0&={\lambda_{\phi t \phi}}^2 F^{\phi t,t \phi}_{-,\phi}-\sum_{\op{O}^-:{\mathbf{V}}}{\lambda_{\phi t \op{O}}}^2 F^{\phi t,t \phi}_{-,\op{O}}+\sum_{\op{O}^+:{\mathbf{V}}}{\lambda_{\phi t \op{O}}}^2 F^{\phi t,t \phi}_{-,\op{O}}+\sum_{\op{O}^-:{\mathbf{S}^{3}}}{\lambda_{\phi t \op{O}}}^2 F^{\phi t,t \phi}_{-,\op{O}}-\sum_{\op{O}^+:{\mathbf{S}^{3}}}{\lambda_{\phi t \op{O}}}^2 F^{\phi t,t \phi}_{-,\op{O}}-2\sum_{\op{O}^-:{\mathbf{I}^-}}\lambda_{t t \op{O}} \lambda_{\phi \phi \op{O}} F^{\phi \phi,t t}_{-,\op{O}},\\
	0&={\lambda_{\phi t \phi}}^2 F^{\phi t,t \phi}_{+,\phi}-\sum_{\op{O}^-:{\mathbf{V}}}{\lambda_{\phi t \op{O}}}^2 F^{\phi t,t \phi}_{+,\op{O}}+\sum_{\op{O}^+:{\mathbf{V}}}{\lambda_{\phi t \op{O}}}^2 F^{\phi t,t \phi}_{+,\op{O}}+\sum_{\op{O}^-:{\mathbf{S}^{3}}}{\lambda_{\phi t \op{O}}}^2 F^{\phi t,t \phi}_{+,\op{O}}-\sum_{\op{O}^+:{\mathbf{S}^{3}}}{\lambda_{\phi t \op{O}}}^2 F^{\phi t,t \phi}_{+,\op{O}}+2\sum_{\op{O}^-:{\mathbf{I}^-}}\lambda_{t t \op{O}} \lambda_{\phi \phi \op{O}} F^{\phi \phi,t t}_{+,\op{O}},\\
	0&=F^{\phi \phi,t t}_{-,1}+\frac{1}{2}\lambda_{t t s} \lambda_{\phi \phi s} F^{\phi \phi,t t}_{-,s}-\frac{1}{2}\sum_{\op{O}^-:{\mathbf{S}^{3}}}{\lambda_{\phi t \op{O}}}^2 F^{\phi t,t \phi}_{-,\op{O}}+\frac{1}{2}\sum_{\op{O}^+:{\mathbf{S}^{3}}}{\lambda_{\phi t \op{O}}}^2 F^{\phi t,t \phi}_{-,\op{O}}\\
	&\qquad+\frac{1}{2}\sum_{\op{O}^-:{\mathbf{I}^-}}\lambda_{t t \op{O}} \lambda_{\phi \phi \op{O}} F^{\phi \phi,t t}_{-,\op{O}}+\frac{1}{2}\sum_{\op{O}^+:{\mathbf{I}^+}}\lambda_{t t \op{O}} \lambda_{\phi \phi \op{O}} F^{\phi \phi,t t}_{-,\op{O}},\\
	0&=F^{\phi \phi,t t}_{+,1}+\frac{1}{2}\lambda_{t t s} \lambda_{\phi \phi s} F^{\phi \phi,t t}_{+,s}+\frac{1}{2}\sum_{\op{O}^-:{\mathbf{S}^{3}}}{\lambda_{\phi t \op{O}}}^2 F^{\phi t,t \phi}_{+,\op{O}}-\frac{1}{2}\sum_{\op{O}^+:{\mathbf{S}^{3}}}{\lambda_{\phi t \op{O}}}^2 F^{\phi t,t \phi}_{+,\op{O}}\\
	&\qquad+\frac{1}{2}\sum_{\op{O}^-:{\mathbf{I}^-}}\lambda_{t t \op{O}} \lambda_{\phi \phi \op{O}} F^{\phi \phi,t t}_{+,\op{O}}+\frac{1}{2}\sum_{\op{O}^+:{\mathbf{I}^+}}\lambda_{t t \op{O}} \lambda_{\phi \phi \op{O}} F^{\phi \phi,t t}_{+,\op{O}}.
\end{align*}}

This example shows the power of \code{autoboot}.
It is almost trivial to add another external operator using \code{autoboot},
whereas it is quite tedious to work out the form of the bootstrap equations by hand.

\begin{figure}[htpb]
	\centering
	\begin{tabular}{cc}
		\begin{minipage}{0.47\hsize}
			\centering
			$\Delta_t=1.234$
			\includegraphics[width=\textwidth]{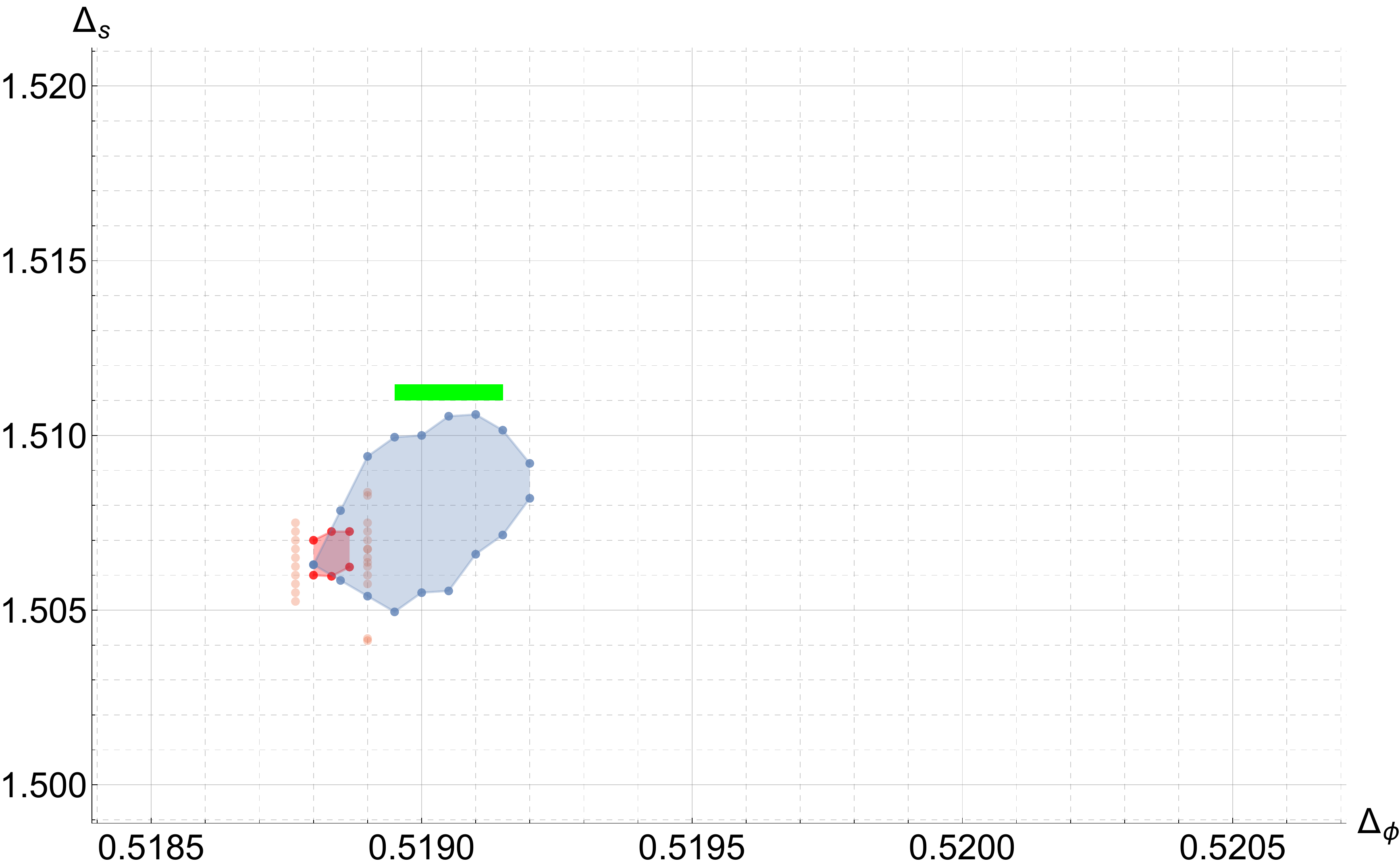}
		\end{minipage}&
		\begin{minipage}{0.47\hsize}
			\centering
			$\Delta_t=1.235$
			\includegraphics[width=\textwidth]{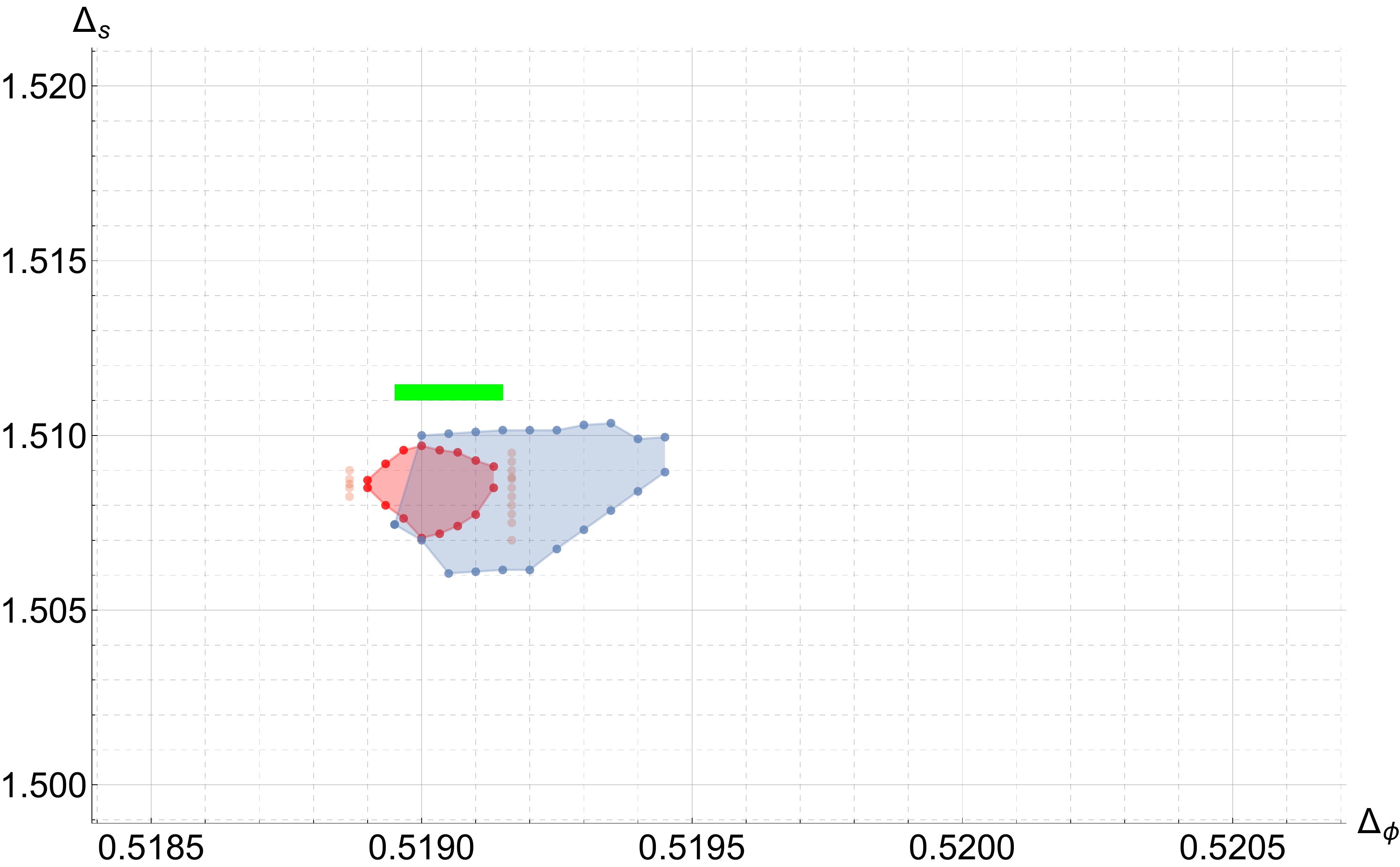}
		\end{minipage}\\\\
		\begin{minipage}{0.47\hsize}
			\centering
			$\Delta_t=1.236$
			\includegraphics[width=\textwidth]{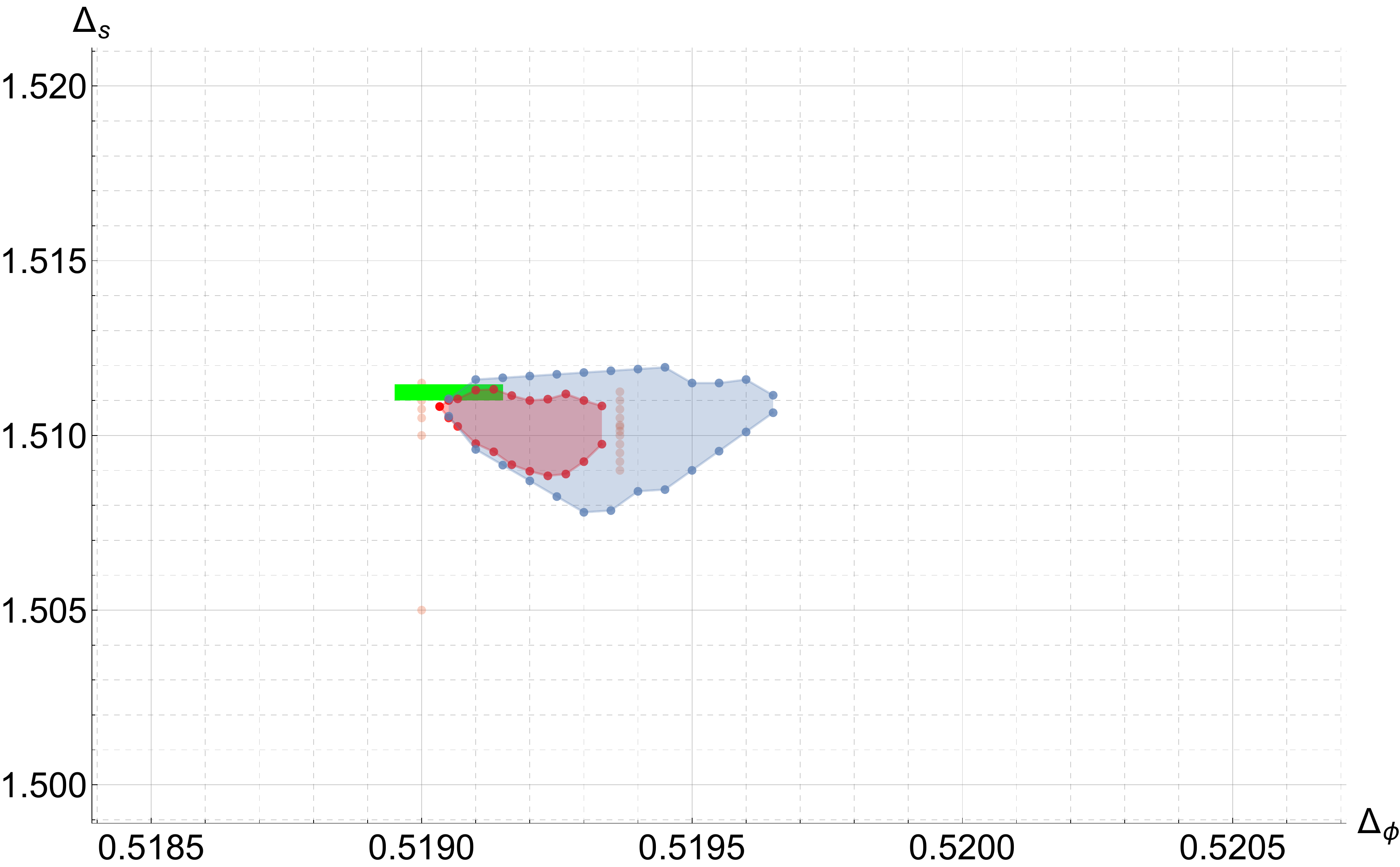}
		\end{minipage}&
		\begin{minipage}{0.47\hsize}
			\centering
			$\Delta_t=1.237$
			\includegraphics[width=\textwidth]{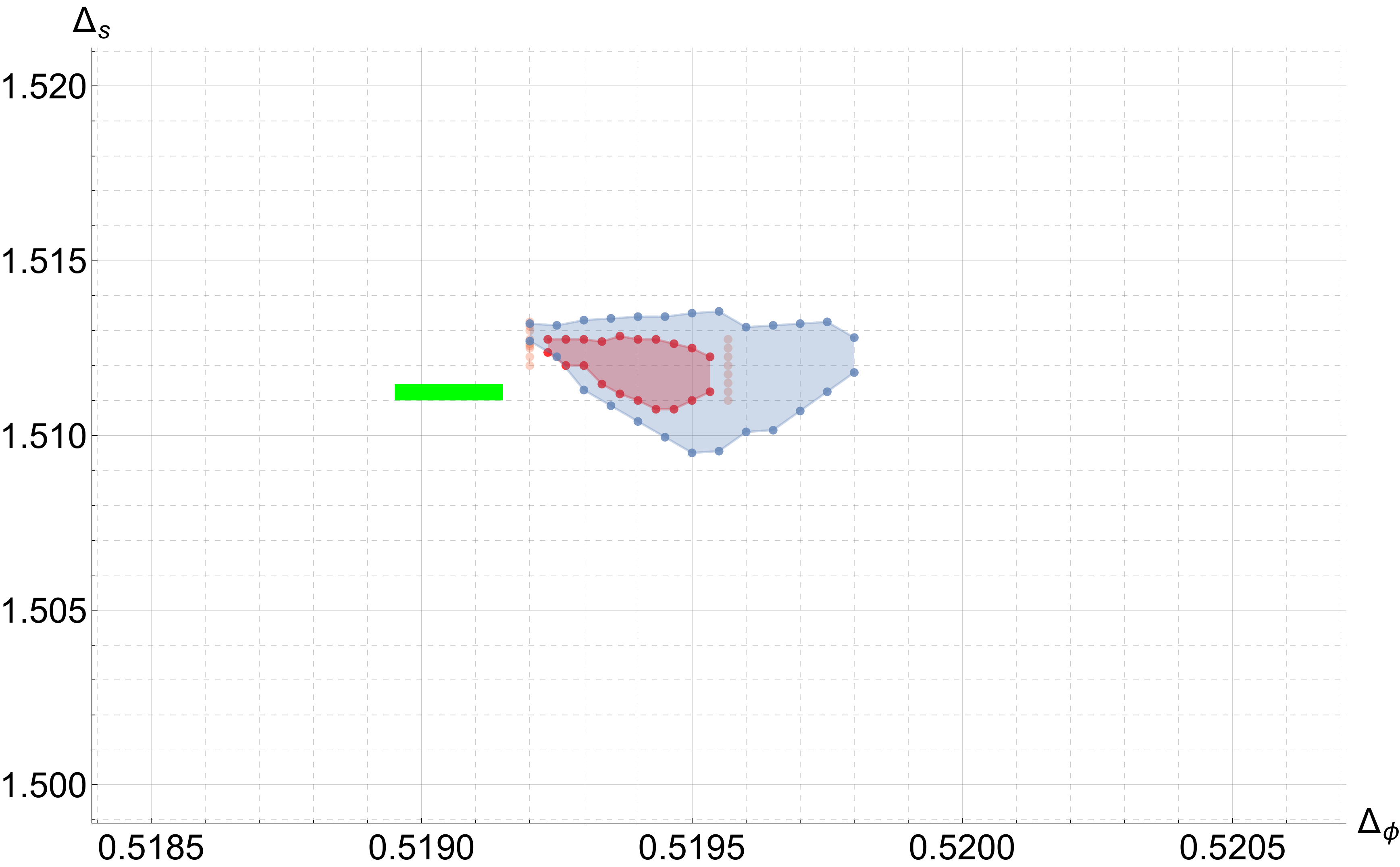}
		\end{minipage}\\\\
		\begin{minipage}{0.47\hsize}
			\centering
			$\Delta_t=1.238$
			\includegraphics[width=\textwidth]{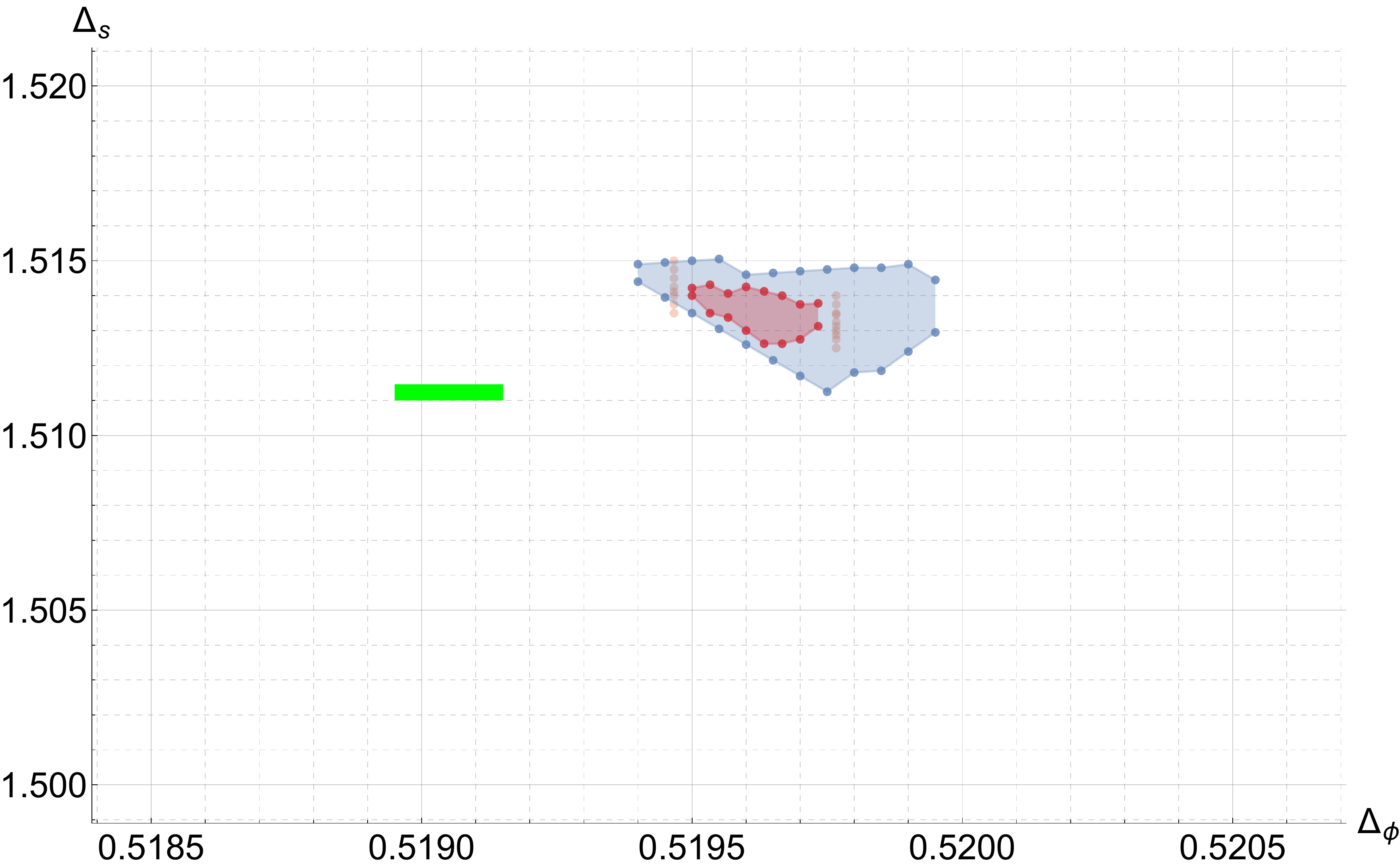}
		\end{minipage}&
		\begin{minipage}{0.47\hsize}
			\centering
			$\Delta_t=1.239$
			\includegraphics[width=\textwidth]{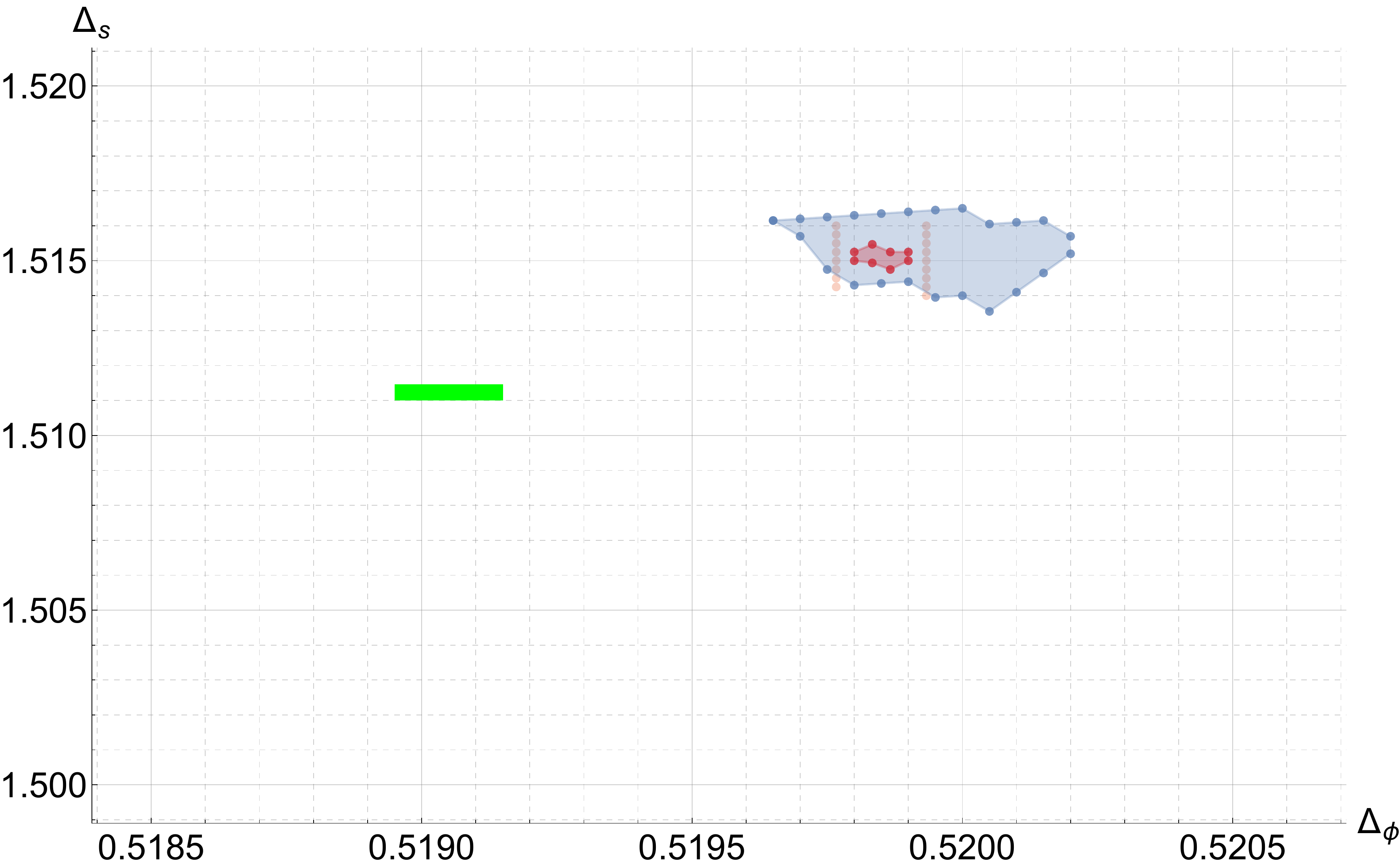}
		\end{minipage}
	\end{tabular}
	\caption{
		The island from the mixed correlator bootstrap of the $O(2)$ model.
		Here, the red regions are our results at $\Lambda=25$;
		the blue regions are those obtained in Appendix B of \cite{Kos:2015mba};
		and the green rectangle shows the Monte-Carlo results of \cite{Campostrini:2006ms}.
	}
	\label{fig:o2}
\end{figure}

\begin{figure}[htpb]
	\centering
	\includegraphics[width=\textwidth]{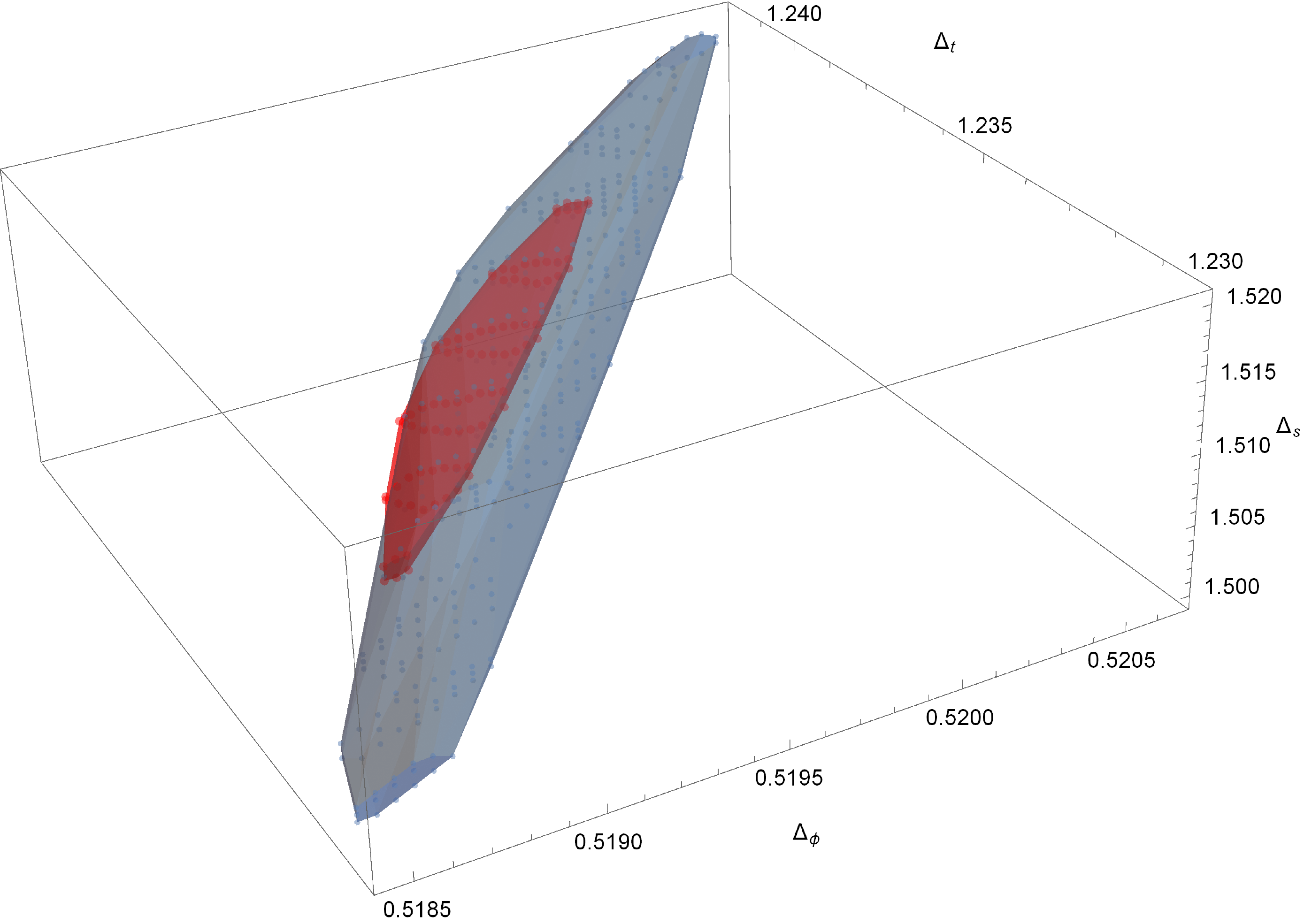}
	\caption{
		Three-dimensional plot of the $O(2)$ island.
		The blue region is the one obtained in Appendix B of \cite{Kos:2015mba},
		and the red region is our result.
	}
	\label{fig:o2-3d}
\end{figure}

We used $\Lambda=25$ and
obtained the island in the $(\Delta_s,\Delta_\phi,\Delta_t)$ space shown in \autoref{fig:o2} and \autoref{fig:o2-3d},
where the results from Appendix B of \cite{Kos:2015mba} are also presented\footnote{
	The authors thank the authors of \cite{Kos:2015mba}, in particular David Simmons-Duffin,
	for providing the raw data used to create their original figures to be reproduced here.
	The authors also thank Shai Chester and Alessandro Vichi for helpful discussions on the computations.
}. Our bound is the following:
\begin{align}
	1.50597 &\le \Delta_s \le 1.51547, \\
	0.5188 &\le \Delta_\phi \le 0.5199, \\
	1.234 &\le \Delta_t \le 1.239.
\end{align}

\begin{figure}[htpb]
	\centering
	\includegraphics[width=\textwidth]{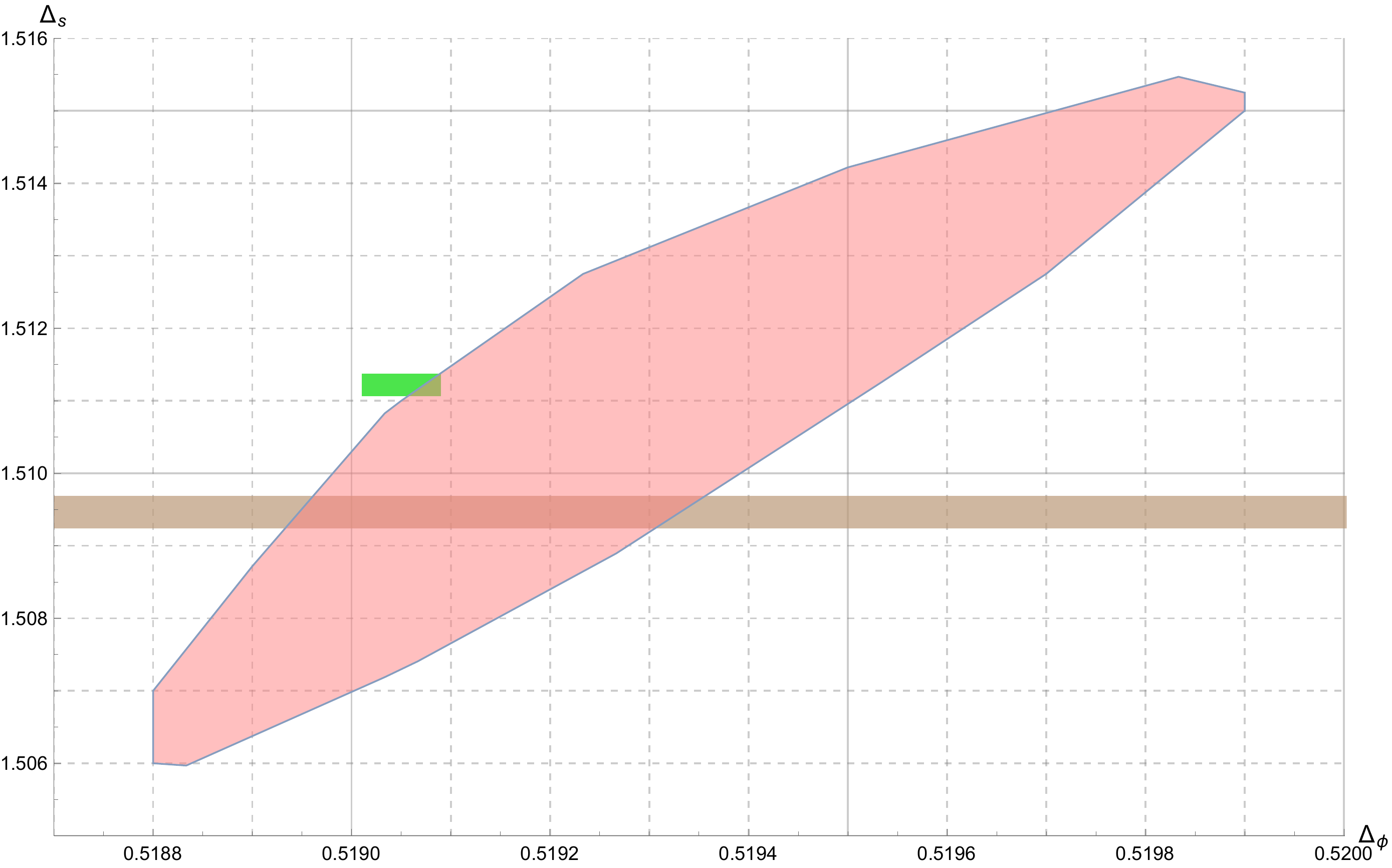}
	\caption{
		The red region is our $O(2)$ island in \autoref{fig:o2-3d} projected on $(\Delta_\phi,\Delta_s)$ space,
		the brown band shows the \ce{^{4}He} data \cite{Lipa:2003},
		and the green rectangle is the results of the Monte Carlo simulations reported by \cite{hasenbusch2019monte}.
	}
	\label{fig:o2-proj}
\end{figure}

In \autoref{fig:o2-proj}, our $\Lambda=25$ island projected on $(\Delta_\phi,\Delta_s)$ is shown
with results from \ce{^{4}He} experiments \cite{Lipa:2003} and the Monte Carlo (MC) studies \cite{hasenbusch2019monte}.
The discrepancy of $\Delta_s$ between experiments and MC is about $8\sigma$ and shows that at least one of them is wrong.
Our rigorous result is not sufficient to determine which is correct, but
a recent study \cite{Chester:2019ifh} revealed smaller island which is consistent only with the MC result:
\begin{align}
	1.51114 &\le \Delta_s \le 1.51158, \\
	0.519066 &\le \Delta_\phi \le 0.519110, \\
	1.23618 &\le \Delta_t \le 1.23640.
\end{align}

\newpage

\section{Conclusion}
\label{sec:conclusion}

As discussed in \autoref{sec:intro}, the conformal bootstrap is a recent method to study the critical points
and compared with previous methods such as the Monte Carlo simulation or $\epsilon$-expansion,
its starting point is not near but just at the critical point.
The main purpose in this thesis is to refine rigorous methods of the conformal bootstrap to study or classify critical phenomena.
For this purpose, we reconsidered each steps in the whole process of the conformal bootstrap
and for each step, developed a generalized package or method faster than existing methods.

We introduced three methods in this thesis: \code{autoboot}, \code{qboot} and hot-starting.
In steps in \autoref{fig:libraries}:
\begin{enumerate}
	\item physical assumption on the CFT,
	\item write bootstrap equations (by \code{autoboot}),
	\item create to a SDP (by \code{PyCFTBoot}, \code{cboot}, \code{qboot}, ...),
	\item solve the SDP (by \code{SDPA}, \code{SDPB}),
\end{enumerate}
our \code{autoboot} and \code{qboot} help the second and third steps,
and hot-starting can be a trick for \code{SDPB} in the last step.
The hot-starting was already used in a recent paper \cite{Chester:2019ifh} after our paper \cite{autoboot},
in which a small island for the $O(2)$ vector model was shown.

Examples in \autoref{sec:examples} shows the correctness of our libraries
and also gives a new method to get rigorous results about irrelevant spectrum, allowed by \code{qboot}.
Irrelevant operators can be measured as a correction to the power law,
for example, the correlation length in the Ising universality class behaves as
\begin{equation}
	\xi \propto \abs{t}^{-\nu}\left(1+b\abs{t}^{\omega\nu}+\cdots\right),
\end{equation}
where $\omega$ corresponds to the next $\mathbb{Z}_2$-even primary $\epsilon'$ as $\Delta_{\epsilon'}=3+\omega$.
The MC simulation also gives predictions on irrelevant spectrum
such as $\omega$, $\omega_A=\Delta_{\sigma'}-3$ (the exponent of the leading $\mathbb{Z}_2$-odd correction) \cite{Pelissetto:2000ek}:
\begin{equation}
	\Delta_{\epsilon'}\approx 3.84(4),\quad \Delta_{\sigma'}\gtrsim 4.5,
\end{equation}
while our results in \autoref{fig:gap-e}, \autoref{fig:gap-s} gives a rigorous upper bound:
\begin{equation}
	\Delta_{\epsilon'}<3.83,\quad \Delta_{\sigma'}<5.3.
\end{equation}
These values also calculated by the EFM in \cite{Simmons-Duffin:2016wlq} as shown in \autoref{tb:ising_efm},
but with non-rigorous errors.

Now we discuss the future directions.
\begin{itemize}
	\item \code{autoboot} can be generalized to all classical lie algebras.
	\item Rewriting \code{autoboot} in \code{C++} allows us to reduce the runtime of \code{NullSpace}, which is a function of \code{Mathematica}
	and one of the most heavy tasks in \code{autoboot}, and to combine with \code{qboot} into a self-contained SDP generator with global symmetry.
	\item The EFM and our method discussed in \autoref{sec:example_ising} can be applied to a general bootstrap equations to
	estimate the (irrelevant) spectrum of a CFT. A natural task is to study the relation between two methods with numerical results.
	\item Using our method in \autoref{sec:example_ising}, we obtained the finite number of spectrum of the Ising model
	in the $\mathbb{Z}_2$-even primary scalars.
	The actual spectrum has infinite scalars, but these finite number of operators is expected to
	`solve' the bootstrap equations approximately.
	This finite spectrum is a good start point for the `truncation method' introduced in \cite{Gliozzi:2013ysa}.
\end{itemize}
The conformal bootstrap is a non-perturbative method to investigate the fixed point of the RG flow.
Its effectiveness in higher dimensions ($d>2$) has been established in piles of recent studies as discussed in \autoref{sec:intro_gen},
and we hope that our methods in this thesis will help next studies to be stacked.

\section*{Acknowledgement}
The author is grateful to his supervisor Prof.~Yuji Tachikawa for introducing the author to the field of the conformal bootstrap,
many advises with clairvoyance and the collaborative work \cite{autoboot} essential to this thesis.
The author also thanks Shai Chester, Walter Landry, David Simmons-Duffin, and Alessandro Vichi for helpful comments to write \cite{autoboot}.
All computation to get the exclusion plots in this thesis was executed on clusters
in the Kavli Institute for the Physics and Mathematics of the Universe.
The author would like to express his sincere gratitude to his family, colleagues and friends.
The author is partially supported by the Leading Graduate Course for Frontiers of Mathematical Sciences and Physics.

% \appendix

\bibliographystyle{ytphys}
\baselineskip=.95\baselineskip
\bibliography{ref}

\end{document}